\documentclass[12pt]{aastex631}
\usepackage{gensymb}
\usepackage{textcomp}
\usepackage{booktabs}
\usepackage{rotating}
\usepackage{multirow}
\usepackage{xcolor}
\usepackage{longtable}
\usepackage{tabularx}
\usepackage{hyperref}
\usepackage{multirow}
\usepackage{graphicx}
\usepackage{amsmath}
\usepackage[rightcaption]{sidecap}

\usepackage[utf8]{inputenc}
\usepackage[T1]{fontenc}

\DeclareUnicodeCharacter{2212}{-}

\newcommand{\modify}[1]{{\color{red}#1}}

\def\mum{\hbox{\,$\mu$m}}
\def\massec{\hbox{\,mas s$^{-1}$}}
\def\etal.{\hbox{\textit{et al.}}}

\def\ie{\hbox{\textit{i.e.}}}

\def\degr{\hbox{$^\circ$}}
\def\sun{\hbox{$\odot$}}
\def\arcsec{\hbox{$^{\prime\prime}$}}
\def\arcmin{\hbox{$^{\prime}$}}
\def\gk{2015~GK$_{56}$}

\newcommand{\calints}{\texttt{calints}}
\newcommand{\kbmod}{\texttt{kbmod}}
\newcommand{\pkbmod}{\texttt{pkbmod}}
\newcommand{\au}{au}   

\newcommand{\likeli}{\mathcal{L}}

\begin{document}

\renewcommand{\arraystretch}{1.2} 

\title{The Luminosity Function of Ultra-Faint Trans-Neptunian Objects Detected by JWST}

\author[0000-0002-0760-1584]{Marielle R. Eduardo}
\affiliation{Department of Physics and Astronomy, University of Victoria, 3800 Finnerty Road, Victoria, BC V8P 5C2, Canada}

\author[0000-0001-5932-9570]{Anastasia N. Morgan}
\affiliation{Department of Astronomy and Planetary Science, Northern Arizona University, Flagstaff, AZ 86011, USA}

\author[0000-0001-6680-6558]{Wesley C. Fraser}
\affiliation{Department of Physics and Astronomy, University of Victoria, 3800 Finnerty Road, Victoria, BC V8P 5C2, Canada}
\affiliation{National Research Council of Canada, Herzberg Astronomy and Astrophysics Research Centre, 5071 W. Saanich Rd. Victoria, BC, V9E
2E7, Canada}

\author[0000-0003-4580-3790]{David E. Trilling}
\affiliation{Department of Astronomy and Planetary Science, Northern Arizona University, Flagstaff, AZ 86011, USA}

\author[0000-0002-8613-8259]{Gary M. Bernstein}
\affiliation{Department of Physics and Astronomy, University of Pennsylvania, Philadelphia, PA 19104, USA}

\author[0000-0003-2434-5225]{John A. Stansberry}
\affiliation{Department of Astronomy and Planetary Science, Northern Arizona University, Flagstaff, AZ 86011, USA}
\affiliation{Space Telescope Science Institute, 3700 San Martin Drive, Baltimore, MD 21218, USA}

\author[0000-0002-6875-1543]{Bryan Hilbert}
\affiliation{Space Telescope Science Institute, 3700 San Martin Drive, Baltimore, MD 21218, USA}

\author[0000-0002-1139-4880]{Matthew J. Holman}
\affiliation{Center for Astrophysics | Harvard \& Smithsonian, 60 Garden Street, Cambridge, MA 02138, USA}

\author[0000-0002-8296-6540]{William M. Grundy}
\affiliation{Department of Astronomy and Planetary Science, Northern Arizona University, Flagstaff, AZ 86011, USA}

\author[0009-0001-3858-2507]{Thomas L. Storer}
\affiliation{Department of Physics and Astronomy, University of Waterloo, 200 University Avenue West, Waterloo, ON N2L 3G1, Canada}

\author[0000-0003-4827-5049]{Kevin J. Napier}
\affiliation{Center for Astrophysics | Harvard \& Smithsonian, 60 Garden Street, Cambridge, MA 02138, USA}
\affiliation{Department of Physics, University of Michigan, Ann Arbor, MI 48109, USA}



\begin{abstract}
We present a definitive discovery of 27 trans-Neptunian objects (TNOs) using the Near-Infrared Camera (NIRCam) aboard the James Webb Space Telescope (JWST). By employing a shift-and-stack technique and a machine learning network geared specifically to identifying false-positive detections in JWST images produced through the shift-and-stack process, we achieved a 40\% detection threshold of $m_{F150W2}=28.8$ mag (corresponding to $m_r\sim29.8$ mag) across a sky area of $0.05 \ \text{deg}^2$. This marks the deepest Solar System survey to date, reaching magnitudes that allow us to explore never-before-seen regions of the TNO size distribution. Our faintest detection has $m_{F150W2}=29.3$ mag and diameter of $\sim10$ km (assuming 15\% albedo). Within our sample, we find that both the Cold and Hot TNO subpopulations exhibit a power-law slope. The distribution of apparent magnitudes of our nominal sample (detections at all epochs) are well fit by a single power law $dN/dm \propto 10^{\alpha m}$ with $\alpha=0.29^{+0.08}_{-0.07}$. The dynamically hot and cold subsamples in our discovery set are consistent with the same power law, suggesting that the planetesimal formation process yields similar slopes despite the differing disk conditions at the presumed $\sim25$ and $\sim45$~au formation regions of the two populations. 


\end{abstract}



\section{Introduction} \label{sec:intro}

The discovery three decades ago \citep{jewitt1993discovery} of the second Trans-Neptunian Object, or TNO\footnote{Often used interchangeably with Kuiper Belt Object, or KBO.}---after Pluto---opened an entirely new window into the origin and evolutionary history of our Solar System. These distant, icy planetesimals beyond the orbit of Neptune have undergone minimal change over the last 4 billion years and therefore serve as natural laboratories for studying early Solar System processes and testing models of planet formation. Dynamically, TNOs can be broadly divided into two categories: the quiescent ``cold'' classicals (CCs), which reside on low-inclination, nearly circular orbits, and the excited ``hot'' population (HTNOs), which instead occupy more eccentric, more highly inclined trajectories.

The size distribution (SD) of TNOs, in particular the cold classicals, which can be inferred from their luminosity function (LF), provides key constraints on the conditions of planetesimal formation as well as the evolutionary processes that shaped them \citep{gladman2001structure, bernstein2004size, fraser2014absolute,fraser_derivation_2008,fraser_kuiper_2008,fraser2010luminosity,kavelaars2021ossos,petit_hot_2023,napier2024decam}. Comparing the observed SD with those predicted by planet formation models helps to constrain the proposed physical processes and underlying initial conditions that shaped the current Solar System. However, the relative faintness and the typical $35-50$~au distances of TNOs limit ground-based surveys to about $m_r \sim 27$ mag ($D \sim 50$~km) \citep{fraser_size_2009, fuentes2009subaru}. The most sensitive search to date was carried out with the Hubble Space Telescope (HST) and reached $m_r \sim 28$ mag ($D \sim 30$ km) \citep{bernstein2004size}. This HST program covered $0.02\,\text{deg}^2$ and detected only three objects, far fewer than the $\sim85$ objects expected from extrapolations of the bright end of the luminosity function. More recently, an HST search for potential extended New Horizons mission targets identified only two objects of diameters $ \sim 40$ km \citep{buie_new_2024}, consistent with extrapolations from \cite{fraser2014absolute} for that survey area of 0.3 square degrees. With so few detections, our knowledge of the TNO SD at small sizes remain poorly constrained, and consequently, our understanding of planetesimal formation and growth in the outer Solar System remains limited.

The scenario of a streaming instability followed by pebble cloud collapse (SI+PCC) has recently emerged as the leading model for  planetesimal formation. In this model, aerodynamic interactions between gas and dust concentrate pebble-sized bodies into dense, self-gravitating clouds, which then collapse rapidly to form planetesimals on the order of $\sim 100$ km in size \citep{youdin2005streaming, schafer2017initial, johansen_forming_2017, nesvorny_binary_2021}. Several lines of evidence support this model, including the exceptionally low bulk densities observed among medium- to large-sized TNOs ($\lesssim1000$ km; \citealt{grundy_mutual_2019, mao_collisions_2021, keane_geophysical_2022}), the high binary fraction across most TNO populations \citep{noll_chapter_2020}, the compositional similarities frequently found between components of binary systems \citep{benecchi_correlated_2009, marsset_col-ossos_2020}, and dynamical studies showing that the SI+PCC mechanism naturally produces a binary orbital distribution consistent with observations \citep{nesvorny_trans-neptunian_2019, grundy_mutual_2019}. Moreover, most recently, \cite{kavelaars2021ossos}, \cite{petit_hot_2023}, and \cite{napier2024decam} found that the SD of TNOs is consistent with an exponentially tapered power law predicted by the SI+PCC mechanism \citep{abod2019mass, li2019demographics,schafer2017initial, simon_mass_2016}. However, \cite{kavelaars2021ossos} and \cite{petit_hot_2023} only reached magnitudes as faint as $H_r \sim 8.3$ ($\sim 75$ km), while \cite{napier2024decam} extended to $H_r \sim 10.5$ ($\sim 27$ km), leaving the faint-end consistency of the SI+PCC model still uncertain.


We conducted a TNO-dedicated search with NIRCam aboard JWST, covering $0.05 \deg^2$ of the sky and achieved a limiting magnitude of $m_{F150W2}=28.8$ mag to explore the smaller end of the TNO size distribution. In this paper, we begin by presenting the observational design in Section \ref{sec:obsstrategy}, detailing the strategies implemented to maximize our detections. Section \ref{sec:dataprocessing} describes the data reduction and processing pipeline, which includes generation and injection of fake sources, image alignment, and subtraction of the sidereal sky. Our methodology for identifying TNO candidates and fitting their orbits is presented in Sections \ref{sec:tnosearch} and \ref{sec:OrbitSelection}, where we explain the tools and criteria used to confidently identify TNOs across multiple epochs. Section \ref{sec:OrbitSelection} summarizes our results, detailing the linking of candidates across three and two epochs, their derived orbital elements and magnitudes, and additional single-epoch detections. In Sections \ref{sec:LF}, \ref{sec:H_dist}, and \ref{sec:MassOfKB}, we turn to the broader implications of our detections. Section \ref{sec:LF} presents our analysis of the luminosity function, identifying the model that best describes our TNO detections. In Section \ref{sec:H_dist}, we present the cumulative absolute magnitude $H$ distribution of the Cold Classical population and compare our results with predictions from SI+PCC models. Section \ref{sec:MassOfKB} then provides an updated estimate of the total mass of the Kuiper Belt. Finally, in Section \ref{sec:DiscussionsAndConclusions}, we place our findings in the context of planetesimal formation and Solar System evolution models, discussing how our results inform our understanding of the formation and evolution of the cold and hot classical populations, as well as the Kuiper Belt as a whole.


\section{Observing Strategy} \label{sec:obsstrategy}
The observing strategy balances several interrelated factors: the area to be surveyed, on-sky location of the survey, observing cadence, integration time, exposure time per epoch, number/timing of epochs, and filter choice. In combination these determined the limiting magnitude (the proxy for minimum TNO size), expected number of detections, and accuracy of the orbit determinations. 

The NIRCam instrument \citep{rieke2023performance}, with its fairly large $8.9$~arcmin$^2$ instantaneous field of view (FOV) and exceptional sensitivity, was the natural choice for the survey. NIRCam has two identical halves (modules), each consisting of a short-wavelength (SW, 0.6 -- 2.35\mum) and a long-wavelength (LW, 2.4 -- 5.0\mum) channel that simultaneously image the same FOV. The FOVs of the two modules are separated by about 40\arcsec. The SW channel in each module is populated with four $2048^2$ HgCdTe detectors with 4\arcsec\ gaps between them, while the LW channel in each module consists of one such detector. The pixel scales in the SW and LW channels are 31 and 64 mas, respectively. The rectangle bounding the FOVs of the two modules is $4.95\arcmin \times 2.14\arcmin$, with the long axis perpendicular to the ecliptic due to roll restrictions imposed by the need to keep the telescope and instruments behind the sun shield.

To maximize sensitivity we selected the F150W2 and F322W2 filters in the short-wavelength and long-wavelength channels. The bandwidth ($\delta \lambda / \lambda$) of these filters is 0.82 and 0.42 respectively, are the widest and hence most sensitive available in each channel. However, each provides nearly diffraction-limited Full-width Half Max (FWHM) Point Spread Functions (PSFs) in each channel. For an initial reconnaisance of the smallest objects ever detected in the trans-Neptunian region, they are optimal. 

\begin{figure}[b]
    \centering
    \includegraphics[width=\textwidth]{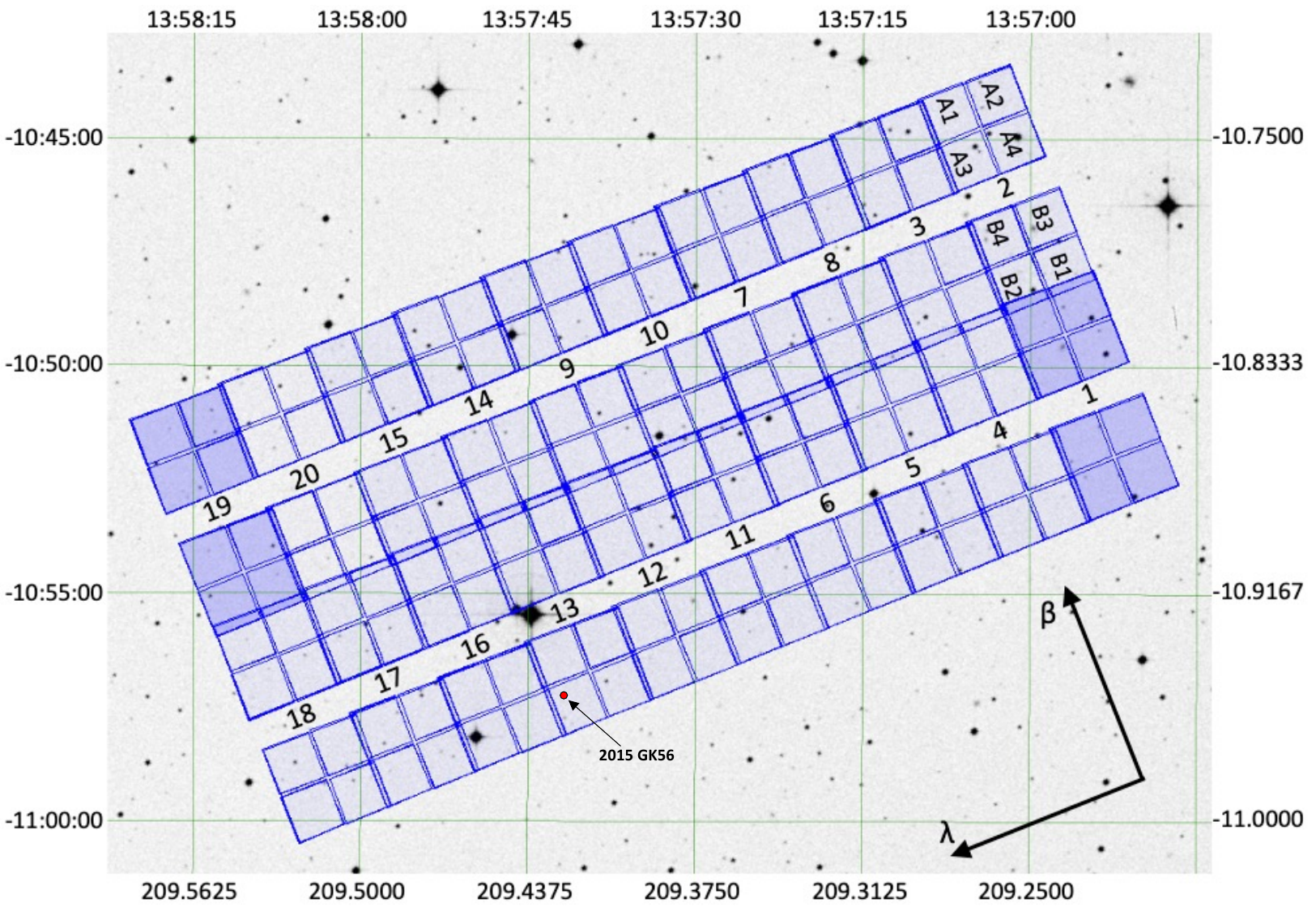}
    \caption{The mosaic used for the observations consisted of 20 tiles, each tile consisting of the footprints of the 8 NIRCam short-wavelength detectors (and the two co-aligned long-wavelength detectors). Darker blue shading shows the footprints of single tiles at left and right ends of the mosaic. The relative positions of the detectors in the A and B modules are shown for the upper-right tile of the mosaic, and highlight the 40\arcsec\ gap between the two modules. Numbers from 1--20 in the mosaic gaps label the tile ordering, equivalent to the JWST visit numbers. The sequence of tiles was repeated in 3 epochs separated by $\approx5$~days. Red circle in Tile 13 is the location of a known TNO, 2015 GK56, which was purposefully included in the survey footprint to evaluate recovery performance. The coordinate labels are in the ICRS equatorial system, with the orientation of the ecliptic coordinate system is shown by the vectors at lower right. The background image is from the Digital Sky Survey.}
    \label{fig:mosaic}
\end{figure}

The observations used sidereal tracking of a field centered at (RA, Dec) of (13:57:33, -10:51:55), which is on the Solar System's invariable plane (1.1\degr\ north of the ecliptic plane at that ecliptic longitude), and in a region of very low stellar density. The number of TNOs to be discovered depends on the area covered by the survey, the true magnitude (or size) distribution of the TNOs (which is not known!), and the detection limit. Our modeling led us to select a mosaic of 20 tiles of the NIRCam FOV as the optimal trade between area and depth for a nominal prediction of the magnitude distribution. Figure \ref{fig:mosaic} illustrates the 20-tile mosaic, and the coverage of the individual detectors for one tile. The full mosaic was observed at each of 3 epochs, with about 5 days between epochs in order to constrain the orbital parameters, as described later. Observations were taken from 24 Jan 2023 through 4 Feb 2023.

Because of the sidereal tracking, the TNOs we sought to discover were subject to trailing in the images. The apparent rate of motion of TNOs in the classical Kuiper belt at 45 au is $\le 0.12$\massec\ ($0.4\arcsec/$hr) when viewed from JWST within $\pm 5\deg$ of $98\deg$ solar elongation (appropriate for our observations, and is about $3\times$ lower as they transition between apparent prograde and retrograde motion. A TNO moving at 0.1\massec\ crosses a 31~mas SW pixel in 310 sec, and moves by the 45~mas full-width at half maximum (FWHM) of the F150W2 point spread function (PSF) in 450 sec. To avoid significant image trailing we used 204 sec integrations, resulting in trailing of $\le 0.7$ SW pixels. Motion of a bright TNO's flux into or out of a pixel during an integration produces non-linear ramps of counts vs time, which can be incorrectly flagged as cosmic ray (CR) hits during science pipeline processing. Our choice of integration time is short enough that only TNOs both brighter and faster than our regime of interest would be flagged as CR hits, allowing us to run the pipeline with CR correction enabled. 

We used exposures with 3 integrations, providing an exposure time of 612~sec (reset overheads mean each such exposure takes 639~sec). Two exposures dithered (by 3\arcsec) were taken at each epoch. Including the time required to execute the dither, the time between the mid-times of the first integration of the 1st exposure and the third integration of the second exposure is 1156~sec, sufficient for a typical TNO's apparent motion to be $\sim60$~mas, \ie\ 2 SW pixels or $1.3\times$ the PSF FWHM. This is easily detectable and yields a useful measure of the rate of motion of each TNO.  One "epoch" consists of a 2-exposure visit to each of the 20 tiles. At a given epoch all 6 integrations were acquired consecutively for a particular mosaic tile, but  observations of adjacent tiles were not always contiguous in time (see Figure~\ref{fig:mosaic}).

\begin{figure}[h!]
    \centering
    \includegraphics[width=0.8\textwidth]{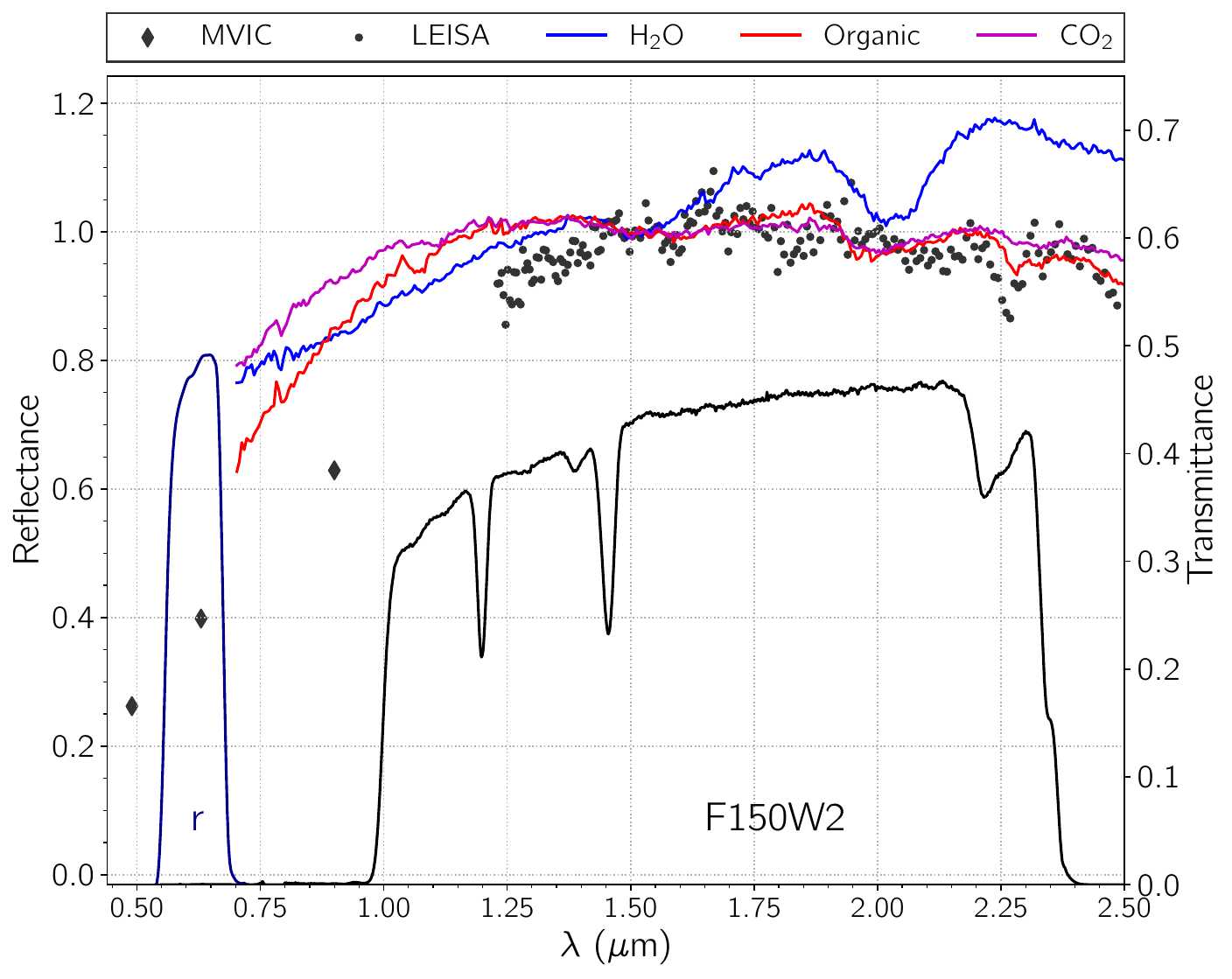}
    \caption{Median reflectance spectra of the Organic, CO$_2$, and H$_2$O surface types identified in the JWST NIRSpec sample. The LEISA and MVIC I/F reflectance data for Arrokoth are shown in black points. All spectra are normalized to $1.5 \ \mu m$. The transmission band passes of the r-band and F150W2 filters are shown in blue and black respectively.}
    \label{fig:AlbedoSpec}
\end{figure}

Taking the visible to near-IR spectrum of Arrokoth from New Horizons spacecraft \citep{grundy2020color} as our template for the reflectance of other small cold classical TNOs (see Figure \ref{fig:AlbedoSpec}), we used the JWST Exposure Time Calculator (ETC) and determined that the ultra wide (1.0 -- 2.35\mum) F150W2 filter gave significantly better sensitivity than any other choice in the SW channel. At the time, the spectra of TNOs (other than Pluto) at $\lambda\gtrsim2.2$ were unknown, so we chose the very wide (2.4 -- 4.0\mum) F322W2 filter in the LW channel. 

To estimate the limiting magnitude of our search, we used the solar-convolved spectrum of Arrokoth and the JWST ETC. The expected $V$-band (Vega) absolute magnitude of a 10~km diameter object with a V-band geometric albedo of 0.15 is 12.7. At a distance of 45~au from JWST and from the Sun, the apparent magnitude of such a TNO is 29.4 mag. In the ETC we normalized the spectrum described above to 29.4 mag at V, and defined the exposure parameters (also discussed above) using 10 groups of the BRIGHT1 exposure pattern, 3 integrations per exposure, and 2 (dithered) exposures. The total effective exposure time per observation is then 1267 sec, and predicted signal-to-noise ratio (SNR) was 11 in the F150W2 filter. As of ETC version 4.0, released in July 2024, that predicted SNR is somewhat lower at 8.4. The expected SNR in a single integration is $2.4\times$ lower (4.4 in the original predictions, 3.5 using ETC 4.0).

The mosaic observation was repeated at three different epochs to allow us to measure the objects' intrinsic velocities as well as their reflex acceleration caused by JWST's orbital motion. We  wanted the total range of apparent motion for TNOs as close as 30 au to be well below the $\approx20\arcmin$ extent of the observed field in the ecliptic direction, so that only a small fraction of detected TNOs will fall off the field during observations.  The apparent ecliptic-plane position of a body at 30 au moves roughly $0.17\arcmin (\Delta t/1\text{day})^2$ from the turnaround point, where $\Delta t$ is the time from turnaround.  To keep this motion to $\approx 1\arcmin$ or less requires $\Delta t<7$~days. Based on these considerations we spaced 3 mosaic observations by $5\pm1$ days, and timed them to occur near quadrature, at $\theta_\text{Sun} \simeq [95\degree, 100\degree, 105\degree]$ ($\pm1\degree$). The 3 sightings we expected for each object discovered would then allow us to determine distance to $\ll1$~au and inclination $i$ to $\ll 1$\degr), and set useful limits on $a$ and $e$ for brighter TNOs. This predicted accuracy is borne out in our results (Section \ref{sec:elements}). We note that given the anticipated few-mas precision of the JWST astrometry, a 10-day arc provides a similar lever on a target’s ephemeris as does a full one-year arc of ground-based observations in typical seeing.

\section{Image Processing} \label{sec:dataprocessing}
    \subsection{Generation of Fake Sources} \label{sec:FakeSources}

To determine detection areas and thresholds, We generate approximately 10,000 synthetic objects that emulate the motion, lightcurve, and color characteristics of real TNOs, distributed within a radius of 0.8\degree\ centered on our observational footprint. Of these, about 900 objects fall within the mosaic area during at least one exposure, and were subsequently implanted into the raw JWST images (we discuss this in Section \ref{sec:ImplantationOfFakeSources}).

The synthetic TNOs are created using a combination of (1) realistic orbital elements informed by known objects, and (2) deliberately ``unrealistic'' distributions to explore hypothetical populations. For realistic objects, semi-major axis \emph{a} and eccentricity \emph{e} are based on data from the Minor Planet Center (MPC)\footnote{\url{https://minorplanetcenter.net}}. We start from the orbital elements of known TNOs and perturb each MPC $a$ and $e$ by adding a random zero-mean deviate to each---fractions of an au for $a$ and a few percent for $e$. 
For unrealistic objects, $a$ is drawn from a logarithmic distribution between 30 and 300 au, and $e$ is drawn from a uniform distribution between 0 and 1.

The absolute magnitudes $H$ for realistic objects are drawn from the power-law distribution given  by \citet{fraser2014absolute}. In contrast, unrealistic objects are assigned $H$ values from a uniform distribution, allowing sufficient statistics to determine the detection efficiency over a broader range of magnitudes. 

We calculate the synthetic objects' apparent sky positions at one-minute intervals throughout the observation period using NASA’s NAIF SPICE toolkit via the \texttt{spiceypy} Python interface
\citep{annex2020spiceypy}. Only objects falling within the mosaic footprint (plus a buffer) during the observation period are retained. 

After determining each object’s trajectory, we model its photometric variability by assigning a sinusoidal lightcurve. Rotation periods are drawn from a Maxwellian distribution spanning 2 to 20 hours, peak-to-peak amplitudes are selected based on the distributions described by \citet{thirouin2019} for cold classical TNOs, and rotational phases are assigned at random. The mean magnitude of each lightcurve is set by the assigned absolute magnitude $H$, and the apparent magnitude $m$ are then computed as a function of the object’s motion and rotational variability. This approach produces realistic TNO brightness variations and allows us to measure our survey sensitivity not only as a function of mean apparent magnitude but also as a function of lightcurve amplitude.

    \subsection{Implantation of fake sources} \label{sec:ImplantationOfFakeSources}

\begin{figure}
\centering
    \includegraphics[width=0.4\textwidth]{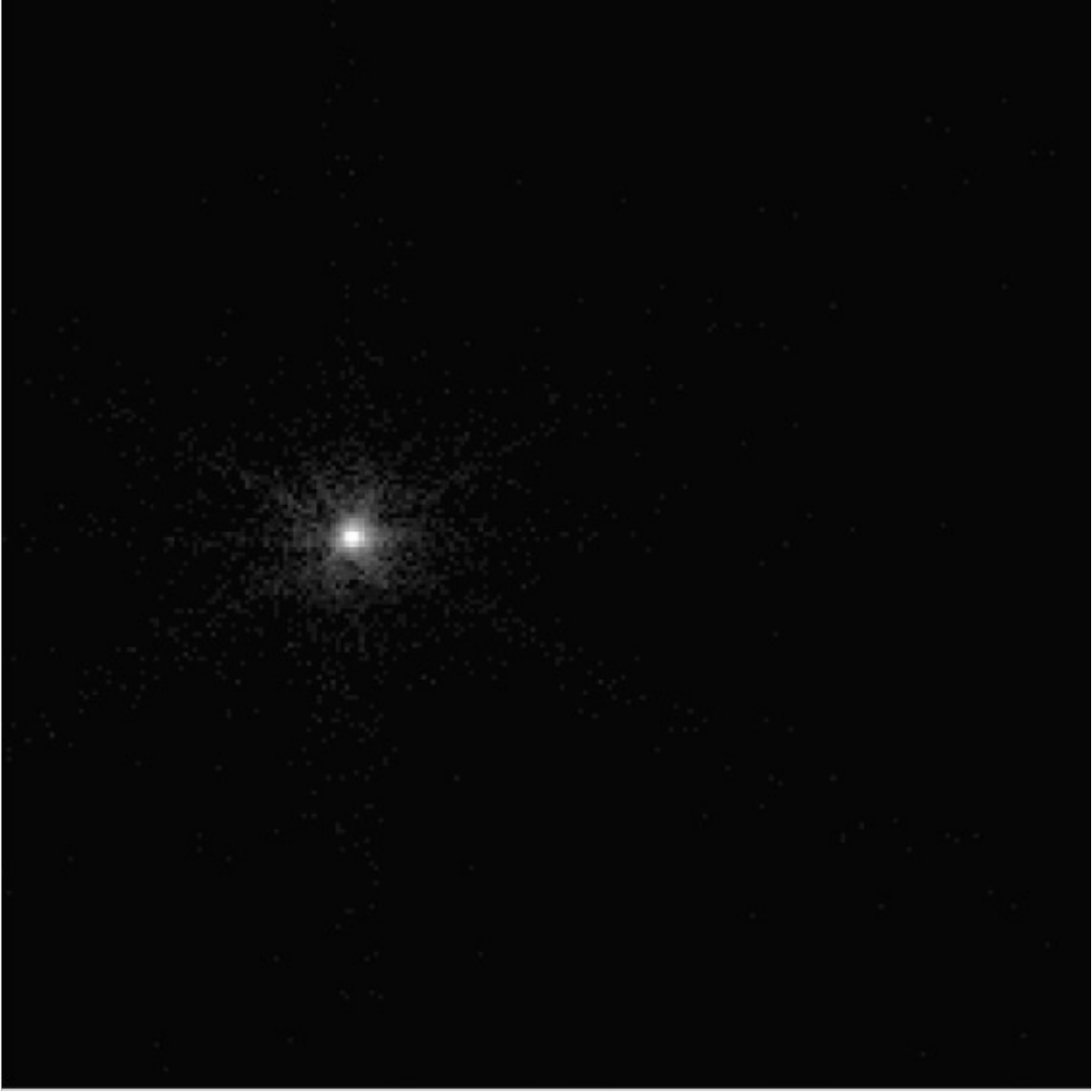}
    \includegraphics[width=0.4\textwidth]{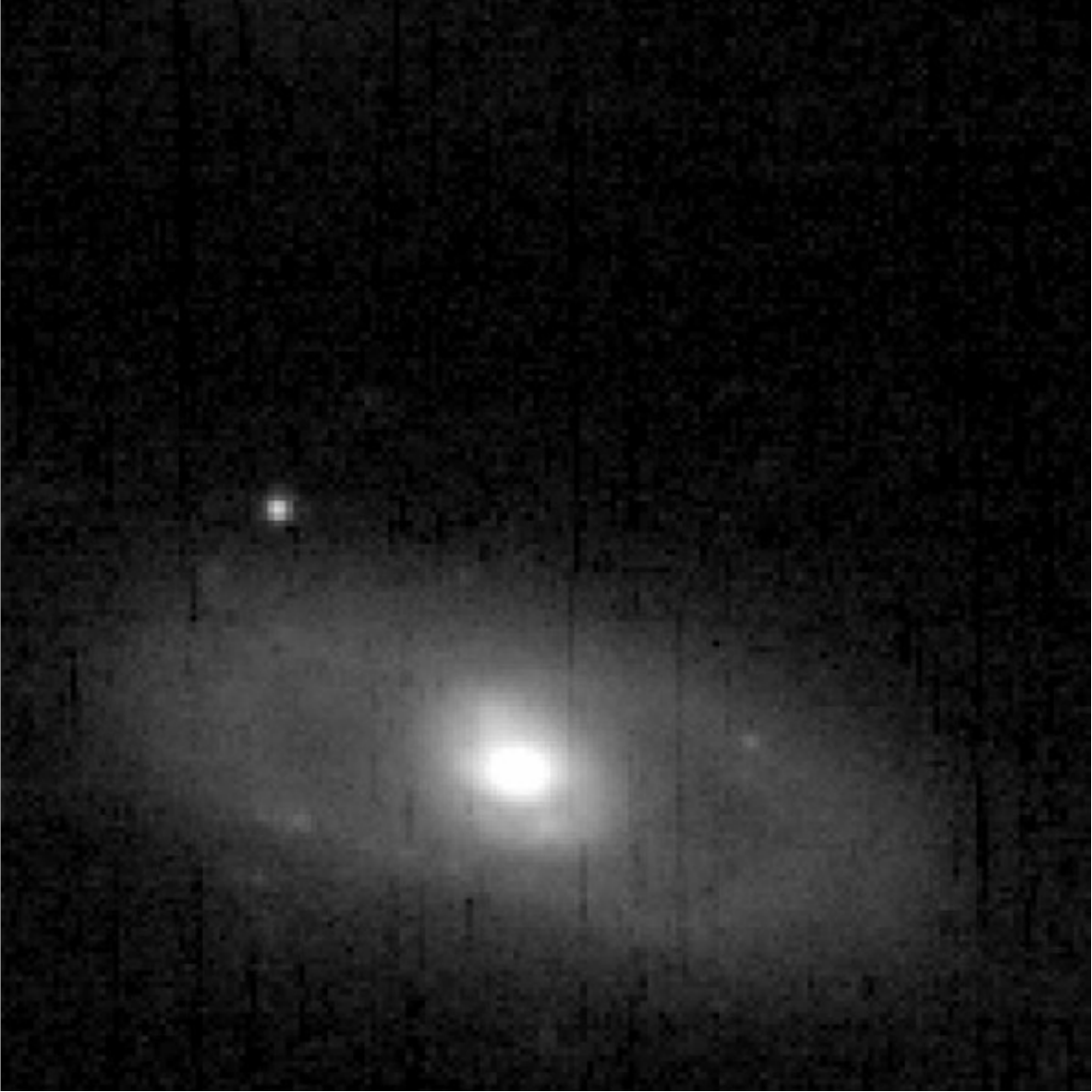}   
    \caption{\textbf{Left:} A cutout of a linearized seed image containing only the artificial TNO. \textbf{Right:} Image output after addition of the seed image to the linearized real JWST image.}
\label{fig:seed_add}
\end{figure}

The process of implantation starts by creating linearized seed images of fake sources with an image simulation software called \texttt{MiRaGe: Multi-Instrument RAmp GEnerator} \citep{hilbertspacetelescope}. Linearized seed images contain only the artificial TNO sources, and give readout signals up the integration ramp that are linear with time. The linearity correction, which compensates for non-linear detector response due to effects such as sensor saturation, is applied in the pipeline so that the linearized image arrays can be accurately added into the linearized real JWST images (see Figure \ref{fig:seed_add}). About 900 artificial sources with a  magnitude range of $m_{F150W2} = 21.5-30.5$ mag and speed of $0.03-8.66\arcsec /\text{hr}$ were implanted across JWST integrations. A typical NIRCam SW detector image contains 6-10 artificial sources.

Once the linearized seed images are added to the linearized real JWST MULTIACCUM integration data---which consist of non-destructive readouts taken throughout an exposure---we proceed to run the slope fitting step from Stage 1 of the JWST Calibration Pipeline. This step fits a line to the signal accumulation in each pixel, converting the raw integration ramps into 2D images of count rates (\ie, the measured signal rate per pixel). The result is a \texttt{rateints} file, containing one count-rate image for each of the three integration in the exposure. 

Next, we run Stage 2 of the JWST calibration pipeline (\texttt{calwebb\_image2}; \cite{bushouse2022jwst}) to create a fully calibrated (\texttt{calints}) image per integration. This stage applies the flat field and performs the flux calibration, such that the output data are in units of MJy/sr. The output products at this point are ``\texttt{calints}'' files, with each file containing all three integrations of an exposure at a single pointing.

    \subsection{Image alignment} 
\label{sec:alignment}
A major task is to map all of the exposures to a self-consistent World Coordinate System (WCS). JWST data use a generalized WCS (gWCS) \citep{dencheva2021spacetelescope}. To improve the astrometric alignment beyond the default MAST\footnote{\url{https://mast.stsci.edu}} products, we used a multi-step process involving several calls to the \texttt{Tweakreg} step of the JWST calibration pipeline \citep{bushouse2022jwst}. \texttt{Tweakreg} improves the WCS of individual images by matching and aligning common source positions between them.

We began by aligning the longwave (LW) module images, which requires fewer detector-to-detector adjustments since each module has only one detector, whereas shortwave (SW) modules have four detectors each. The initial alignment focuses on a subset of mosaic tiles (Tiles 10, 14, and 15) which show significant misalignments between epochs 2 and 3. Different guide stars were used in epoch 3 of these tiles compared to the first two epochs, so uncertainties in the absolute guide star locations propagate into differences in the WCS between the epochs. We first call \texttt{Tweakreg} individually on these tiles, in order to improve the epoch-to-epoch alignments.

After the data within each tile are well aligned, we proceed to align the tiles with respect to each other. 
We first perform a relative alignment of pairs of  tiles in the North/South direction of the $2\times10$ mosaic, \ie \ within a column (Figure  \ref{fig:mosaic}). The detector overlap between tiles is largest in this direction, allowing \texttt{Tweakreg} to identify 8--16 sources in each overlapping region. Aligning with these sources typically results in a reported RMS of 0.007--0.008\arcsec.

Once the tiles within each column have been aligned, we use a separate call to \texttt{Tweakreg} to align the columns to one another. With the tiles for a given column aligned, we effectively double the area of overlapping sky available when aligning adjacent columns. RMS values reported by \texttt{Tweakreg} are
generally 0.008--0.030\arcsec, based on the 15--20 sources in the overlap region between each column pair.

After completing the relative tile-to-tile alignment, we use \texttt{Tweakreg} to align the enitre LW mosaic to the GAIA DR3 catalog in a single calculation. \texttt{Tweakreg} reports an RMS of $0.03\arcsec$ for this step.

With the relative and absolute alignment of the LW mosaic  complete, we align the SW mosaic to the LW mosaic using a single call to \texttt{Tweakreg}  aligning the SW files to a source catalog from the aligned LW mosaic. For this task, \texttt{Tweakreg} identifies  130--195 sources per tile, and calculated RMS values of 0.01--0.03\arcsec\ per tile, when fitting for shift, rotation, and skew.

\subsection{Background Subtraction}

After completing the WCS alignment, we proceed to subtract the $1/f$ noise from the images using a code\footnote{\url{https://github.com/chriswillott/jwst/blob/master/image1overf.py}} specifically designed for processing and analyzing JWST data. In general, we subtract the $1/f$ noise independently from each of the four amplifiers in a given image. In some cases, however, large diffraction spikes or large extended sources result in a poor $1/f$ correction for a given amplifier. In these cases, we subtract the $1/f$ signal in all four amplifiers simultaneously. 

The next step was to subtract the “background” from each of the images, where in this case, “background” refers to all sidereal sources. After subtracting the background the only sources that should remain in the data are solar system objects and any sources with time-variable brightness. Working one tile at a time, we use the exposures from any two of the epochs to create the background image that will be subtracted from the third epoch’s data. In order to create the background images, we use Stage 3 of the JWST calibration pipeline \citep{bushouse2022jwst}. This stage of the pipeline is not designed to run on \texttt{calints} files. Therefore, we split the images from the three integrations within each \texttt{calints} file into three separate \texttt{cal} files. In a single call, we run the pipeline on all of the \texttt{cal} files from two of the epochs, being sure to turn off the \texttt{Tweakreg} step since the files were previously aligned, as discussed above. By including all files for two epochs, all moving targets will not be in the same location from one file to the next, and will therefore be flagged and ignored when the images are resampled onto the output pixel grid. The result is a “background” image containing only sidereal targets; all moving targets have been removed. 

\begin{figure}
\centering
    \includegraphics[width=0.4\textwidth]{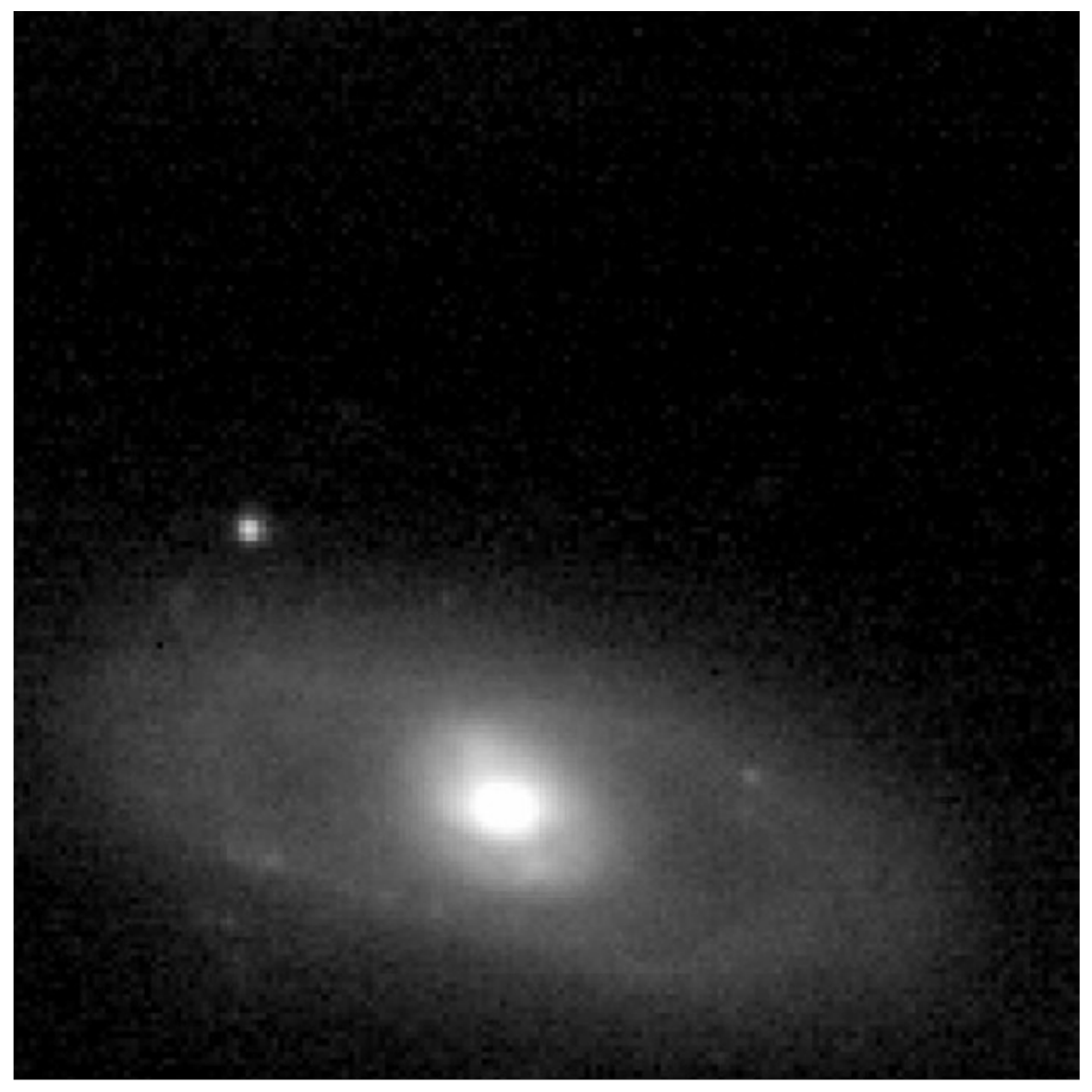}
    \includegraphics[width=0.4\textwidth]{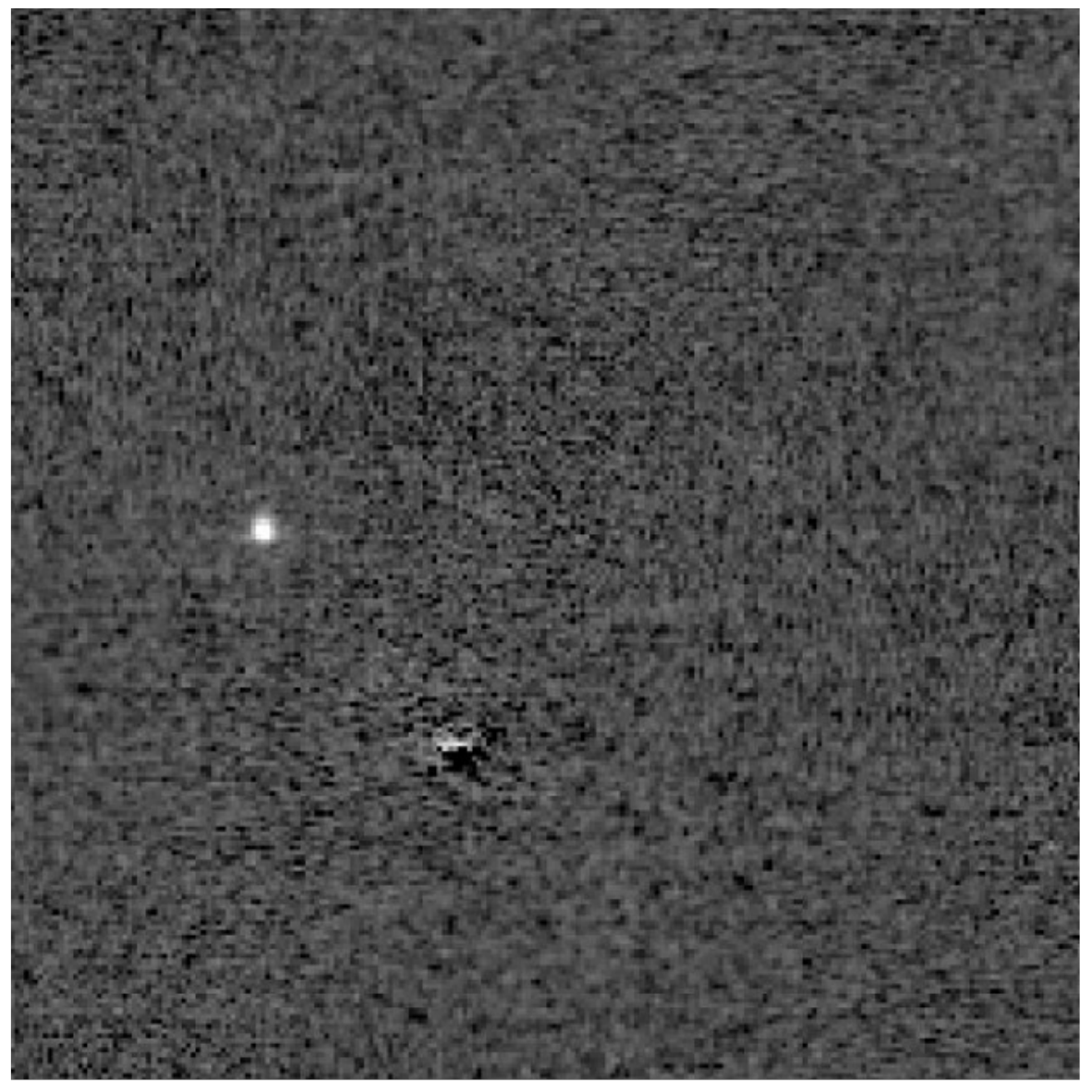}   
    \caption{\textbf{Left:} A cutout of a JWST image featuring moving point source next to a bright galaxy. \textbf{Right:} same image cutout after the subtraction process.}
\label{fig:subtraction}
\end{figure}

We then call the Stage 3 pipeline separately on each \texttt{cal} file from the third epoch. In this case, operating on a single file each time, no sources are flagged and ignored. As part of the pipeline call, we specify that the input image be resampled onto the same pixel grid as the background image described above. The background image is then subtracted from the resampled individual image, creating an image where sidereal targets with a constant signal have been removed, and only moving targets or time variable sources remain. We follow this recipe for all tiles, and also for each combination of epochs, resulting in a background-subtracted version of each integration for all tiles and epochs (see Figure \ref{fig:subtraction}).

\section{Search Process} \label{sec:tnosearch}
The search process in our survey is composed of five steps: (1) Shift-and-stack, (2) Machine learning vetting, (3) Single-epoch scene modeling, (4) Linking, and (5) Orbit refinement and selection. 
In the shift-and-stack stage, we perform an initial search across the image stacks of each single epoch to identify candidate objects exhibiting sky motions consistent with a TNO. The machine learning vetting step refines the candidate list by applying a binary classification of the shift-and-stack feature output as consistent with a TNO (real) vs a false positive (spurious). During scene modeling, we verify whether the image pixel data are consistent with a model of a moving point source superimposed on a static sky background. This step also provides optimal estimates of the source's flux and position as a function of time, along with their associated uncertainties. The linking step finds all combinations of one track from each epoch that are consistent with a single TNO orbit. Finally, the orbit refinement and selection determine the best-fit orbital parameters for each linked treatment by a joint fit to all three epochs' pixel data, producing best-fitting orbital parameters and fluxes, and goodness-of-fit measures that are used to select a sample with no false positives and high completeness.

\begin{figure}[h]
  \centering
  \includegraphics[width=\textwidth]{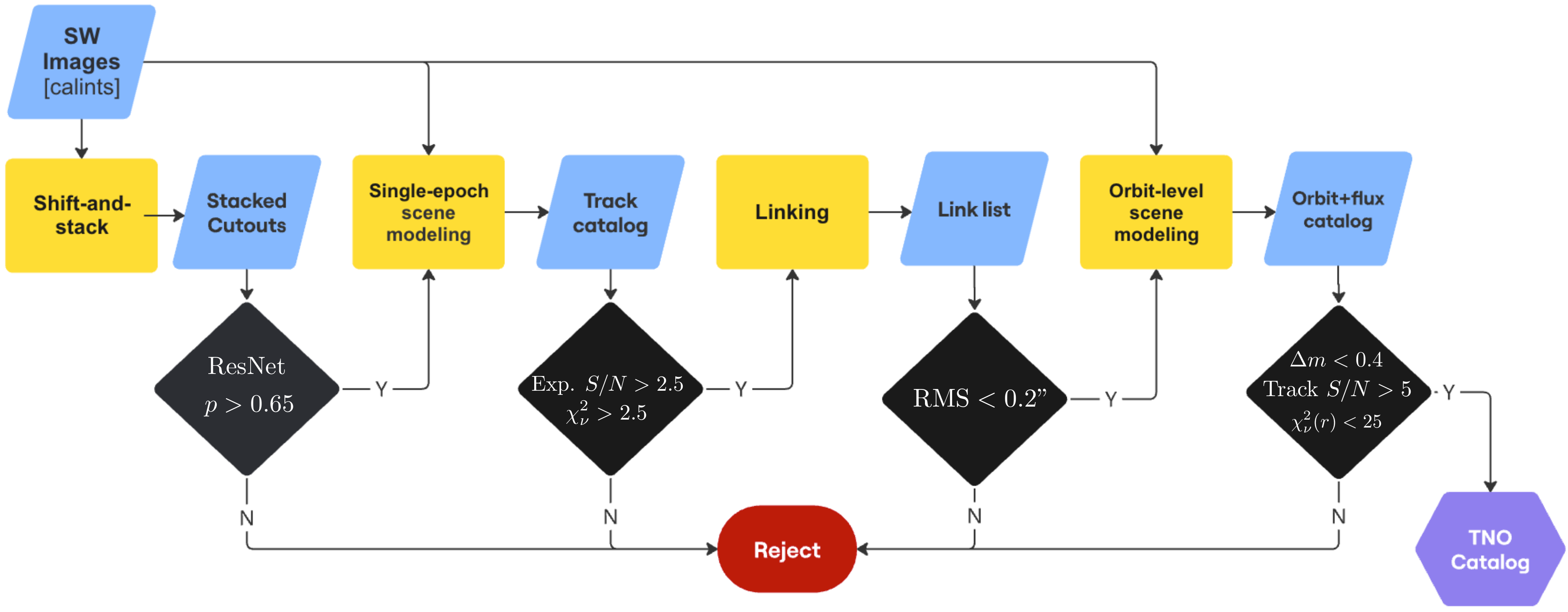}
  \caption{A chart depicting the processing workflow from calibrated and subtracted imagery to a set of detected KBOs of high purity. Black rectangles indicate stages where candidate sources (singular or linked multiple sources) are rejected. Major process stages and the resultant data products are depicted by the yellow rectangles and blue trapezoids, respectively.}
  \label{fig:search_diagram}
\end{figure}

A schematic diagram outlining the sequence and interrelation of these steps is presented in Figure \ref{fig:search_diagram}, and we discuss each of them in the following subsections.

   \subsection{Shift and Stack} \label{sec:KBMOD}
The implanted, subtracted, and pixel-aligned \calints~ files were searched using the digital tracking, or shift-and-stack. We followed the approach implemented in \kbmod~ \citep{Whidden2019}, but rewritten to allow for some improvements. We made use of the same likelihood function as in \citet{Whidden2019} and indeed that used in \citet{bernstein2004size}, but with use of the JWST NIRCam PSF model. The two \calints~ exposures were separated into their 3 separate integrations, preserving their variance planes. Mask planes were generated marking all bad pixels identified by the earlier processing, resulting in 6 separate images for the shift-and-stack.  

A grid of linear shifts was used in the search with the grid encompassing bound orbits outside $\sim15$~au, corresponding to rates of motion of $-2.6 \leq v_{\alpha} \leq 2.6 \mbox{"/hr}$ and $-1.3\leq v_{\beta} \leq 1.3 \mbox{"/hr}$ for the first epoch, and $-1.3 \leq v_{\alpha} \leq 1.3 \mbox{"/hr}$ and $-1.3\leq v_{\beta} \leq 1.3 \mbox{"/hr}$ for the second and third epochs. The grid spacing was $0.16 \mbox{"/hr}$, resulting in a maximum possible mis-shift of less than 0.5 pixels. The result of the shift-stack was a list of candidates with estimates of position and velocity vectors, and an estimate of source brightness, $B_{\textrm{ml}}$ determined from exact solution of the likelihood model \citep{Whidden2019}. 

After the initial shift-stack, any candidates with negative maximum likelihood fluxes were rejected. Candidates then underwent numerous filters to reject bogus sources. First, a brightness filter was applied whereby at a given detection location, the candidate's maximum likelihood estimated brightness value, $B_{\textrm{ml}}$  was varied from 0.5 to 1.5 times, sampled on a grid of 9 evenly spaced values. At each brightness, the likelihood was calculated exactly. If the maximum likelihood of those 9 different evaluations was either at the minimum or maximum of the grid, then the source was deemed to not behave as expected for a real moving source, and is rejected. Real moving sources show behavior like assumed in the likelihood model, and so have valid brightness estimates $F_{\textrm{ml}}$ and show a peak in likelihood near their true fluxes. Junk sources tend not to, and so can be removed by this filter. Increasing the density of the samples beyond 9 evaluations did not eliminate any additional candidates.


The remaining candidates underwent a so-called predictive filter which was designed to take advantage of the fact that the majority of good detections (real or implanted) were detected at more than one rate. Given a source detected at a grid rate that most closely matches its true rate of motion, the median position of that source at a discrepant rate can simply be estimated from the difference between the best grid rate discrepant rates, and the centroid of the source at the best grid rate. The predictive filter\footnote{This predictive filter is what gives the shift-stack software package its name; \pkbmod, which is available here \url{https://github.com/fraserw/pkbmod}} then considered all rates, searching for sources with lower SNR that are found near the predicted positions with a tolerance of 1.5 pixel radius, and any possible sources were collected and identified as a cluster of detections of a single source. Those detections were removed and the process repeated on the next remaining highest likelihood source, and so on until all candidates had been considered. Any source that could not be associated with a cluster was rejected from further consideration. While it is possible that the predictive filter could mis-associate valid detections of nearby clouds, this was a very rare occurrence because the probability of two objects sharing very similar positions and velocities that could trick the filter was extremely low; a manual check showed that this never occurred, even for our implant list. 

Finally, the peak SNR source in each cluster then underwent a filter that varied the position of a source to $\pm 3 $~pixels around its highest likelihood point, akin to the brightness variation filter discussed above. Then, any source with a likelihood greater than 5 was kept. These are the shift-stack candidates, and still have a bogus:good ratio of more than 100:1. 

   \subsection{Machine Learning Vetting} \label{sec:MLsearch}

While the shift-and-stack method provides a fast and efficient way of searching over moving object candidates across multiple images, it still requires a significant amount of human effort for visually vetting real moving detections, even after the image subtraction process. To reduce the number of such false positive outputs from shift-and-stack, we integrated machine learning techniques to perform an initial filtering of bad sources from the candidate list.

We explored different types of neural network architecture each trained to perform a binary classification task, determining whether a source should be accepted as good or rejected as bad. Our experiments included standard convolutional neural networks (CNNs) with a range of complexities and depths to evaluate how architectural choices influenced classification performance. While those models delivered reasonable results, we achieved a significant improvement in performance when using an ensemble of residual networks (ResNets; \cite{he2016deep}). The ResNet architecture we used here is similar to the one employed in \cite{fraser2024candidate}, which constructs a \textit{master} network made up of multiple ensembles (or \textit{branches}) of simpler networks. Our chosen ResNet model consistently outperformed all previous CNN-based models in terms of detection depth and peak detection efficiency, while also significantly reducing the number of trainable parameters. 

\begin{figure}[h!]
    \centering
    \includegraphics[width=\textwidth]{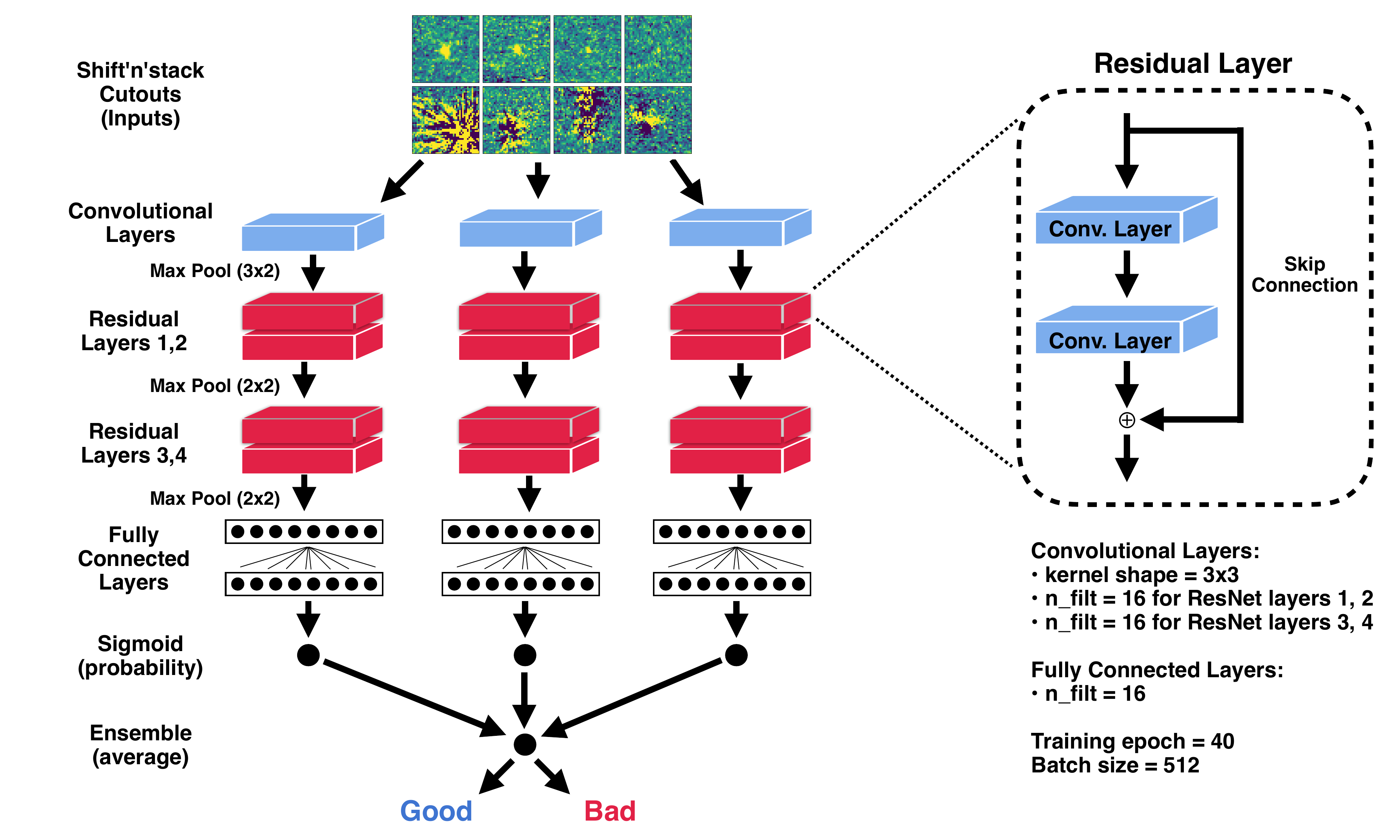}
    \caption{The network adopted for this project is an ensemble of three ResNets. Each \textit{branch} of the ensemble has four residual layers utilizing 16 filters, represented in red blocks. Each red block contains two convolutional layers. The outputs from the fourth residual layer are fed into two fully connected layers, which ultimately lead to a sigmoid probability layer. The sigmoid outputs from all three models are then averaged to generate a final probability score for binary classification. For a detailed explanation of ResNet operations, we refer the reader to \cite{he2016deep}.}
    \label{fig:Resnet}
\end{figure}

Our adopted ensemble consists of three identical multilayer ResNets (see Figure \ref{fig:Resnet}), each independently initialized and trained on the same dataset. By combining multiple models into one ensemble, we improve the network's predictive performance, robustness, and accuracy. This approach is based on the idea that a group of models working together can often make better decisions or predictions than a single model. The final prediction probability of the ensemble is calculated by averaging the output probabilities from each individual ResNet.

\begin{figure}[h!]
    \centering
    \includegraphics[width=\textwidth]{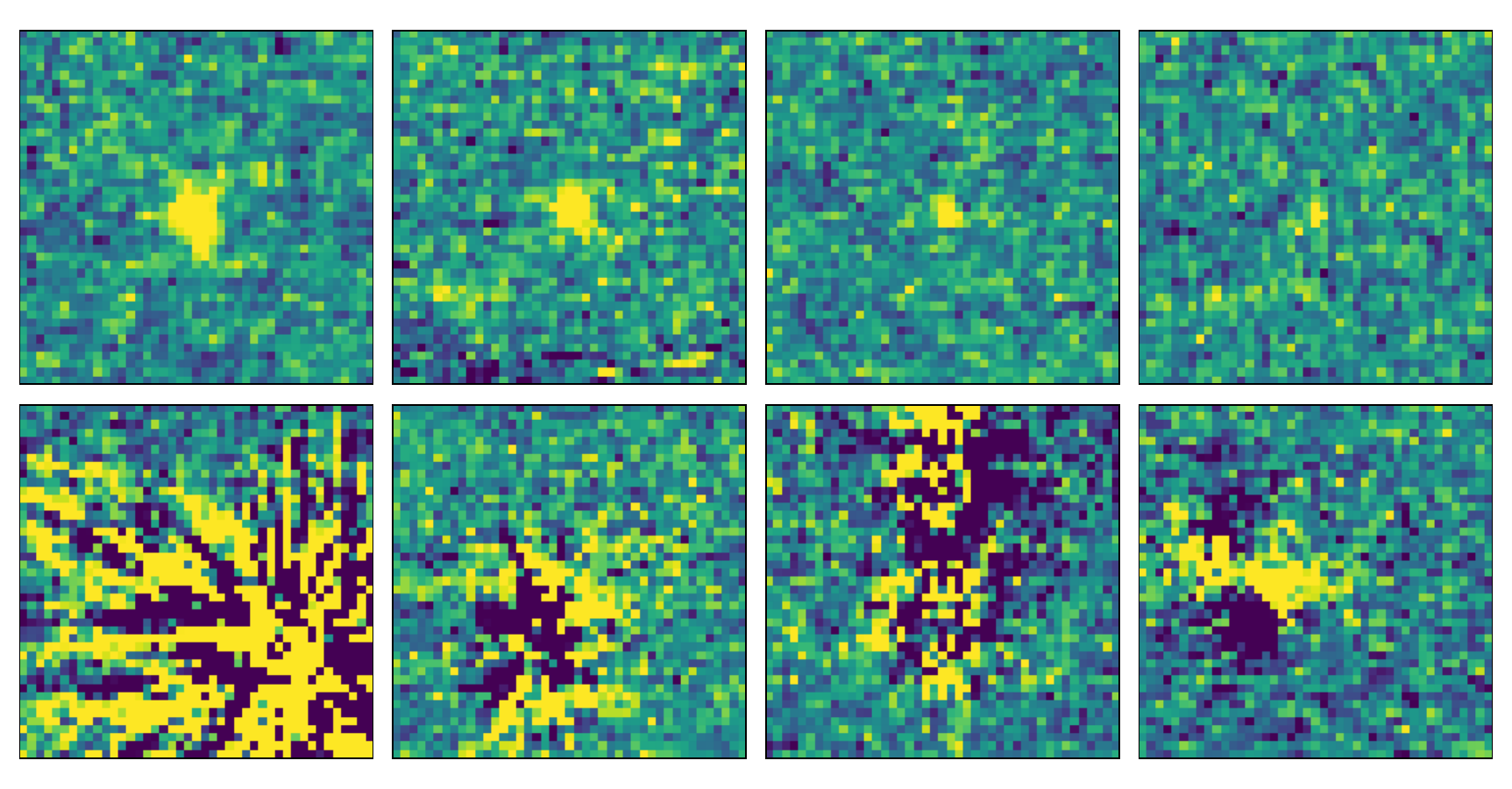}
    \caption{The upper panel displays normalized examples of source stamps (with normalized pixel values) classified as \textit{good} by our selection process, while the lower panel shows those labeled as \textit{bad}. Objects with shift-and-stack shifts that closely match their movement appear as nearly round sources in the stacked images. In contrast, bad sources are typically a result of poorly matched shift-and-stack shifts or residual artifacts left by very bright sources after image subtraction.}
    \label{fig:good_bad_label}
\end{figure}

We trained all our neural networks using supervised learning, where the model learns to identify features in manually labeled training images. Developing an effective training procedure was a challenging and iterative task, and the approach that ultimately proved successful is described as follows. A shift-and-stack detected source was labeled as \textit{good} if it was located within 1 pixel of a known implanted source and  its rate of motion matched the implant’s to within $0.3\arcsec/\text{hr}$. All other sources that did not meet this criteria were labeled as \textit{bad} (see Figure \ref{fig:good_bad_label}). However, the shift-and-stack output was overwhelmingly dominated by junk/bad detections, with a ratio of approximately 350:1 compared to good sources -- far too imbalanced for effective training. To address this, we built our training dataset by including all good detections and a randomly selected subset of bad detections at a ratio of 1.5:1 relative to the good sources. We find that increasing the bad to good ratio further would cause the model to become overly biased toward rejecting candidates, significantly reducing its ability to correctly identify true positives. Additionally, we applied sample weights of 1 for good sources and 0.8 for bad candidates, which helped achieve more balanced and accurate network performance. Furthermore, to increase the sample size and introduce necessary variability for training, we applied several augmentation techniques to our images. Specifically, each source image was flipped horizontally and vertically, rotated 90 degrees clockwise and counterclockwise, and duplicated with a one-pixel shift in four directions: left, right, up, and down. Such augmentations would help our model to learn to recognize features in our images regardless of orientation. Our sample size increased to 56,850 after augmentation, where 95\% were used for training and the remaining 5\% were used for validation.

We explored multiple hyperparameter settings and ultimately settled on the configuration shown in Figure \ref{fig:Resnet}. Then, using 10 different training sets (each using the same set of good sources but a newly randomized selection of bad sources, as described above), we trained 10 different iterations of that network. Among the 10 resulting models, we selected the one that achieved the best performance, defined by the faintest detection depth and highest peak detection efficiency. Once the optimal model was chosen, we used it to categorize all outputs from our shift-and-stack pipeline. We set the probability acceptance threshold to 0.65, a value we manually determined to balance sensitivity and precision before predictions became overly contaminated by false positives. Appendix \ref{sec:ML_criteria} provides a graphical summary of network and threshold selection.

The performance of the network ensemble was impressive. Our shift-and-stack pipeline produced over a million detections -- some artificial implants, some real sources, but the majority were junk. Our chosen network successfully filtered this to roughly 50,000 candidate sources across 20 NIRCam tiles and 3 epochs, though this came with a loss of about $\sim0.2$ mag in total depth compared to the original shift-and-stack results. Without this significant reduction in false positives, the subsequent vetting process would have been impractical. The sources identified as good were subsequently subjected to scene modeling to assess whether their image data is consistent with the profile of a moving point source.

Lastly, a primary concern with using machine learning was overfitting, a common issue where a model recognizes specific details of the training data instead of learning the underlying patterns. In this case, the model will label a source used in training more confidently than it would a source not used in training. To investigate this, we created a duplicate of our ensemble and trained each version on completely separate datasets, then validated each model using the dataset on which the other was trained. This cross-validation approach allowed us to assess whether the model’s behavior and performance remained consistent when exposed to data outside their training distribution. We found that both ensembles produced similar results in terms of similar detection depth and peak efficiency, suggesting that our chosen network was effectively learning the general features of the data and that overfitting was not a significant issue in our setup.


   \subsection{Scene modeling---single epoch} \label{sec:SceneModelling}
\subsubsection{Scene modeling technique} 
\label{sec:smp1}

Once the ResNet classification has identified a candidate TNO, we perform scene-modeling photometry (SMP) on the original (unregistered) images to confirm that the image data are consistent with a model of a moving point source atop a static sky background. The SMP process also yields optimal estimates of the flux and position vs time of the moving source, and the uncertainties on these quantities.

The SMP procedure is similar to that used by \citet{Pedro}.  For a candidate single-epoch track in some region of the search field, we locate all exposures that covered this region and extract similar ``postage stamps'' of $\approx25\times25$ pixels from each exposure centered on the nominal center of the TNO track proposed by the shift-and-stack software.  Using $j$ as the index over exposures,\footnote{In this context, ``exposure'' means one count-rate image from a single NIRCam integration.} and using $(m,n)$ to represent the pixel indices within the exposure, we model the value in the image as
\begin{equation}
    \hat f_{jmn} = b_j 
    + \sum_{\mu,\nu} f_{\mu\nu} P_j(u_{jmn}-u_{\mu\nu},v_{jmn}-v_{\mu\nu}) + f^T_j P_j\left[u_{jmn}-u^T(t_j),v_{jmn}-v^T(t_j)\right].
\label{eq:smp1}    
\end{equation}

In this equation, $(u,v)$ are sky-plane coordinates---all coordinates are taken in a gnomonic projection about $(\alpha,\delta)=(209\fdg40, -10\fdg87).$  The pixels of the image have central coordinates  $(u_{jmn},v_{jmn}),$ known from the WCS of each exposure. Each stamp also has a known point-spread function (PSF) $P_j(u,v),$ and an unknown constant background level $b_j$.  We presume that the sky consists of static point sources on a grid indexed by $(\mu,\nu)$ at sky positions $(u_{\mu\nu},v_{\mu\nu}).$  The fluxes $f_{\mu\nu}$ of the background are free parameters of the model.

The functions $u^T(t), v^T(t)$ are the candidate sky motion of the target TNO, and the fluxes $f^T_j$ are its observed light curve.  When fitting a model to a single observing epoch, we assume linear motion:
\begin{eqnarray}
    u^T(t) & = & u_0 + \dot u(t-t_0) \nonumber \\
    v^T(t) & = & v_0 + \dot v(t-t_0) 
    \label{eq:uvlinear}
\end{eqnarray}
We assign a reference time $t_0$ to the track, and then the parameters $\{u_0,v_0,\dot u, \dot v\},$ the position and rate, are free.

The SMP fits this model to the data, maximizing the log-likelihood of the data, which under an assumption of Gaussian uncertainties $\sigma_{jmn}$ on each pixel datum, becomes minimization of the usual
\begin{equation}
\chi^2 \equiv \sum_{jmn} \left(\frac{f_{jmn}-\hat f_{jmn}}{\sigma_{jmn}}\right)^2.
\end{equation}
Pixels with error flags set are omitted from the calculation.
The free parameters in this model include the exposure backgrounds $\{b_j\},$ the static scene $\{f_{\mu\nu}\}$, the target light curve $\{f^T_j\},$ and the ephemeris parameters $\{u_0,v_0,\dot u, \dot v\}.$  All except the last group enter the model linearly, and hence their values that minimize $\chi^2$, and their uncertainties, can be found algebraically for a given choice of ephemeris.  We minimize $\chi^2$ over the (nonlinear) ephemeris parameters using a Newton-Raphson iteration, and derive a $4\times4$ covariance matrix for the ephemeris parameters by linearizing the model.

In practice, we usually derive the scene for a track candidate in epoch 1 by using the exposures from epochs 2 and 3 (similarly for candidates in epochs 2 and 3).  
 Then we fit the stamps from epoch 1 while hold the scene $\{f_{\mu\nu}\}$ fixed while varying the background and TNO models. The $\mu\nu$ grid is square (on the sky) with a pitch of 1 NIRCam pixel. Some nearly-degenerate modes of the scene solution are nulled.

PSF models are obtained from the \texttt{WebbPSF} package provided by STScI.\footnote{\url{https://github.com/spacetelescope/webbpsf}} The WCS models are described in Section~\ref{sec:alignment}.

\subsubsection{Quality control and cuts: single-epoch tracks}
\label{sec:trackqc}
The SMP analysis yields the optimal values and uncertainties on the TNO fluxes $f^T_j$ and the astrometry, and in addition several quantities useful for discriminating true TNO tracks from spurious shift-and-stack triggers, such as noise peaks, unmasked detector defects, image persistence in the detector, and subtraction errors.  The main characteristic that we will use to mark a track as valid vs spurious is that a TNO will have undetectable variation in flux during the 3 integrations making up one of the two dithered exposures in the epoch.  For each dither with exposures $j=\{1,2,3\},$ we use the fluxes $f_j$ and flux uncertainties $\sigma_j(f)$ to calculate
\begin{eqnarray}
    \bar f & \equiv & \frac{\sum_j f_j/\sigma^2_j(f)}{\sum_j 1/\sigma^2_j(f)} 
    \label{eq:meanflux}\\
    \sigma^2_f & \equiv & \frac{1}{\sum_j 1/\sigma^2_j(f)} \\ 
    \chi^2_\nu(f) & \equiv & \frac{1}{2}\sum_j (f_j-\bar f)^2/\sigma^2_j(f) 
    \label{eq:chi2f} \\
    S/N & \equiv & \frac{\bar f}{\sigma_f}
    \label{eq:snf}
\end{eqnarray}

\begin{figure}
    \centering
    \includegraphics[width=\textwidth]{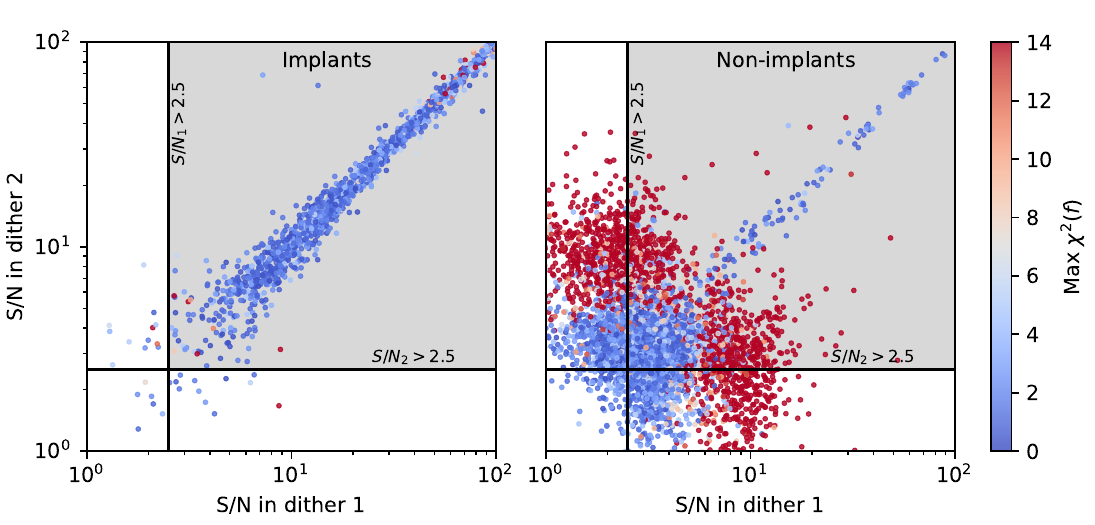}
    \caption{The criteria used to select ``good'' single-epoch tracks are illustrated.  The axes of each panel are the total $S/N$ of the flux detection (Equation~\ref{eq:snf}) in the two dithered exposures of the track's epoch.  Each point is the result of a scene-model fit to one KBMOD detection of a moving source.  The color of the point codes the worse of the internal flux consistency measures $\chi^2_\nu(f)$ of the two dithers.  Our cuts are chosen from the distribution of the values for the implanted objects, shown at left: we require $S/N>2.5$ in \emph{both} dithers (the gray region), and also $\chi^2_\nu(f)<7$ in \emph{both} dithers, \ie\ only points with a blue tinge.  The right-hand side shows the distribution of the non-implanted, potentially real objects.  We have subsampled the tracks with $S/N<10$ to avoid overcrowding the figure, nonetheless the flood of noise or artifact tracks at $S/N<5$, but with acceptable $\chi^2_\nu(f),$ is apparent.}
\label{fig:track_cuts}
\end{figure}

Figure~\ref{fig:track_cuts} illustrates how we use the resultant $S/N$ and $\chi^2_\nu(f)$ values to eliminate most of the spurious tracks with minimal loss of real TNOs' tracks.  We demand that both dithers yield detections of the source with $S/N>2.5$ using the dithers' common rate of motion, \ie\ the source cannot disappear.  We also demand that the reduced $\chi^2$ of the three fluxes indicate loosely consistent flux during each 10-minute exposure: $\chi^2_\nu(f)<7.$  The left side of the Figure shows that this encompasses nearly all tracks that are position-matched to an implanted object---and some of the ones that fall outside the selection region are seen upon visual inspection to be coincidences of an implant near a noise or artifact track.  The right-hand plot shows that the number of spurious tracks increases exponentially at low~$S/N,$ even though most of the brighter ones can be excluded by the $\chi^2_\nu(f)$ criterion (red tinged).  Please note that we never examined any plots with the implanted objects removed until \emph{after} the selections were made.  We wanted to remain blind to the number of detectable real TNOs while making analysis choices, so we only examined plots with the implants and observed tracks mixed together.

\begin{figure}[htb]
    \centering
    \includegraphics[width=0.75\textwidth]{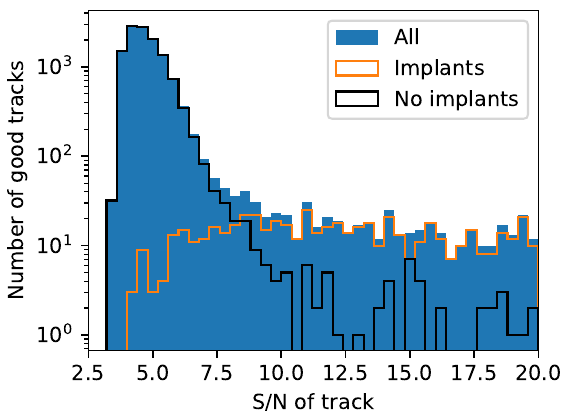}
    \caption{We plot the distributions of total $S/N$ for all tracks that pass the cuts from Section~\ref{sec:trackqc} on the per-dither $S/N$ and $\chi^2_\nu(f)$.  These are shown for the implants, the non-implanted tracks, and the total. Note the logarithmic $y$~axis. The exponential increase in non-implanted tracks for $S/N<10$ indicates the onset of noise detections, which dominate the total track counts for $S/N<7.5$}
    \label{fig:track_counts}
\end{figure}

One diagnostic that we calculate but do \emph{not} use to cull the sample is $\chi^2_5,$ defined to be the total $\chi^2$ of the data relative to scene model, summed over all the valid pixels in all integrations contained within a $5\times5$ pixel box centered on the best-fit TNO position.  A high value of $\chi^2_5$ relative to the number of included pixels indicates that the data are a poor fit to the scene model, suggesting an artifact.  It is also true, however, that bright TNOs are expected to have $\chi^2>1$ per pixel: the PSF models are not perfect, and the cosmic-ray rejection algorithms for the NIRCam up-the-ramp sampling can be spuriously triggered by a bright moving source, biasing the output rate images.  We also might obtain high $\chi^2$ from binary TNOs. We therefore do not exclude tracks based on their $\chi^2_5$ values.  Instead we use this as a collective indicator that our models and uncertainties are sensible, finding that $\left\langle \chi^2_5 \right\rangle$ per pixel is indeed near unity for tracks with $S/N\lesssim30,$ rising for TNOs brighter than this.  In Figure~\ref{fig:track_cuts} one can see some implants at high $S/N$ failing the cut on $\chi^2_\nu(f),$ likely caused by PSF mismatches and spurious triggers of the cosmic-ray clipping algorithm.  We therefore apply the $\chi^2_{\nu}(f)$ cut only for $S/N<30.$ Since sources this bright are rare, and brighter than a pure noise fluctuation can produce, and easily classified as real/spurious by eye,
we can remove this cut and still obtain a highly reliable final catalog.
    
Figure~\ref{fig:track_counts} plots the distribution of the total $S/N$ of the 14,000 unique tracks that pass these two cuts (total $S/N$ being the quadrature sum of the two dithered exposures' $S/N$ values).  The exponential increase at $S/N\lesssim7.5$ is the behavior one expects from noise detections over a  wide search area.  We have chosen cuts for \kbmod, the ResNet classifier, and track $S/N$ that allow our sample to be dominated by noise detections at single-epoch track level.  The goal is to reliably detect real TNOs to the faintest possible flux.
The spurious tracks will be weeded out by their failure to link with tracks from the other two epochs in a plausible TNO orbit, as investigated later in this section.

\subsubsection{Completeness}\label{sec:completeness}
\begin{figure}[h!]
    \centering
    \includegraphics[width=\textwidth]{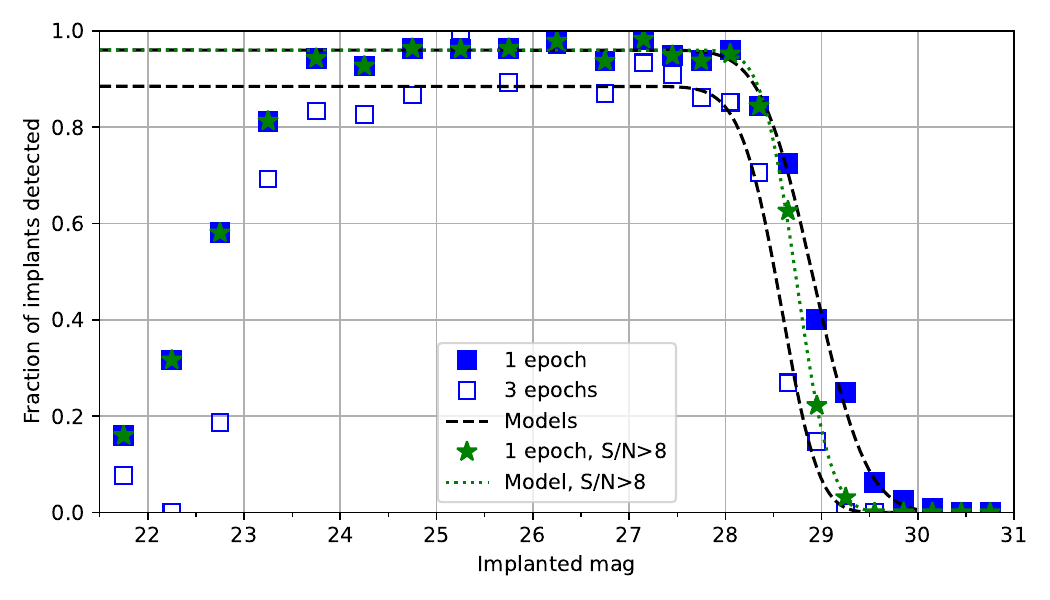}
    \caption{The filled blue squares plot the fraction of objects implanted onto valid regions of the NIRCam SW detectors that yield a fitted track that passes the quality cuts described in Section~\ref{sec:trackqc}, as a function of injected magnitude.  The upper dashed black line plots the completeness function $p_{\rm det}$ from Equation~(\ref{eq:pdet}) with parameters that maximize the likelihood of the implant results.  The analytic function is seen to be a good representation of the implant results.  
    The lower black curve plots the cube of the upper completeness function, which is the expectation for independent detection probabilities in all three epochs, and agrees with the implant results plotted as open blue squares. The green stars and the dotted green line are the measured implant recovery rate and fitted model when a cut of $S/N>8$ is imposed on the tracks, as is done in creating sample B of TNOs in Section~\ref{sec:selectAB}. Lastly, we stress here that the decline in efficiency at the bright end results from the tuning applied during the shift-and-stack stage of the search, rather than from overtraining of the ResNet.}
    \label{fig:track_eff}
\end{figure}

The probability with which a TNO will be detected as a good track during a single epoch, if it falls on a detector during both dithers, is quantified using the implants. It is well fit by the functional form
\begin{equation}
    p_{\rm det}(m) = \frac{p_0}{2}\, {\rm erfc}\left(\frac{m-m_0}{w}\right)
\label{eq:pdet}
\end{equation}
with the bright-end efficiency $p_0=0.96,$ the magnitude of half that efficiency at $m_0=28.92,$ and transition width $w=0.61$~mag. Figure~\ref{fig:track_eff} plots this curve, along with binned rates of detection for the implants.  This suggests that about 4\% of sources are lost from overlaps with bright sources or defects on the detectors.  
It should be noted that the PSF model is most likely responsible for the decrease in detection efficiency at the brightest end of our implanted magnitude range. That is, inaccuracies in the PSF combined with the tolerances in the predictive filter tended to reject the brightest implants at a slightly worse rate than for fainter source for which the tolerances were tuned.

We also check the probability that an implant that falls on a detector in all of the three epochs is detected as a valid track in all three epochs.  The binned success fractions are shown as open symbols in Figure~\ref{fig:track_eff}, and are well fit by the cube of the single-epoch $p_{\rm det}$ function in Equation~(\ref{eq:pdet}), which is what we expect if the detection process is statistically independent in each epoch. 

In Section~\ref{sec:selectAB} we will create a sample of detections constrained to obtain $S/N\ge8$ in single epochs.  The green stars in Figure~\ref{fig:track_eff} plot the measured recovery efficiency of implants subject to the condition.  The recovery rate is well fit by Equation~\ref{eq:pdet}, with parameters $p_0=0.96, m_0=28.74,$ and $w=0.40$~mag.

\subsection{Lightcurve bias}

\begin{figure}[h]
\centering
    \includegraphics[width=0.52\textwidth]{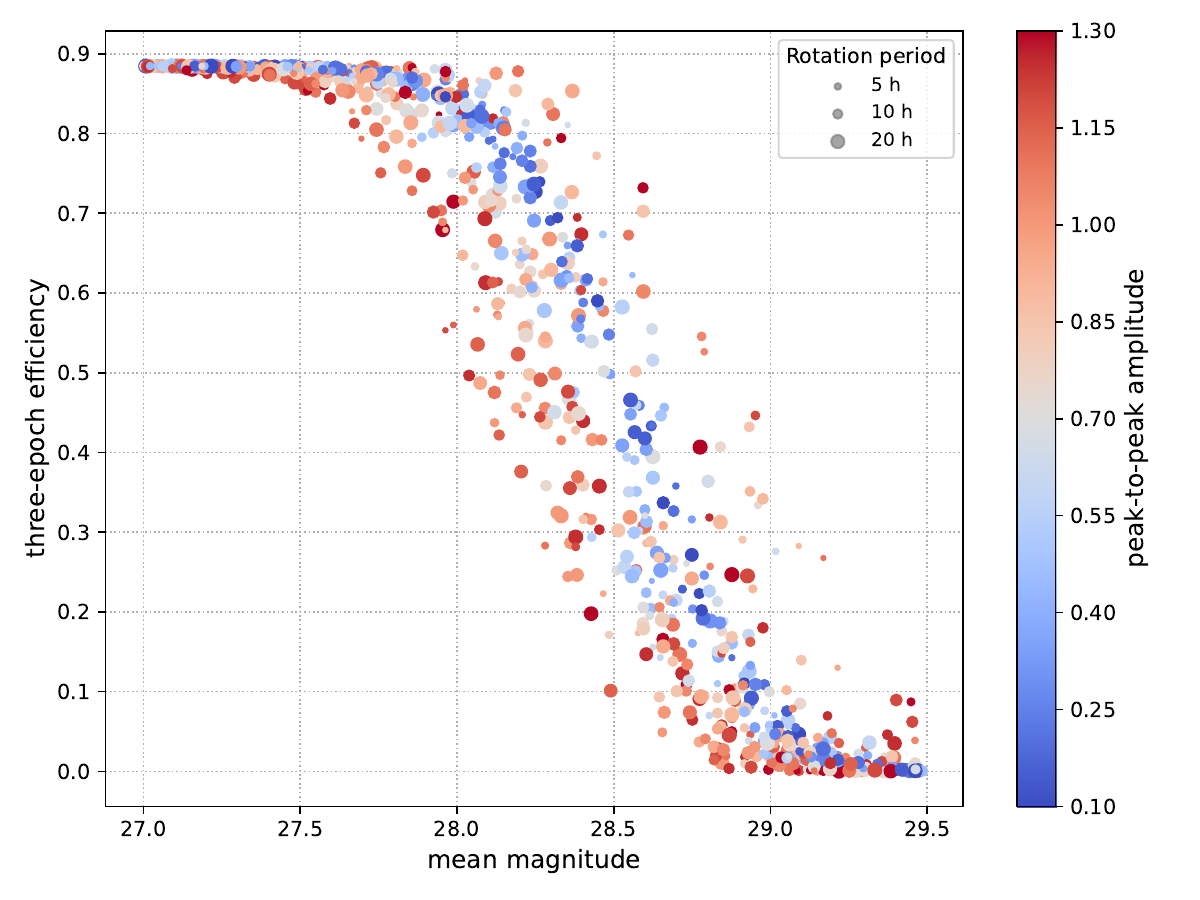} 
    \includegraphics[width=0.47\textwidth]{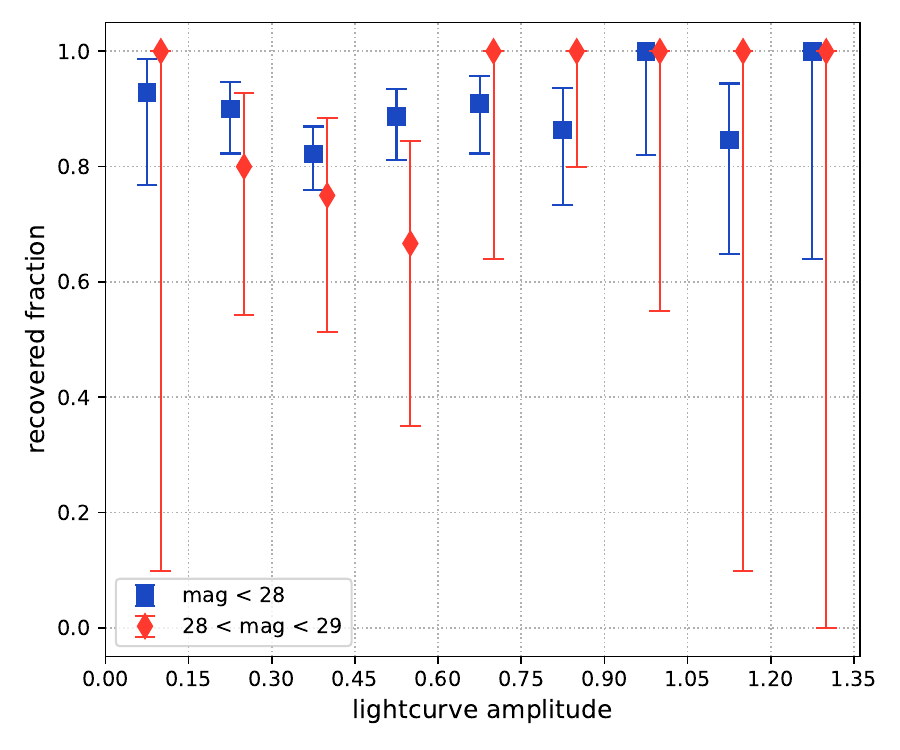} 
    \caption{\textbf{Left:} Three-epoch detection efficiency for simulated objects with a mean magnitude greater than 27 mag. The data points are color-coded according to their peak-to-peak brightness variation, $\Delta m$. The sizes of the markers are proportional to the objects’ spin rates, which span from 2 to 20 hours. \textbf{Right:}  Fraction of recovered implants over three epochs for objects with implant magnitudes between $28–29$ mag (green diamonds), where the detection efficiency begins to decline. For reference, blue squares show the recovery fraction for implants with magnitudes $< 28$ mag. Lightcurve amplitude is binned in $0.15$-mag intervals, and error bars indicate $1\sigma$ Poisson uncertainties on the number of missed objects in each bin.  }
\label{fig:efficiency2}
\end{figure}

TNOs with larger amplitude variations are more likely to be missed at certain rotational phases, even if their mean brightness is above the survey's limiting magnitude. To test whether this affects our survey, we simulate 1000 objects of varying rotational periods, peak-to-peak amplitudes $\Delta m$, and magnitudes. For each simulated object, we calculate the three-epoch efficiency value and plot them as a function of their lightcurve mean magnitude and peak-to-peak amplitude (see left of Figure \ref{fig:efficiency2}). Fainter objects with high $\Delta m$ (red circles) and whose cube efficiency are high were likely recovered when they are at brighter phases. In parallel, we also calculate the fraction of recovered implants over our three epoch survey as a function of their amplitude as seen right of Figure \ref{fig:efficiency2}. Typically, we would expect a drop in the recovered fraction for objects with higher amplitudes, as their brightness occasionally falls below the detection threshold. However, the recovery fraction is notably high for the bins at 1.05, 1.20, and 1.35 although with large uncertainties due to very small number of implanted sources in these bins. Consequently, we do not find any variation of our survey completeness with lightcurve amplitude; however, the insufficient number density of our implanted objects limits us from robustly quantifying this effect in our survey.



\subsubsection{Decoupled errors for linking}
\label{sec:decoupled}
While SMP yields the best-fit position $(u_0,v_0)$ and rate $(\dot u, \dot v)$ for a single TNO track, with their full covariance matrix, the linking software is simplified if we convert these 4 variables into two ``pseudo-detections'' at times $t_1,t_2$ with assigned positions $(u_1,v_1)$ and $(u_2,v_2).$  These positions are chosen so as to yield the same position and rate originally found, and the times are selected such that the covariance between $(u_1,v_1)$ and $(u_2,v_2)$ is near zero.  Each epoch is thus converted into two pseudo-observations, each with its own position errors, that can be treated as statistically independent measurements by the linking code, while carrying the same information as the position/rate format.

   \subsection{Cross-epoch linking} \label{sec:OrbitFitting}

At this stage, we have pairs of pseudo-detections that can be treated as independent observations and carry the position/velocity information from a single track.  The challenge is to determine which tracks, observed at distinct epochs, correspond to the same real or implanted object.  This is known as {\it linking.}

The observed angular coordinates of the candidate TNO in the sky plane are given by
\begin{eqnarray}
u^T(t) &  = & { {x(t^\prime) - x_E(t)} \over {z(t^\prime) -
z_E(t)}} \nonumber \\
v^T(t) &  = &  { {y(t^\prime) - y_E(t)} \over {z(t^\prime) -
z_E(t)} },
\label{eq:exact}
\end{eqnarray}
where $t^\prime=t-\Delta t$, and $\Delta t$ is the light travel time from the object to the observer.

Each track's pseudo-observations provide four quantities (the two sky positions). However, a full Keplerian orbit requires six parameters. The two missing elements are effectively the object's distance from the observer and its line-of-sight velocity at a reference time. We adopt the \texttt{HelioLinC} approach \citep{Holman.2018}, which involves asserting values for the two unknown quantities, computing the corresponding orbit for each track, and then identifying clusters of tracks from distinct epochs that have yielded similar orbits. When the assumed values are correct, the resulting orbits for tracks that correspond to the same object should differ only due to the measurement uncertainties of the observations.

We adopt a version of the coordinate system of \citet{Bernstein.2000}. The $z$ axis points in the direction of (RA,Dec)$=(209.4\degree,-10.87\degree),$ approximately the center of our search region.  The $x$ and $y$ axes lie in the sky plane normal to $z$.  Rather following the ecliptic,  we align the $x$ axis with the direction of increasing RA, and the $y$ axis then points in the direction of increasing declination. The reference time is set to JD 2459974.5 (TDB), which is near the midpoint of the observation period.

For each pseudo-detection, we look up the position of JWST at its time of observation and transform into our coordinate system. To account for the limitations of the PSF model for bright sources, we add $0.0005\arcsec$  in quadrature to their positional uncertainties. Then, we use the parameterization of \citet{Bernstein.2000},  based on the components of the position and velocity of the target at the reference time: $\alpha = {x_0}/{z_0}$, $\beta = {y_0}/{z_0}$, $\gamma = 1/{z_0}$, $\dot\alpha = {\dot x_0}/{z_0}$, $\dot\beta = {\dot y_0}/{z_0}$, and $\dot\gamma = {\dot z_0}/{z_0}$.
With these, $\alpha$ and $\beta$ are the components of the angular position of the object at the reference time,  $\dot \alpha$ and $\dot \beta$ are angular rates of motion in the inertial coordinate system, $\gamma$ is a measure of distance to the object, and $\dot \gamma$ is a scaled radial velocity.   

In terms of these parameters,  the observation $\theta_x$ is given by:
\begin{equation}
        \theta_x(t)  =  { { \alpha + \dot\alpha t^\prime + \gamma g_x(t^\prime) - \gamma x_E(t) }
        \over { 1 + \dot\gamma t^\prime + \gamma g_z(t^\prime) - \gamma z_E(t) } }
\label{posexpression}
\end{equation}
where $t^\prime \approx t - \frac{1}{c \gamma}$ is the light-time corrected time of the observation.  There is a corresponding equation for the fitted observation $\theta_y$.

  Equation~\ref{posexpression} can be rearranged to yield a simple expression for the linear motion of the object:
\begin{equation}
        { { \alpha + \dot\alpha t^\prime }} = 
                    \theta_x \left[ 1 + \dot\gamma t^\prime + \gamma g_z(t^\prime) - \gamma z_E(t)\right] 
                    \\ 
                     -\gamma g_x(t^\prime) + \gamma x_E(t)
\label{rearrange}
\end{equation}
where $\theta_x$ and $\theta_y$ are observed quantities, and the observatory position ($x_E$, $y_E$, $z_E$) is known precisely.  We note that in Equation \ref{rearrange} the transverse components of the gravitational perturbation, $g_x(t^\prime)$ and $g_y(t^\prime)$, are much smaller than $g_z(t^\prime)$ and can be ignored. 

Finally, to investigate possible orbital configuration, first, we assume a constant acceleration due to the total mass of the solar system in the direction of the barycenter. This an excellent approximation for the time span of these observations, as the acceleration is very small \citep{Bernstein.2000}. Then, we loop over pairs of $\gamma$ and $\dot\gamma$, where $\gamma$ is uniformly spaced between 0 and $0.06/\mathrm{au}$ with 721 values, and $\dot\gamma$ is uniformly spaced between $-10^{-4}~\mathrm{rad/{day}}$ and $10^{-4}~\mathrm{rad/{day}}$ with seven values. This range of values encompasses bound orbits from inside the orbit of Uranus to infinity. We established the spacing of the values empirically.

This rapid algebraic search for valid orbit fits is applied to all potential triplets of pseudo-observation pairs from three distinct JWST observing epochs (\texttt{HelioLinC} uses clustering algorithms that do not actually require examining every distinct triplet).  From the $\sim14,000$ single-epoch tracks passing the qaulity cuts in Section~\ref{sec:trackqc}, we find $\sim21,000$ unique 3-epoch triplets that are consistent with a bound orbit in the sense of having RMS residuals $\le0.2\arcsec$ between the measured positions and the positions predicted by the orbit model.  The vast majority of these include at least one track that yielded $S/N<6$ across its JWST epoch.  The $\le0.2\arcsec$ cut on orbit-fitting RMS is intentionally very loose and admits many false-positive linkages.  The next section describes how these candidates are culled to eliminate false positives at high confidence while retaining as many real TNOs as possible.

There are 417 triplets of implanted tracks detected (some implanted objects are detected more than once in a given epoch, leading to multiple ways to link one per epoch).  Of these, the linking process described herein finds all but two.  In fact these two cases are very faint implants, which we suspect are actually being mismatched to nearby spurious detections, and thus should \emph{not} yield good links. We conclude that the linking process has negligible false negatives.

\section{Orbit Refinement and Selection} \label{sec:OrbitSelection}

\subsection{Three-epoch linkages}
\label{sec:3epoch}
The previous section's linkage procedure produces a list of candidate TNO orbits, each consisting of one track in each of the three epochs.  This section describes how each of these candidate orbits is refined, and selection cuts are made to exclude spurious linkages while retaining real sources.  This procedure parallels the treatment of single-epoch tracks in Section~\ref{sec:SceneModelling}: first, each orbit candidate is subjected to scene modeling. Next, quality cuts are defined
based on the results for implanted sources, such that the false negative rate for implanted objects is kept minimal while rejecting all readily identifiable spurious linkages.
Finally, the full population of  3-epoch orbit candidates---which is a mixture of real TNOs, implanted TNOs, and spurious detections, or mixtures of these types, with the investigators blind to which is which---is passed through the quality cuts.  A rapid upturn in number of accepted triplets at the faint end would be the telltale sign of false positive detections, and we find no such upturn. We end up with a list of validated three-epoch TNO detections, and can unblind the results by removing implanted sources from this list.

The SMP for full orbits again models each integration's image (3 integrations per dither $\times$ 2 dithers per epoch $\times$ 3 epochs) as the sum of a time-invariant background scene, plus a point-source TNO with time-varying flux, following an ephemeris $[u^T(t),v^T(v)].$ The main difference between the single-epoch SMP and the full-orbit SMP is that 
 we replace the linear apparent motion of Equations~(\ref{eq:uvlinear}) with an ephemeris parameterized by the initial ICRS barycentric Cartesian state vector $({\bf x}_0,{\bf v}_0)$ of the TNO.  The TNO motion is integrated under the gravity of the Sun and giant planets (assumed to be constant over our 10-day observing period) and the observatory's state vector is retrieved from JPL Horizons to permit calculation of apparent coordinates vs time of observation using Equations~(\ref{eq:exact}).  The orbit is fit to exposures of all three epochs simultaneously, and yields a maximum-likelihood initial state vector and a covariance matrix for its 6 elements.  A separate TNO flux is fit to each NIRCam integration.  Another difference from the single-epoch SMP process is that there are three distinct static background images: the background for the location of the TNO in epoch $j$ is ascertained by using images of the same sky region in the other 2 epochs.  A total of 54 postage stamps are thus extracted and fit for each candidate TNO orbit.

\begin{figure}
\centering
\includegraphics[width=0.5\textwidth]{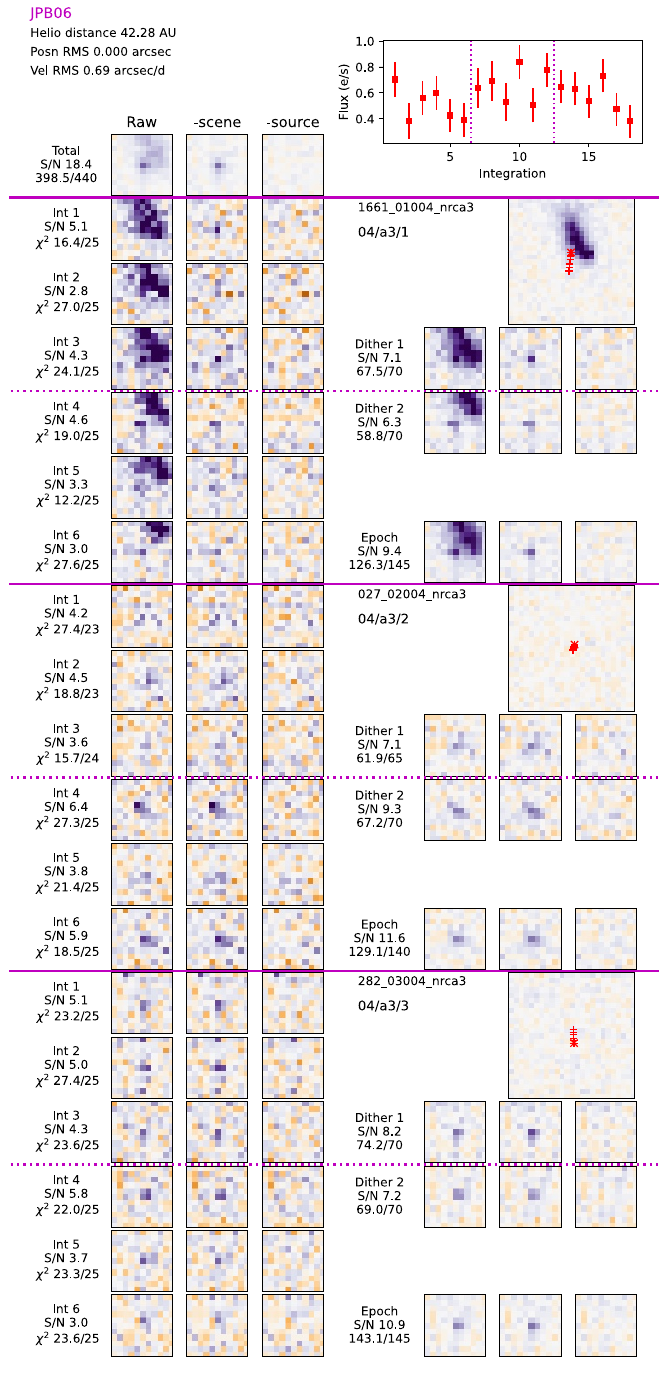}
\caption{Illustration of the orbit-level SMP fit to the newly discovered TNO JPB06 at $m_{AB}=28.6$~mag.  Each row shows a set of three cutouts from the NIRCam SW data spanning 0\farcs34.  The left is the calibrated image, the center subtracts the model for the background, and the right also subtracts the model of the TNO moving along the best-fit orbit.  Ideally the right-hand images are pure noise.  Between each pair of purple lines, the images from one epoch's 6 individual integrations are stacked at the left.  At the right are (top to bottom) the background scene derived for the region; the means of the 3 integrations in each of the 2 dithered exposures; and the mean of all 6 exposures, \textit{i.e.} ``shift-and-stack'' images. The signal-to-noise ratio of this source, and the $\chi^2/\textrm{DOF}$ of the model fit in the  $5\times5$ pixels centered on the TNO, are shown for each individual integration and stack.
At the top left are the results of averaging all 18 integrations of the target, and at top right a plot of the fitted flux vs integration number.  The accuracy of the background$+$source model is apparent, as is the gain in detection significance as more exposures are combined.}
\label{fig:jpb06}
\end{figure}


 Figure~\ref{fig:jpb06} illustrates the result of the full 3-epoch SMP process for one newly discovered TNO. The auxiliary files available with this paper contain similar diagrams for all of the 3-epoch TNOs,  as well as the numerical arrays comprising all relevant image cutouts and models (see Appendix~\ref{sec:products}).

 The outputs of the multi-epoch SMP are the estimated fluxes $f_j$ at each of the 18 integrations and their uncertainties $\sigma_j(f)$, plus a best-fit ICRS barycentric state vector at the reference time, and its covariance matrix.  The main diagnostic criteria for distinguishing real from spurious linkages is whether the fitted orbit is indeed the best description of the data in the three original epochs.  For spurious linkages, the three individual-epoch fits are likely to better fit the pixel data than constraining the TNO to follow a physical orbit across 3 epochs.

Three quantities are used to select genuine TNOs from among the $\approx$21,000 three-epoch linkages returned by the algorithms of Section~\ref{sec:OrbitFitting}.  To derive the first, we calculate the flux at each epoch using Equation~(\ref{eq:meanflux}) to average over the epoch's 6 integrations, then convert this flux into an $AB$ magnitude using the calibrations provided by the Space Telescope Science Institute.  We then define $\Delta m = m_{\rm orbit}-m_{\rm track}.$  The magnitude derived from fluxes returned by SMP for that epoch when all three epoch are fit simultaneously to an orbit is $m_{\rm orbit},$ and $m_{\rm track}$ is the magnitude measured by SMP when fitting a linear motion to just the single track (Section~\ref{sec:SceneModelling}).  A value $\Delta m>0$ will result if the PSF being used by SMP for the full-orbit fit is not well centered on the object's flux, which is a sign that the three tracks do not actually arise from a TNO on a Keplerian orbit.  We take ${\rm max}(\Delta m)$ over the three linked epochs as one quality indicator.

The second selection asks whether the apparent rate of linear motion ${\bf r}_{\rm track}=(\dot u, \dot v)$ fit to each individual track via Equation~(\ref{eq:uvlinear}) agrees with the rate ${\bf r}_{\rm orbit}$ implied by the orbit derived as best fit in the 3-epoch SMP. The SMP for each epoch returns a $2\times 2$ covariance matrix $C_r$ for the components of ${\bf r}_{\rm track}.$ We take the maximum over the 3 epochs $j$ of the statistical significance of the disagreement between the single-epoch rate of motion and the orbit fit's rate of motion:
\begin{equation}
    \Delta\chi^2_\nu({\bf r}) \equiv {\rm max}_j \left[\left({\bf r}_{{\rm track},j}-{\bf r}_{{\rm orbit},j}\right)^T C_{r,j}^{-1} \left({\bf r}_{{\rm track},j}-{\bf r}_{{\rm orbit},j}\right)/2\right].
    \label{eq:dchirate}
\end{equation}
A high $\Delta\chi^2_\nu({\bf r})$ value implies that the motion observed between the three epochs does not agree with the motion observed within each epoch, contrary to behavior of a real TNO.

The final diagnostic property calculated for each orbit is ${\rm min}_j(S/N_j),$ the minimum among epochs $j$ of the $S/N$ of detection in that epoch.  The $S/N$ comes from Equation~(\ref{eq:snf}), using the $f$ and $\sigma_f$ values derived for the epoch's 6 integrations by the SMP fit to the full orbit.  A orbital linkage to a noise peak, or other spurious track is likely to yield a very low ${\rm min}(S/N)$.

\begin{figure}[p]
    \centering
    \includegraphics[width=\textwidth]{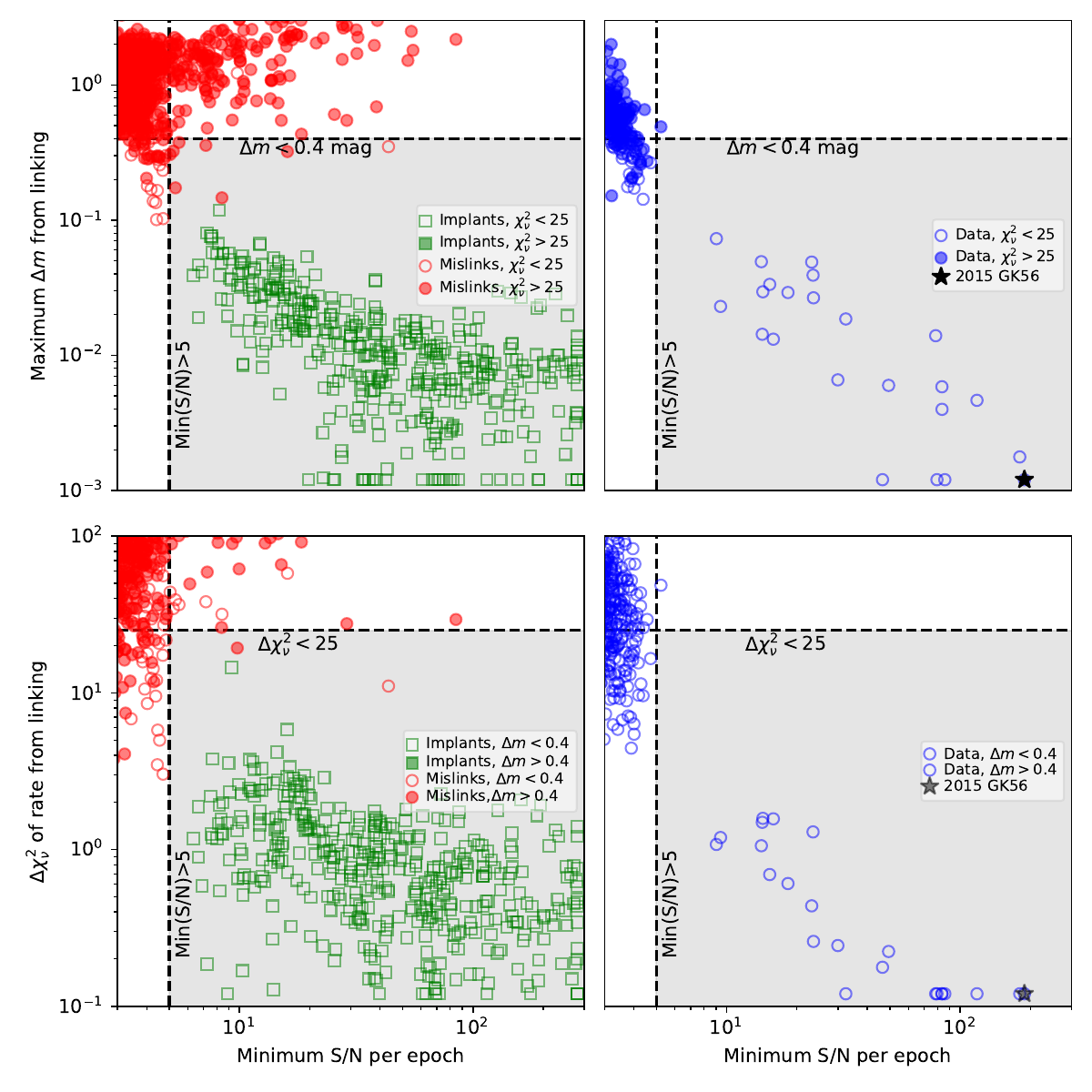}
    \caption{The top row of panels show the positions of 3-epoch linkages in the plane of diagnostic quantities ${\rm min}{(S/N})$ (the lowest $S/N$ of the putative TNO found among the three epochs) and $\Delta m$ (the worst epoch's magnitude loss incurred by forcing a fit to an orbit) described in Section~\ref{sec:OrbitSelection}.  The left panel plots candidate orbits that include 3 tracks from the same implanted source as the green squares, and tracks that include only 1 or 2 tracks from a given implant---\ie\ spurious linkages---as the red circles.  The right panel shows results for linkages that do not include any implanted sources.  The symbols are filled if the selection $\Delta\chi^2_\nu({\bf r})<25$ on the orbital vs track rates of motion is failed for any epoch.  The gray region shows the cuts ${\rm min}(S/N)>5$ and $\Delta m<0.4$ that we use to separate genuine TNO linkages from spurious linkages.  On the left, the combination of the three cuts is 100\% effective in recovering correct linkages of implanted TNOs, with only one false positive of a mislink (open red circle inside the gray region) which is readily rejected by eyeball inspection.  The right panel show that 23 non-implant linkages, including \gk, are clearly isolated by these criteria out of the thousands of candidates.  The bottom row shows the same process in the plane of ${\rm min}(S/N)$ vs $\Delta\chi^2_\nu({\bf r})$, with symbols filled if they fail the $\Delta m$ cut.
}
\label{fig:orbit_cuts}
\end{figure}

Figure~\ref{fig:orbit_cuts} plots the locations of the linked triplets in the plane of ${\rm min}(S/N)$ vs $\Delta m$ in the top row, and ${\rm min}(S/N)$ vs $\Delta\chi^2_\nu({\bf r})$ in the bottom row.  At left we plot the results for correct linkages of 3 epochs of tracks linked to the same implanted source.  We see that if we define selection cuts
\begin{eqnarray}
{\rm min}(S/N) & > & 5 \nonumber \\
\Delta m & < & 0.4\,{\rm mag} \nonumber \\
\Delta\chi^2_\nu({\bf r}) & < & 7
\label{eq:cuts}
\end{eqnarray}
then we successfully retain all of the 413 correct linkages of implants.  The red circles are linkages that contain only 1 or 2 tracks linked to any implanted source, \ie\ they are mis-links either between 2 different implants, or between implanted and non-implanted sources, and hence are definitively mis-linkages of accidentally aligned detections.  We see that only one of the 7000 of these known mislinks passes the cuts of Equation~\ref{eq:cuts} (open red circle within the gray regions), and this one is easily identified as a mislink upon visual examination of its SMP images.  

Once these cuts are established at values that guarantee low false-negative rates for the implants, we are ready to apply them to the full population of 3-epoch linkages and decide if these cuts also avoid false positives from spurious linkages.  The plots in the right-hand column of Figure~\ref{fig:orbit_cuts} show that 23 linkages are well separated from the many thousands of other linkages in these diagnostic plots and lie within the locus of known implants.  These are our confidently detected 3-epoch TNOs.  Please note that the right-hand plots were not actually made until after the cuts in Equations~(\ref{eq:cuts}) were decided by examining plots that marked the known mis-linkages, but did not distinguish implanted triplets from non-implanted triplets.  Confidence that we have no false positives arises from the clear separation of the loci of known mis-links vs the correctly linked implants.  After unblinding by removing all triplets including any implants, a similar separation into 2 loci is seen in the right-hand panel of Figure~\ref{fig:orbit_cuts}.

\begin{figure}[h!]
    \centering
    \includegraphics[width=0.8\textwidth]{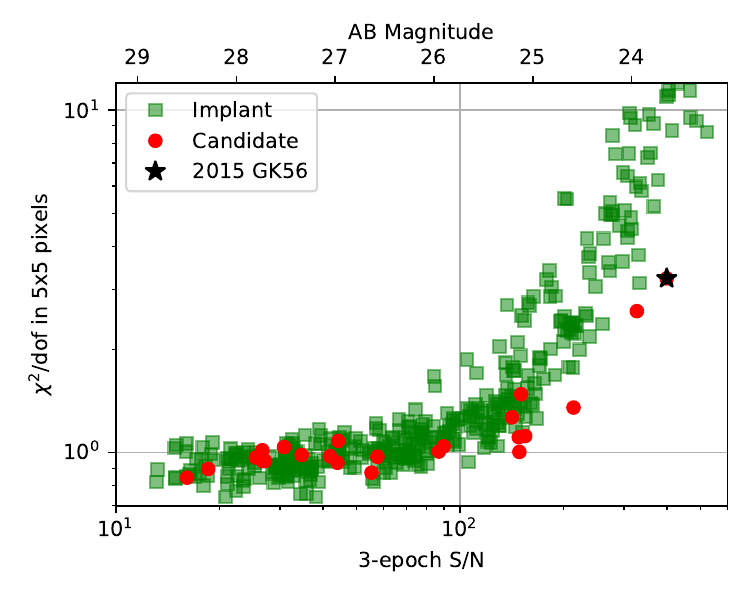}
    \caption{The $\chi^2$ per pixel of the SMP fit to the pixel data is shown for all 3-epoch orbit fits that pass the selection cuts in Equation~(\ref{eq:cuts}).  Only valid pixels in $5\times5$-pixel boxes around the source are used.  Implanted objects are in green squares, newly discovered real-sky linkages are in red, and \gk\ is the star.  The statistic behaves as expected, with values slightly below unity at low $S/N$, rising for $S/N\gtrsim30$ sources for which the mismatch between the assumed PSF and the real or implanted PSF cause significant model error.}
    \label{fig:orbit_chi}
\end{figure}

Note that our selection criteria do \emph{not} require a valid orbit to have fluxes that match in any way across epochs or dithers, since that would induce a bias against highly variable TNOs.  We also do not invoke any cuts on the reduced $\chi^2_\nu$ of the SMP model's fit to the pixel data.  We do calculate the $\chi^2_5$ value per pixel of the full-orbit SMP models, but we do not invoke cuts on this quantity when making our TNO selection, for fear of eliminating potential binary TNOs or very bright sources from our sample.  Figure \ref{fig:orbit_chi} plots the $\chi^2_5$-per-pixel values of all of linkages that pass our cut, both implanted sources and the 23 real-sky triplets.  We find the expected behaviors: low-$S/N$ sources cluster around a value of $\approx 0.9,$ slightly below unity because the actual number of degrees of freedom of the fit is below the number of pixels.  Also, the values rise for higher-$S/N$ sources because of mismatches between the true PSF and the model we used to fit.  Notice that the bright implanted sources have higher $\chi^2_5$ values than the real ones at the bright end (including \gk), because the PSF model used during detection was closer to the true PSF than it was to the PSF assumed for implantation.

The 23 real-sky 3-epoch linkages that pass selection cuts arise from 19 distinct TNOs. One of them is \gk, and 18 are new discoveries. Four of the new TNOs were detected in 2 different tiles in one of the observing epochs, thus generating two successful 3-epoch linkages per TNO.  The 18 newly discovered TNOs have time-averaged fluxes corresponding to AB magnitudes of $24.5\le m_{\rm F150W2}\le28.6.$
All 19 of these objects are listed in Table~\ref{tab:objects} along with their key photometric and dynamical properties.

\subsection{Two-epoch linkages} \label{sec:2epoch}

We next search the list of non-implanted single-epoch tracks for pairs that are consistent with a common TNO orbit.  There are three motivations for this.  First, it is expected that due to the gaps and edges of the NIRCam-imaged field, there should be a ratio of $\approx0.3:0.4:1$ TNOs that appear in 1, 2, and 3 epochs' images, respectively.  This ratio was determined from the implanted orbits, and is only mildly dependent on the barycentric distance $d$ and inclination of the sources over the range $30<d<60$~\au. The discovery of such objects is therefore expected and serves as a crude cross-check on our 3-epoch linkage rates.

The second motivation is that TNOs detected in only 1 or 2 epochs have scientific value as measurements of the abundance and/or IR colors of small TNOs, even if our knowledge of their orbital state is less complete. 

The third motivation for examining all potential 2-epoch linkages stems from concerns that the ResNet’s is overtrained on the PSF of implanted objects, which differs slightly from the on-sky PSF of actual TNOs. Such overtraining could potentially lead the network to misclassify or reject genuine detections of bright TNOs with sufficient $S/N$ to distinguish the two PSFs. 


\subsubsection{Search among ResNet rejections}
We proceed by collecting the $uv$-plane positions and rates for the $\approx4000$ single-epoch tracks that pass SMP quality cuts and exhibited $S/N>5,$  the lower limit placed on successful linkages in Equation~(\ref{eq:cuts}).  We examine all pairs of these good, non-implanted tracks that occurred in distinct epochs to find any that would fit a bound TNO orbit with distance $d>6.7$~au. This is accomplished using the model of Equation~(\ref{posexpression}), assuming the line-of-sight motion $\dot\gamma=0,$ and scanning across inverse distance $\gamma.$  The range of $\gamma$ (if any) that yields a $p>0.01$ fit to both epochs' positions and rates are recorded.

For each of the $\approx5000$ 2-epoch track pairs meeting this criterion, we next search through all KBMOD detections from the third epoch for any track whose position and velocity are broadly consistent with an orbit passing through the 2-epoch pair.  This is done \emph{regardless of the ResNet score} for the third track.  In this way we should locate any KBMOD track, regardless of ResNet score, that forms a plausible TNO 3-epoch triplet with two ResNet-approved, $S/N>5$ tracks.\footnote{Note that searching the large collection of ResNet-rejected KBMOD detections is feasible because the requirement to link with 2 ResNet-accepted epochs enormously reduces the available phase space for candidates.  It would be infeasible to bypass the ResNet selection for the entire search.}
If this process is thorough, then given any 2 of the tracks in one of the 23 known real 3-epoch orbits, we should locate the third.  Indeed we re-discover the third epoch in all 69 possible cases.

After removing all of the new triplets that are duplicates of one of the 23 known triplets, we perform single-epoch and three-epoch SMP on each of the 277 remaining triplets.  Applying the quality cuts described in previous sections yields five newly discovered 3-epoch triplets, each by construction containing one track that was rejected by the ResNet cuts.  One of these, at $m_{AB}=24.5,$ is a 4th detection of the previously known triplet JPB04.  Two others are new 3-epoch TNOs, JPB01 and JPB05, at $26.5<m_{AB}<27,$ where overtraining of the ResNet might be an issue.  The detections that the ResNet rejected have $S/N>30,$ and they are flawless under visual inspection.  The ResNet successfully classifies all of the implanted TNOs in this brightness range as real.  It is justified, therefore, to include JPB01 and JPB05 in our sample of 3-epoch TNOs, and still consider the implants' results as a valid measure of the detection efficiency of the survey, even though the implants were subjected to the ResNet cut but these two failed it---we chalk this up to a difference of ResNet efficiency at high $S/N$ between real and implanted sources.

The other two links of two ResNet-approved tracks to a ResNet-rejected track in the third epoch are JPB02 and JPB03.  These are fainter than any ResNet-approved 3-epoch TNO, at $m_{AB}\ge29.$ In this regime we know that there are many spurious detections and our discovery efficiency is limited by the ability of ResNet to distinguish spurious from real tracks.  It is therefore unsurprising that turning off the ResNet filter for one epoch yields fainter linkages---but we should \emph{not} use these in our analyses because (a) they may not be real, and (b) our implant-derived detection efficiency assumes the ResNet filter is in place.  We include these two sources in Table~\ref{tab:objects} and the auxiliary data files as examples of fainter TNOs that \emph{could} be real, but are too faint for us to reliably distinguish from spurious detections or use to infer population statistics.

\subsubsection{Follow-up of $S/N>8$ ``orphan'' tracks}
Returning to the question of whether there are 2- or 1-epoch TNO detections among our good, non-implanted tracks, we examine by eye the 111 single-epoch tracks that are not implants, are not apparitions of one of the 21 known 3-epoch TNOs, pass all quality cuts, and have $S/N>8$ from SMP fitting.  At this $S/N$ level, one can readily distinguish real from spurious sources by eye, and Figure~\ref{fig:track_counts} shows that this is the $S/N$ level at which there are probably similar numbers of real and spurious tracks passing the single-epoch cuts.
For $S/N<8,$ we do not attempt to distinguish between real and spurious detections that lack confirmation from linking to 2 other epochs.  Figure~\ref{fig:track_counts} shows that the number of such ``orphan'' tracks grows very rapidly as $S/N$ decreases (doubling for $S/N<7$), which means that the fraction of real sources is dropping rapidly.  We thus do not have confidence that we could exclude false positives from a sample of 2-epoch linkages that involve $S/N<8$ tracks.

Among the $\approx50$ orphan $S/N>8$ tracks that visual inspection confirms as real, we search for all pairs that can plausibly be fit to a bound orbit of a body at distance $>15$~\au.  These pairs are listed in Table~\ref{tab:objects}, where they are recognizable as missing data from one of the three epochs. With the observed positions from two epochs, we can strongly constrain a linear combination of the TNO's distance and its transverse velocity in the ecliptic plane ($\dot x$).  The degeneracy between these two quantities is only weakly broken by the rates of motion observed within each of the two detected epochs.  The apparent position of the TNO at any given time in our 10-day window is thus constrained to lie within $<1\arcsec$ of a line segment on the sky that is parameterized by the TNO's distance $d.$  We thus can determine, for any of our JWST exposures, for what ranges of $d$ (if any) the TNO would have appeared on a detector in that exposure.  In each case where the missing 3rd detection could have appeared on a detector, we searched for it again both by eye, and by re-running KBMOD with a lower detection threshold over this limited  range of sky position and rate.  The results of these searches are detailed in Appendix~\ref{sec:oddballs}.  Six of the 7 two-epoch TNOs have good geometric reasons for being missed in their the third epoch---the orbit could be in a gap or off the edge of the detectors, or, in the case of JPB24, atop a bright galaxy (we recovered this track).  The only mysterious case is JPB25, which was imaged in a high-quality region of a detector in epoch 2 for any plausible distance, but further searching for $S/N>4.5$ tracks turned up nothing.  An extreme--but not impossible-- $2\times$ flux dimming during this epoch could explain its disappearance (e.g., contact binary shapes \citep{showalter_statistical_2021}).

The net result of the search for TNOs is thus as follows:
\begin{itemize}
    \item 21 TNOs (including \gk) and 1 centaur, that have reliable detections in all three JWST epochs.
    \item 2 TNOs, JPB02 and JPB03, that link three epochs only by bypassing the ResNet in one epoch, and are too faint to be considered reliable.
    \item 7 TNOs that were detected at $S/N>8$ in two epochs and undetected otherwise.
    \item 25 single-epoch detections at $S/N>8$ that do not link with detections in another epoch.
\end{itemize}

\begin{figure}[h]
\centering
    \includegraphics[scale=0.53]{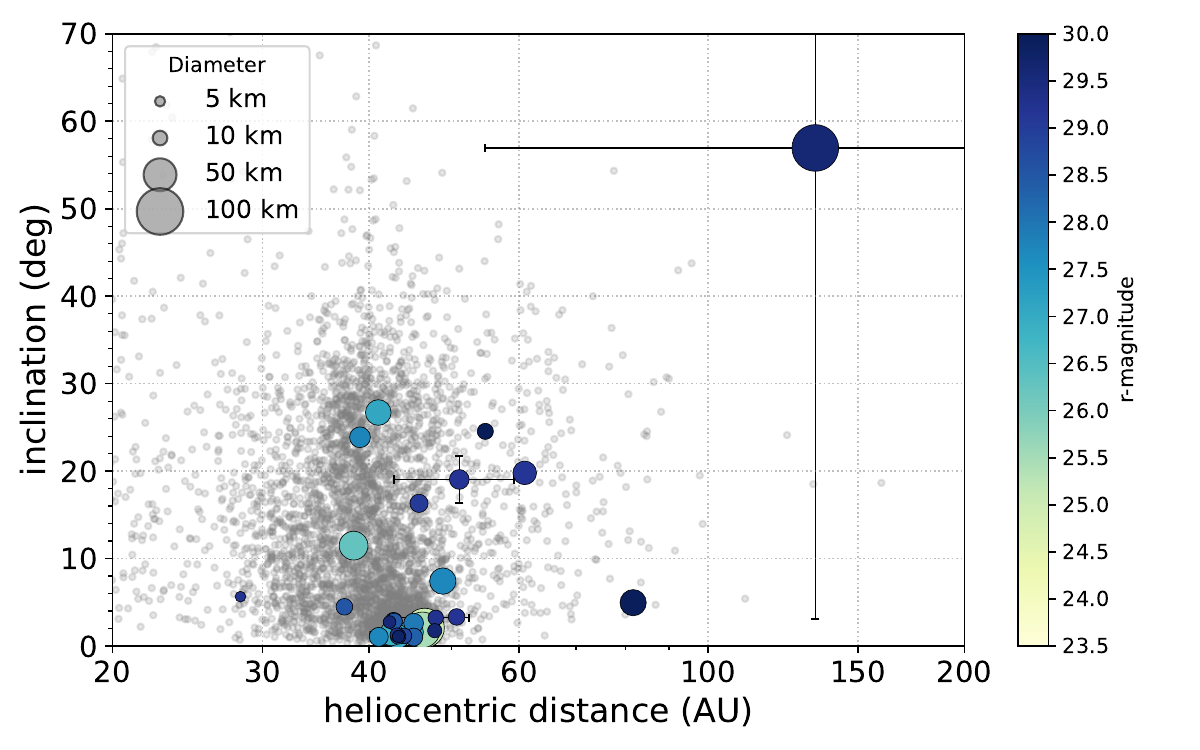} 
    \includegraphics[scale=0.53]{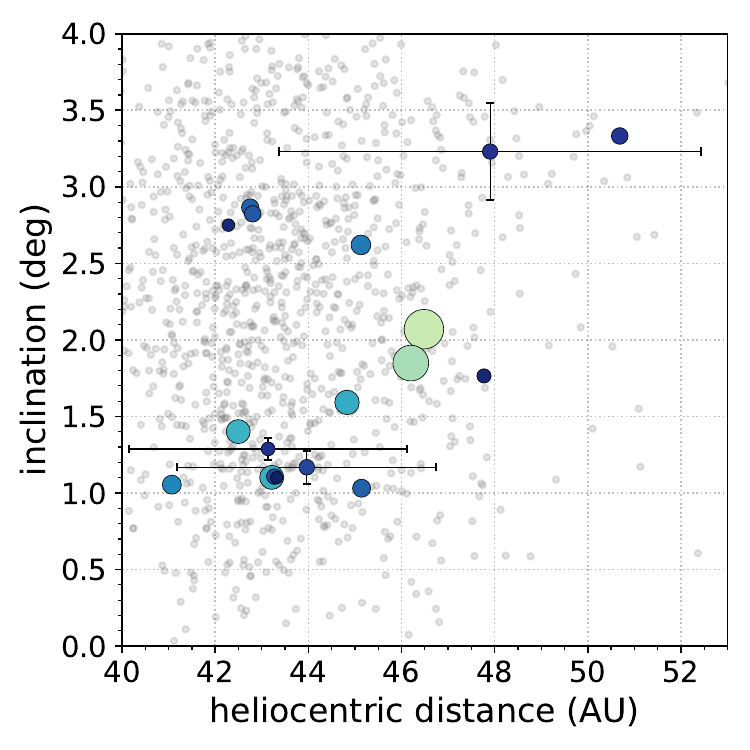} 
    \caption{\textbf{Left:} Inclination and heliocentric distances of 30 objects detected in three or two epochs of this survey, color-coded by their brightness. The marker sizes are scaled to their diameters in kilometer by assuming a 15\% albedo for the cold classicals and 8\% for the hot classicals. Heliocentric distances and inclinations are taken from Table~\ref{tab:objects}; for objects with a range of values, the median is plotted and error bars represent the minimum and maximum of the range. \textbf{Right:} Zoomed-in view of the left panel, highlighting the detections in the cold classical kernel. In both panels, gray circles represent known TNOs from the MPC.}
\label{fig:final_dets}
\end{figure}

Summaries of the available photometric and dynamical information on the 30 2- or 3-epoch detections are given in Table~\ref{tab:objects} and shown in Figure \ref{fig:final_dets}, and for the 25 singlets in Table~\ref{tab:singlets}.  While the singlets are certainly real sources, we do not have sufficient information to deem them to be outer-solar-system bodies.  Visual inspection shows that 12 of them (marked as ``static'' in Table~\ref{tab:singlets}) lie atop images of bright galaxies and stars, and all of these have measured rates of motion that are within $3\sigma$ of being zero.  It thus seems likely that these are all variable stars in the Milky Way or other galaxies.  Another (JPB39) has an apparent rate of motion that requires it to be an outer asteroid.  The dynamical circumstances of the other 12 sources are indeterminate, but we include their information in the \modify{Table \ref{tab:singlets}} and in the auxiliary data files for reference.

\begin{figure}
\includegraphics[width=\textwidth]{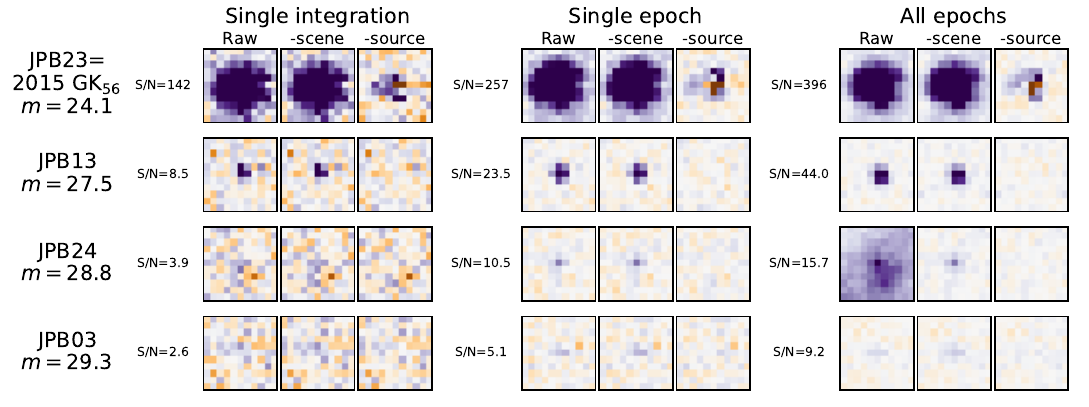 }
\caption{Each row shows representative images of TNOs from the F150W2 search images.  Each triplet of cutouts has the original image, the image with background model subtracted, and the image with the TNO model also subtracted. The groups, from left to right, are: images from a single 120~s integration; from a stack of 6 such integrations in the first observing epoch; and from a stack of all 3 epochs' images. The top row is \gk, which was  intentionally targeted and is the brightest moving object in the field, and shows significant model residuals due to the finite accuracy of our PSF model.  The second row is JPB13, which is $23\times$ fainter and still confidently detected even in single integrations.  Third is JPB24, our faintest confidently detected source, $3\times$ fainter than JPB13, which requires stacking over an epoch to be clearly visible.  The bottom row is JPB03, another $1.6\times$ fainter, which failed the ResNet filter for one epoch and is in a brightness regime where our discrimination between real and spurious sources is too weak to estimate the real population accurately.}
\label{fig:gallery}
\end{figure}

Figure~\ref{fig:gallery} shows the F150W2 images of TNOs spanning the range of magnitudes found in this survey, from \gk, previously discovered in ground-based images, at $m=24.1,$ to JPB03, which is $>100\times$ fainter at $m=29.3,$ where we cannot assure freedom from false positives. No objects this faint produced three ResNet-approved tracks passing single-epoch SMP cuts.

%
\begin{longrotatetable}
\begin{deluxetable}{ccccccc|cccc}
  \tablewidth{0pt}
  \tabletypesize{\small}
\tablecaption{Summary of each object found in $\ge2$ epochs of the
  JWST data. \label{tab:objects}}
\tablehead{
  \colhead{Name} &
  \colhead{$m_{\rm F150W2}$} &
  \colhead{$f_{322}/f_{150}$} &
  \colhead{$d_{\rm bary}$} &
  \colhead{$i$} &
  \colhead{$e_{\rm min}$} &
  \colhead{$a_{\rm min}$} &
  \multicolumn{4}{c}{$m_{\rm F150W2}$, date / single-epoch SNR\tablenotemark{b}} \\
samples\tablenotemark{a} & (AB mag) &  & (au) & & & (au)  & Epoch 1 & Epoch 2 & Epoch 3 &  duplicate }
\startdata
JPB23 & $24.06\pm0.00$ & $0.18\pm0.00$ & 46.48 & 2.07\arcdeg & 0.05 & 44.3 
    & $23.80\pm0.00$ & $24.10\pm0.00$ & $24.34\pm0.01$  & \nodata  \\
$(=$2015 GK$_{56})$ C & & & & & &  & -4.941 / 245.4 & -0.447 / 241.8 & +5.643 / 182.2 &   \\[5pt]
JPB04 & $24.47\pm0.00$ & $0.17\pm0.00$ & 46.20 & 1.85\arcdeg & 0.06 & 43.7 
    & $24.48\pm0.01$ & $24.49\pm0.01$ & $24.49\pm0.01$ & $24.43\pm0.01$  \\
AC & & & & & &  & -5.438 / 174.9 & -0.861 / 200.0 & +5.156 / 176.7 & -5.462 / 178.2* \\[5pt]
JPB22 & $25.29\pm0.01$ & $0.31\pm0.00$ & 38.37 & 11.48\arcdeg & 0.03 & 37.1 
    & $25.31\pm0.01$ & $25.28\pm0.01$ & $25.27\pm0.01$  & \nodata  \\
A & & & & & &  & -5.438 / 115.1 & -0.861 / 123.2 & +5.156 / 123.6 &   \\[5pt]
JPB27 & $25.84\pm0.01$ & $0.17\pm0.01$ & 42.49 & 1.40\arcdeg & 0.02 & 43.5 
    & $25.77\pm0.01$  & \nodata & $25.90\pm0.01$  & \nodata  \\
BC & & & & & &  & -5.438 / 97.0 &  & +5.156 / 86.4 &  \\[5pt]
JPB20 & $25.89\pm0.01$ & $0.17\pm0.01$ & 43.21 & 1.10\arcdeg & 0.03 & 44.5 
    & $25.89\pm0.01$ & $25.85\pm0.01$ & $25.90\pm0.01$ & $25.91\pm0.01$  \\
AC & & & & & &  & -4.964 / 82.0 & -0.470 / 81.9 & +5.506 / 88.0 & -0.541 / 82.8 \\[5pt]
JPB21 & $25.98\pm0.01$ & $0.15\pm0.01$ & 44.83 & 1.59\arcdeg & 0.02 & 44.2 
    & $25.97\pm0.01$ & $25.98\pm0.01$ & $26.00\pm0.01$ & $25.99\pm0.01$  \\
AC & & & & & &  & -5.365 / 90.4 & -0.518 / 84.0 & +5.530 / 87.8 & -0.541 / 77.5 \\[5pt]
JPB19 & $26.07\pm0.01$ & $0.36\pm0.01$ & 41.02 & 26.70\arcdeg & 0.01 & 41.6 
    & $26.10\pm0.01$ & $26.07\pm0.01$ & $26.05\pm0.01$  & \nodata  \\
A & & & & & &  & -5.581 / 79.5 & -0.932 / 77.6 & +5.009 / 82.9 &   \\[5pt]
JPB01 & $26.70\pm0.01$ & $0.18\pm0.02$ & 41.07 & 1.05\arcdeg & 0.07 & 44.4 
    & $26.72\pm0.02$ & $26.70\pm0.02$ & $26.69\pm0.02$  & \nodata  \\
AC & & & & & &  & -4.869 / 50.6 & -0.375 / 51.2* & +5.715 / 54.2 &   \\[5pt]
JPB17 & $26.71\pm0.01$ & $0.12\pm0.01$ & 48.83 & 7.43\arcdeg & 0.07 & 45.8 
    & $26.73\pm0.02$ & $26.67\pm0.02$ & $26.72\pm0.02$  & \nodata  \\
A & & & & & &  & -4.893 / 48.8 & -0.399 / 50.1 & +5.691 / 53.6 &   \\[5pt]
JPB18 & $26.76\pm0.01$ & $0.34\pm0.02$ & 39.04 & 23.87\arcdeg & 0.25 & 52.1 
    & $26.71\pm0.02$ & $26.86\pm0.02$ & $26.73\pm0.02$  & \nodata  \\
A & & & & & &  & -4.964 / 53.1 & -0.470 / 45.2 & +5.620 / 48.1 &   \\[5pt]
JPB05 & $26.94\pm0.02$ & $0.13\pm0.02$ & 45.13 & 2.62\arcdeg & 0.02 & 44.3 
    & $26.73\pm0.02$ & $27.15\pm0.03$ & $26.97\pm0.03$  & \nodata  \\
AC & & & & & &  & -4.869 / 45.5 & -0.375 / 33.3 & +5.715 / 40.0* &   \\[5pt]
JPB16 & $27.28\pm0.02$ & $0.13\pm0.02$ & 42.75 & 2.86\arcdeg & 0.04 & 44.3 
    & $27.33\pm0.03$ & $27.24\pm0.03$ & $27.27\pm0.03$  & \nodata  \\
AC & & & & & &  & -5.557 / 31.7 & -0.979 / 32.4 & +5.034 / 33.9 &   \\[5pt]
JPB15 & $27.34\pm0.02$ & $0.14\pm0.03$ & 45.14 & 1.03\arcdeg & 0.02 & 44.1 
    & $27.35\pm0.03$ & $27.38\pm0.04$ & $27.27\pm0.03$  & \nodata  \\
AC & & & & & &  & -5.462 / 30.9 & -0.885 / 29.2 & +5.131 / 33.3 &   \\[5pt]
JPB13 & $27.49\pm0.02$ & $0.10\pm0.02$ & 42.80 & 2.82\arcdeg & 0.00 & 42.9 
    & $27.65\pm0.05$ & $27.54\pm0.04$ & $27.43\pm0.04$ & $27.35\pm0.04$  \\
AC & & & & & &  & -4.917 / 23.7 & -0.422 / 24.3 & +5.667 / 27.8 & -0.399 / 26.2 \\[5pt]
JPB14 & $27.54\pm0.03$ & $0.22\pm0.03$ & 37.43 & 4.49\arcdeg & 0.38 & 60.7 
    & $27.48\pm0.04$ & $27.53\pm0.04$ & $27.61\pm0.05$  & \nodata  \\
A & & & & & &  & -4.869 / 24.6 & -0.375 / 23.9 & +5.715 / 23.6 &   \\[5pt]
JPB12 & $27.79\pm0.03$ & $0.18\pm0.04$ & 43.27 & 1.11\arcdeg & 0.01 & 42.8 
    & $27.72\pm0.05$ & $27.92\pm0.06$ & $27.74\pm0.05$  & \nodata  \\
AC & & & & & &  & -4.964 / 19.8 & -0.470 / 18.2 & +5.620 / 21.1 &   \\[5pt]
JPB29 & $27.84\pm0.04$ & $0.09\pm0.07$ & 41.2--46.9 & 1.3\arcdeg--1.1\arcdeg & 0.02--0.14 & 40.4--54.2
    & $27.81\pm0.06$  & \nodata & $27.87\pm0.06$  & \nodata  \\
BC & & & & & &  & -4.823 / 19.3 &  & +5.762 / 18.6 &  \\[5pt]
JPB11 & $28.04\pm0.03$ & $0.26\pm0.04$ & 45.79 & 16.32\arcdeg & 0.18 & 38.6 
    & $28.03\pm0.07$ & $27.96\pm0.06$ & $27.93\pm0.05$ & $28.28\pm0.08$  \\
A & & & & & &  & -4.917 / 15.5 & -0.422 / 16.4 & +5.667 / 19.6 & +5.506 / 13.7 \\[5pt]
JPB10 & $28.15\pm0.04$ & $0.33\pm0.06$ & 60.93 & 19.80\arcdeg & 0.16 & 52.4 
    & $28.06\pm0.07$ & $28.10\pm0.07$ & $28.31\pm0.08$  & \nodata  \\
A & & & & & &  & -5.533 / 15.6 & -0.956 / 15.9 & +5.034 / 13.9 &   \\[5pt]
JPB30 & $28.17\pm0.05$ & $0.37\pm0.07$ & 42.8--60.2 & 16.3\arcdeg--22.2\arcdeg & 0.06--0.15 & 40.5--70.5
     & \nodata & $28.10\pm0.07$ & $28.24\pm0.08$  & \nodata  \\
B & & & & & &  &  & -0.956 / 15.0 & +5.059 / 13.7 &  \\[5pt]
JPB08 & $28.19\pm0.04$ & $0.14\pm0.06$ & 50.69 & 3.33\arcdeg & 0.14 & 59.1 
    & $28.23\pm0.07$ & $28.21\pm0.08$ & $28.12\pm0.07$  & \nodata  \\
AC & & & & & &  & -5.485 / 14.2 & -0.885 / 14.2 & +5.131 / 14.8 &   \\[5pt]
JPB09 & $28.23\pm0.04$ & $0.28\pm0.06$ & 28.26 & 5.65\arcdeg & 0.12 & 25.2 
    & $28.25\pm0.07$ & $28.19\pm0.07$ & $28.24\pm0.07$  & \nodata  \\
A & & & & & &  & -5.365 / 15.1 & -0.518 / 15.5 & +5.530 / 15.4 &   \\[5pt]
JPB28 & $28.26\pm0.05$ & $0.19\pm0.08$ & 43.4--48.2 & 3.6\arcdeg--3.2\arcdeg & 0.03--0.17 & 42.2--58.3
    & $28.32\pm0.08$ & $28.19\pm0.07$  & \nodata  & \nodata  \\
BC & & & & & &  & -4.964 / 13.8 & -0.470 / 15.2 &  &  \\[5pt]
JPB26 & $28.29\pm0.05$ & $0.27\pm0.06$ & 40.5--48.8 & 1.3\arcdeg--1.2\arcdeg & 0.09--0.26 & 37.0--65.8
    & $28.45\pm0.10$ & $28.20\pm0.07$  & \nodata & $28.24\pm0.07$  \\
BC & & & & & &  & -4.894 / 11.3 & -0.056 / 15.2 &  & -4.775 / 15.0 \\[5pt]
\\
JPB06 & $28.58\pm0.06$ & $0.29\pm0.08$ & 42.28 & 2.75\arcdeg & 0.04 & 44.0 
    & $28.72\pm0.12$ & $28.43\pm0.09$ & $28.61\pm0.10$  & \nodata  \\
AC & & & & & &  & -5.509 / 9.1 & -0.932 / 11.4 & +5.083 / 11.1 &   \\[5pt]
JPB25 & $28.60\pm0.06$ & $0.41\pm0.11$ & 54.7--279.4 & 3.3\arcdeg--164.4\arcdeg & 0.05--0.08 & 57.6--258.4
     & \nodata & $28.73\pm0.13$ & $28.40\pm0.09$ & $28.72\pm0.12$  \\
B & & & & & &  &  & -0.494 / 8.6 & +5.131 / 12.2 & +5.595 / 8.7 \\[5pt]
JPB07 & $28.60\pm0.07$ & $0.16\pm0.08$ & 47.77 & 1.76\arcdeg & 0.05 & 45.7 
    & $28.56\pm0.11$ & $28.60\pm0.12$ & $28.66\pm0.12$  & \nodata  \\
AC & & & & & &  & -5.365 / 9.6 & -0.518 / 9.4 & +5.530 / 8.9 &   \\[5pt]
JPB24 & $28.79\pm0.07$ & $-0.09\pm0.11$ & 43.32 & 1.10\arcdeg & 0.00 & 43.4 
    & $28.65\pm0.10$ & $29.00\pm0.17$ & $28.73\pm0.11$  & \nodata  \\
BC & & & & & &  & -5.509 / 10.2 & -0.932 / 7.4* & +5.083 / 9.7 &   \\[5pt]
JPB02 & $28.97\pm0.08$ & $0.36\pm0.12$ & 81.67 & 4.96\arcdeg & 0.31 & 62.2 
    & $28.98\pm0.14$ & $28.88\pm0.13$ & $29.05\pm0.15$  & \nodata  \\
 & & & & & &  & -5.509 / 7.8* & -0.932 / 9.0 & +5.083 / 7.1 &   \\[5pt]
JPB03 & $29.34\pm0.12$ & $-0.07\pm0.16$ & 54.78 & 24.54\arcdeg & 0.65 & 33.1 
    & $29.30\pm0.21$ & $29.37\pm0.22$ & $29.34\pm0.19$  & \nodata  \\
 & & & & & &  & -5.388 / 5.5 & -0.541 / 5.4 & +5.530 / 5.9* &   \\
\enddata
\tablecomments{
The distance from the solar system barycenter is $d_{\rm bary}$; $a_{\rm min}$ and $e_{\rm min}$ are the smallest possible values, \textit{i.e.} the case of zero radial velocity; the LW/SW flux ratio is defined as unity for equal $AB$ magnitudes. Ranges for orbital parameters within $2\sigma$ of the maximum likelihood are given when astrometry is available for only two epochs, otherwise the uncertainties are small.
Dates of observations are MJD$-$59974.}
\tablenotetext{a}{Membership in characterized sample A or B is noted by the matching letter.  ``C'' indicates a member of dynamically cold group.}
\tablenotetext{b}{The signal-to-noise ratio is that measured for the single-epoch track, not forced to fit an orbit.  An asterisk after the SNR indicates that the track was not discovered through normal procedures and cuts.  See Appendix~\ref{sec:oddballs} for explanations.}
\end{deluxetable}
\end{longrotatetable}

\begin{deluxetable}{ccccc}
  \tablewidth{0pt}
  \tabletypesize{\small}
\tablecaption{Unlinked detections}
\label{tab:singlets}
\tablehead{\colhead{Name} & \colhead{$m_{\rm F150W2}$} &
  \colhead{$f_{322}/f_{150}$} & \colhead{Rate} & \colhead{Static?} \\
 & (AB mag) & & (\arcsec/day) & }
\startdata
JPB53 & $26.56\pm0.02$ & $0.22\pm0.03$ & $(+3.65,+6.94)\pm (0.13,0.13)$  & \\
JPB47 & $27.13\pm0.04$ & $0.45\pm0.06$ & $(-0.28,-0.51)\pm (0.23,0.22)$  & * \\
JPB44 & $27.28\pm0.06$ & $0.35\pm0.10$ & $(-0.45,+0.24)\pm (0.35,0.35)$  & * \\
JPB41 & $27.43\pm0.05$ & $0.37\pm0.07$ & $(-0.18,-0.07)\pm (0.33,0.32)$  & * \\
JPB49 & $27.53\pm0.04$ & $0.56\pm0.07$ & $(+0.40,+0.29)\pm (0.27,0.26)$  & * \\
JPB52 & $27.68\pm0.06$ & $0.59\pm0.12$ & $(+0.37,+0.34)\pm (0.37,0.37)$  & * \\
JPB40 & $27.87\pm0.06$ & \nodata             & $(+7.93,-2.37)\pm (0.37,0.34)$  & \\
JPB42 & $27.89\pm0.06$ & $1.33\pm0.12$ & $(+0.25,+0.25)\pm (0.35,0.36)$  & * \\
JPB45 & $28.29\pm0.10$ & $0.80\pm0.20$ & $(-1.16,-0.02)\pm (0.62,0.59)$  & * \\
JPB34 & $28.31\pm0.13$ & $0.43\pm0.15$ & $(-1.60,-1.05)\pm (0.68,0.67)$  & * \\
JPB32 & $28.35\pm0.11$ & $0.27\pm0.20$ & $(-1.39,-1.26)\pm (0.62,0.62)$  & * \\
JPB39 & $28.37\pm0.09$ & $0.35\pm0.12$ & $(+37.13,-14.63)\pm (0.66,0.64)$  & \\
JPB31 & $28.37\pm0.10$ & $0.46\pm0.15$ & $(-1.28,-0.11)\pm (0.51,0.51)$  & * \\
JPB48 & $28.43\pm0.09$ & $0.39\pm0.12$ & $(-0.20,-1.39)\pm (0.56,0.56)$  & \\
JPB43 & $28.52\pm0.12$ & $0.44\pm0.20$ & $(+0.45,+0.02)\pm (0.75,0.74)$  & * \\
JPB55 & $28.58\pm0.11$ & $-0.19\pm0.15$ & $(-2.36,+2.93)\pm (0.71,0.70)$  & \\
JPB46 & $28.60\pm0.12$ & $0.26\pm0.14$ & $(+4.05,-7.44)\pm (0.76,0.78)$  & \\
JPB37 & $28.63\pm0.11$ & $0.20\pm0.15$ & $(+4.87,+6.75)\pm (0.74,0.73)$  & \\
JPB35 & $28.65\pm0.11$ & \nodata             & $(+5.79,-1.52)\pm (0.74,0.66)$  & \\
JPB36 & $28.70\pm0.11$ & $0.15\pm0.16$ & $(-7.90,-1.13)\pm (0.61,0.59)$  & \\
JPB51 & $28.73\pm0.13$ & $1.13\pm0.34$ & $(-1.41,-0.28)\pm (0.77,0.76)$  & * \\
JPB54 & $28.77\pm0.12$ & $0.02\pm0.17$ & $(+4.25,-4.38)\pm (0.77,0.75)$  & \\
JPB50 & $28.86\pm0.13$ & $-0.10\pm0.17$ & $(+1.82,-6.49)\pm (0.88,0.86)$  & \\
JPB38 & $28.88\pm0.13$ & $0.39\pm0.18$ & $(+0.63,+1.60)\pm (0.96,0.96)$  & \\
JPB33 & $28.93\pm0.13$ & $0.30\pm0.22$ & $(+2.07,-0.51)\pm (0.77,0.79)$  & \\
\enddata
\tablecomments{Single-epoch detections that are not plausible links to
  any other detections.  The ``Static?'' column is marked if the
  object appears atop a static source (galaxy or star), indicative of
  it being a variable star rather than a solar system body.}
\end{deluxetable}
%

\subsection{Orbital characteristics}
\label{sec:elements}
Detection of a TNO in three distinct epochs over the 10-day JWST observing period yields precise estimates of 5 of the 6 degrees of freedom in the TNO's state vector.  The apparent angular acceleration of the TNO is the reflex of the Earth's acceleration transverse to the line of sight, and determines the observer-TNO distance to high precision.  The transverse angular position and velocity are very well measured as well, and combine with the distance to determine the full Cartesian space position, and the 2 transverse components of its velocity.  Any line-of-sight velocity, however, has an effect on the observed positions that is smaller than the systematic errors in establishing the JWST astrometric solutions (see Section~\ref{sec:alignment}), and is therefore either undetectable or unreliable in our solutions.  \textit{All of the orbital solutions used in this work set the line-of-sight velocity to zero.}  The orbit-level SMP for the 3-epoch TNOs yields an estimate of this restricted state vector, and its uncertainties, which are reported in the auxiliary data files.  State vectors and derived quantities are given for the nominal epoch JD 2459974.5 (TDB).

Since the line-of-sight velocity is almost the same as the radial velocity from the solar system barycenter, the barycentric distance and angular momentum of a TNO are well determined, but we can only set a lower bound on its total energy.  We report in Table~\ref{tab:objects} the barycentric distance $d_{\rm bary}$ of the TNO at the nominal time  and the ecliptic inclination angle $i$ of the TNO, both of which are well constrained. The semi-major axis $a$ and eccentricity $e$ are, however, degenerate because of the unknown line-of-sight velocity $v_\parallel,$ so we do not report them.  We can, however, ascertain the \emph{minimum} possible values of these elements, which occur for $v_\parallel=0,$ as 
\begin{eqnarray}
    e_{\rm min} & = & \left|1 - \frac{r v_\perp^2}{GM}\right|,  \\
    a_{\rm min} & = & r \left(2-\frac{rv_\perp^2}{GM}\right)^{-1},
 \end{eqnarray} 
where $r$ is the barycentric radial location, $v_\perp$ the non-radial component of velocity, and $M$ the solar system mass.  These quantities are reported in Table~\ref{tab:objects} and the auxiliary data files (Appendix~\ref{sec:products}), and have unimportantly small measurement errors for the 3-epoch TNOs.

For the 2-epoch TNOs, there is, as noted above, an additional degeneracy among the state vector elements because the weak measure of the angular acceleration leaves the radial distance $d_{\rm bary}$ is weakly constrained.  For JPB27, this degeneracy is broken by clear detection of the body on the contemporaneous HST WFC3 images of the field reported in \citet{ana}. For the other 5 2-epoch objects, we can determine $a_{\rm min},$ $e_{\rm min},$ and the expected sky position and rate of the body in any observation by asserting a value of $d_{\rm bary}.$  We then scan the range of $d_{\rm bary}$ to determine the following:
\begin{enumerate}
    \item The $d_{\rm bary,ML}$ value that produces the minimum $\chi^2$ of the difference between the orbit model and the measured positions and rates at the two epochs.
    \item The bounding values $(d_{\rm bary,min},d_{\rm bary,max})$ that produced $\chi^2$ values that are 4 larger than the minimal $\chi^2$ value from step (1).  This is a not-rigorous $\pm2\sigma$ confidence region.
    \item We clip $(d_{\rm bary,min},d_{\rm bary,max})$ to the range that yields a bound orbit.
    \item We eliminate from the $(d_{\rm bary,min},d_{\rm bary,max})$ interval any values of $d_{\rm bary}$ for which the TNO would have fallen onto a usable part of a NIRCam SW detector in the missing epoch.  This part of the inference depends upon an assumption that each of these objects was bright enough in its third epoch to be detectable at near 100\% efficiency.
\end{enumerate}
Appendix~\ref{sec:oddballs} reports the constraint methods applicable to each 2-epoch TNO.  The columns for $d_{\rm bary}, i, a_{\rm min},$ and $e_{\rm min}$ in Table~\ref{tab:objects} for the 2-epoch TNOs report the range of values obtained across the inferred possible range of $d_{\rm bary}.$  The auxiliary data described in Appendix~\ref{sec:products} contain all of the dynamical properties at $d_{\rm bary,min}, d_{\rm bary,ML},$ and $d_{\rm bary,max}$.

An additional detail about the determination of orbital elements: for those TNOs that are detected on 2 distinct tiles in a given epoch, we make use of only one observation per epoch when determining the state vector and elements.  This is because the systematic uncertainty in astrometric registration of different tiles (Section~\ref{sec:alignment}), combined with the small time interval between the observations and very high precision of the position determinations, can force the orbit to fit a spuriously high rate of motion between these two observations.  We obtain a more reliable result by omitting one, preferably retaining a measure that is on the same tile as the other epochs' detections.  

\subsection{Long-wave data}
\label{sec:longwave}
Every NIRCam exposure collected data in filter F150W2 on the short-wave (SW) detectors, and in filter F322W2 on the long-wave (LW) detectors.  We do not search the LW images for TNOs since the SW images will reach higher $S/N$ for every TNO.  No implantation of synthetic sources is done for the LW data since we do not need to characterize discovery efficiencies.  We can, however, measure LW fluxes for all objects found in the SW data.  We do this in a ``forced photometry'' mode of the SMP algorithms, whereby the celestial coordinates of the moving TNO are taken from the orbit fit to the SW data, and only the LW fluxes are left as free parameters. Thus a full F322W2 lightcurve is obtained for every detected TNO, albeit at a lower $S/N$ which leaves many of the objects without significant flux detections even when stacking all exposures together.  Rather than assigning magnitudes to these low-$S/N$ fluxes, we present in 
Tables~\ref{tab:objects} and \ref{tab:singlets} the ratio of each object's F322W2 flux to its flux in F150W2, in units such that a ratio of 1 implies equal AB magnitudes in each band.

Processed data tables giving SW and LW pixel data and derived orbital and photometric properties for all of the objects listed in Tables~\ref{tab:objects} and \ref{tab:singlets} are made publicly available with this paper at DOI 

\subsection{Characterized samples}
\label{sec:selectAB}

To constrain the magnitude or size distributions of TNOs, we need to define selection criteria for observed TNOs such that (1) we are confident that our detected source list is free of false positives, and (2) we can derive, from our implanted sources, the detection efficiency of the survey for the given sample.  We create two disjoint samples that should satisfy these conditions.

\textbf{Sample A} is defined as those TNOs detected in 3 epochs by KBMOD, passing ResNet cuts, passing the single-epoch SMP cuts of Section~\ref{sec:trackqc}, and passing full-orbit SMP cuts of Equations~(\ref{eq:cuts}).  In other words, those objects found and linked through the ``normal procedures'' described in Sections~\ref{sec:tnosearch} and \ref{sec:3epoch}, where we demonstrate they should be free of false positives.  This includes 18 TNOs, after we exclude the targeted \gk.  To these sources, we add JPB01 and JPB05, because they failed the ``normal procedure'' only because of a likely overtraining bias of the ResNet against high-$S/N$ real sources.  Sample A thus contains 20 3-epoch detections. 

\textbf{Sample B} is defined as those sources with NIRCam detections in only two of the three JWST observing epochs, both with $S/N>8$ in the SMP analyses of their single-epoch tracks.  The absence of false positives in this sample is assured because of the relative ease of distinguishing real from spurious sources by eye at the higher $S/N$ threshold, and also because there are few enough of these high-$S/N$ single-epoch detections (only $\approx50$, see Figure~\ref{fig:track_counts}) that the chances of two of them spuriously being linked into a common TNO orbit are small.  We detect 7 TNOs meeting the sample B conditions.

The reasons that a TNO might be missed in one epoch and end up in sample B instead of A would be:
\begin{enumerate}
    \item Geometry---the object is simply not within the field of view of any images from the missing epoch.  This is definitely the case for two of the sample B members, and plausible for three others.
    \item Hiding---the object was lost to poor subtraction of a brighter overlapping source.  This was the case for JPB24, which we did later locate in the 3rd epoch, but we leave it in sample B.
    \item Light curve---the object faded below the $S/N>5$ threshold applied to each epoch for sample A during one of the epochs.  This may have been the case for JPB25.
\end{enumerate}

The search efficiency, or equivalently the effective search solid angle, for both samples can be derived from the implant results.  For an implanted or real TNO to be detected in some epoch through the normal procedures, two events must occur:
first, the TNO must fall onto one of the NIRCam SW detectors.  Second, if on the image, the source must be detected by KBMOD, be approved by the ResNet, and yield acceptable SMP $S/N.$  We can use the implants to determine the probabilities of these two events separately.

The geometric probability of a TNO landing on an imaged region in 2 or 3 different epochs can be determined numerically simply by counting the number of implants that do so.  If we assume that the orbital distribution is independent of the flux distribution for small TNOs, we can use \emph{all} of the implants, regardless of whether they are detected, to calculate the effective area on the sky of the survey for 2- and 3-epoch capture of TNOs by NIRCam.  Doing so for the implants at distances $35<d_{\rm bary}<55$~\au\ and inclinations $i<5^\degree$ yields effective areas of $A_1=29, A_2=56,$ and $A_3=139$~arcmin$^2$ for being imaged onto a NIRCam SW detector in 1, 2, and 3 epochs, respectively.  (We do not count apparitions within 15 pixels of the edge of the detector.) For the ``hot'' implants with $i>5\degree$ in the same distance range, the values are 60, 81, and 128~arcmin$^2$---a slightly larger chance of being imaged at all because their apparent motion is larger over the 10-day survey, but also a larger chance of appearing in a gap or falling off the array in at least one epoch.

In Section~\ref{sec:completeness}, we derived a probability $p_{\rm det}(m)$ of any apparition of a TNO of magnitude $m$ creating a track that passes quality cuts, and a variant we will call $p_8(m)$ of this track having $S/N>8.$  Under the validated assumption of independent detection probability in each epoch, and ignoring light curve variations of a given TNO, the effective area of sample A at magnitude $m$ becomes $A_3 p_{\rm det}^3(m).$

The effective area of sample B is a bit more complex as a TNO can enter the sample in two ways.  First, it can appear in only two epochs, and be detected at $S/N>8$ in both, yielding effective area of $A_2 p_8^2(m).$  Alternatively it can appear on the NIRCam array in three epochs, achieve $S/N>8$ on two of them, but fail to be detected in the third epoch (else it would be in sample A).  The effective area for this detection channel is hence $3 A_3 p_8^2(m)\left[1-p_{\rm det}(m)\right].$  

At magnitudes $m<28$ where $p_{\rm det}=p_8=0.96,$ the above formulae imply the expected ratio of effective areas for sample B to sample A (and hence the expected ratio of number of detections) would be $\left\langle N_B/N_A\right\rangle=3(1-p_8) + A_2/p_8 A_3=0.54$ and 0.78 for the cold and hot objects, respectively.  Our search yielded $N_A/N_B=5/12$ and $2/8$ for these two dynamical groups, both within the 91\% probability region of expected count ratios under Poisson statistics.  The effective areas suggest that the number of single-epoch TNO detections at $m<28$ should be $<30\%$ of the number of $m<28$ TNOs in samples A$+$B, and a larger fraction at $m>28,$ once we exclude the likely static variables from the singlet list.  This is roughly what we find.

\section{Binary Detections}

The search presented above did not reveal any evidence of binarity in the detected objects. This may be the result of a bias to the preferential detection of singletons from our pipeline. For example, the ResNet training set contained only singleton implants, and so it may be possible that the ResNet preferentially classed partially resolved binary sources as false. 

To determine if this were the case, the ResNet was retrained using a training set with both binary and single TNO implants. Using a dataset containing binary implants in place of every singleton injected into the original search, we rerun the full calibration process (Section \ref{sec:dataprocessing}). To avoid injection into the identical pixels, each artificial binary was offset by $0.05\arcsec-0.32\arcsec$ ($ \approx 1.2-10.3$ SW pixels) from the location of the singleton implant. The primary binary sources were implanted with the same brightness as used for the singletons. The secondary components were assigned random brightness and spatial offsets, drawn uniformly from $1\%–100\%$ of the primary's brightness and $0.05\arcsec–0.20\arcsec$ ($\approx 1.2–6.5$ SW pixels), respectively. After the full calibration, subtraction, and shift-and-stack steps, the ResNet was retrained like discussed in Section \ref{sec:MLsearch}, but using detected implanted sources from both the singletons of the original search, and the binary implants.

After classification using this new ResNet model, all candidate non-implant sources were visually inspected. This new search only revealed the exact set of detections as found in the original search, and did not detect any binary sources.

An experiment was performed to assess the significance of a null binary set. In particular, we asked the question: if objects in this pencil-beam survey shared the same single-binary ratio and binary properties as the Solar System Origins Legacy Survey dataset (SSOLS; \cite{porter_detection_2024}), how many binaries should we have detected? To that end, we assume that binaries from this survey would have a similar primary-secondary brightness ratios and separation distributions ($\Delta r$, in units of the system's hill sphere) as the SSOLS dataset. We also assume that any hidden binaries share identical densities with their SSOLS counterparts. Under these assumptions, the apparent primary-secondary separation $\Delta r'$ of a hypothetical binary in our survey with brightness $m_{F150W2}$ is given by

\begin{equation}
    \Delta r' = \Delta r \left(\frac{m'}{m}\right)^{2/3}
    \label{eq:binarysep}
\end{equation}

\noindent
where $\Delta r$ is the apparent primary-secondary separation of the SSOLS binary, $m$ and $m'$ are the masses of the primary bodies from the SSOLS and this survey, respectively. Both $m$ and $m'$ are inferred directly from their apparent brightnesses assuming $(F606W-F150W2)=0.8$. 

For each object $i$ detected in our survey, we determined $\Delta r'_{ij}$ from Equation~\ref{eq:binarysep} for each binary $j$, observed by SSOLS. Then using the observed primary-secondary brightness ratio $\Delta m_j$, we determined the probability $P_{ij}$ of detection of such a system from the outcomes of the ResNet trained on the binary+singleton sample. 

Additionally, we also performed a simple numerical experiment to evaluate the consistency with the SSOLS dataset. For each detection $i$ in this survey, a random number was drawn; if it was less than or equal to the SSOLS observed binary fraction (24 resolved binaries in 204 targets), it was treated as a binary. For these binary detections, three additional random numbers were drawn to simulate three observation epochs, and a binary was flagged as detected if any of these numbers were less than the detection probability $P_{ij}$. We repeated this procedure across all survey detections for 5,000 trials to determine if in one iteration of the experiment, a binary was detected in at least one of three epochs. In only 9.5\% of cases was a binary TNO numerically detected, meaning that no detections would be expected in the vast majority of cases. We thus conclude that that the lack of binary TNO detections in this survey is fully consistent with scaled expectations from the SSOLS dataset and should not be interpreted as a change in binary properties of the fainter sources of this survey.

We highlight that the majority of weight in the above calculation is held by the three brightest detections of our survey, with $P_{ij}=0$ for all other targets. At such small sizes, only very widely separated binaries (with respect to their hill spheres; \citet{parker2011characterization}) would be detected.
\section{The Luminosity Function} \label{sec:LF}


 The goal of this section is to identify the functional forms of the differential surface density that are most consistent with the survey data. Given an assumed distribution $\Sigma(m)$ and a survey's areal coverage $\Omega$, the expected number of detections is given by:
\begin{equation}
    \bar{N}\left(\theta\right) = \Omega \int\eta(m) \ \Sigma(m,\theta) \  dm
\end{equation}
where the detection efficiency of the survey, $\eta(m)$, describes the probability of successfully detecting an object with mean magnitude $m$ that lies within the survey's geometric area, $\Omega$. The $\eta$ function is the detection probability described in Section~\ref{sec:completeness}. The vector $\theta$ holds any free parameters of the $\Sigma$ function. 
%

Given some assumed functional representation, the shape of the luminosity function can be quantified by applying a maximum likelihood fit to the observed data. With this approach we obtain parameter values that best describe the observed distribution of magnitudes in the data. So for a single survey with detections indexed by $i$ we have


\begin{equation}
    \mathcal{L}(\theta) \propto e^{-\bar{N}(\theta)} \  \prod_{i=1}^N  \  \Sigma(\hat{m}_i, \theta) \ \eta(\hat{m}_i)
    \label{eq:likelihood}
\end{equation}


We note that the assumptions leading to Equation \ref{eq:likelihood} differ somewhat from those adopted in previous studies \citep{gladman_pencil-beam_1998, gladman2001structure, petit_kuiper_2006, fraser_derivation_2008, fraser_size_2009, fraser2010luminosity, fraser2014absolute, napier2024decam}, which were all based from a statistical approach presented in \cite{loredo_accounting_2004}. Our final likelihood form was derived based on two assumptions: First, the measured magnitude $\hat{m}$ of each source is a good approximation of its true magnitude $m$, and second, the detection of a source does not necessarily result in a real and valid measurement of its brightness (i.e., the search process is non-deterministic). While for the observations presented here, we were able to measure a valid brightness for every detection, one can realize certain circumstances where this is not always be true. A detailed derivation of this likelihood function is presented in Appendix \ref{sec:derivation_likelihood}.

Despite our assumptions and the likelihood form that results from those assumptions, we tested all variations of the above assumptions. That is, we also fit using the form by \cite{loredo_accounting_2004} that omits the efficiency function in the second term, and we also tested both variations when assuming the observed likelihood is treated as a gaussian variable. In all cases, the resultant fits were nearly identical. For example, the FullA sample resulted in best fit alphas of $\alpha = 0.27^{+0.07}_{-0.06}, 0.26^{+0.07}_{-0.06}, 0.26^{+0.07}_{-0.06}, 0.27^{+0.07}_{-0.06}$ for the likelihood forms A ($\hat{m} \approx$ $m$, with $\eta(m)$ factor), B ($\hat{m} \sim$ gaussian variable, with $\eta(m)$ factor), C ($\hat{m} \approx$ $m$, no $\eta(m)$ factor), and D ($\hat{m} \sim$ gaussian variable, no $\eta(m)$ factor), respectively. We conclude that for this project, the adopted likelihood form makes little difference.



\subsection{Sample Selection}
 
As mentioned in Section \ref{sec:selectAB}, we created two disjoint detection samples to ensure the catalog is free of false positives and to measure the survey’s detection efficiency using implanted sources. For the following luminosity function analysis, we perform two separate analyses using these datasets: one based on all three-epoch detections (sample A), and another based on the combined set of three-epoch and two-epoch detections (sample A+B).

\begin{figure}[h]
    \centering
    \includegraphics[width=\textwidth]{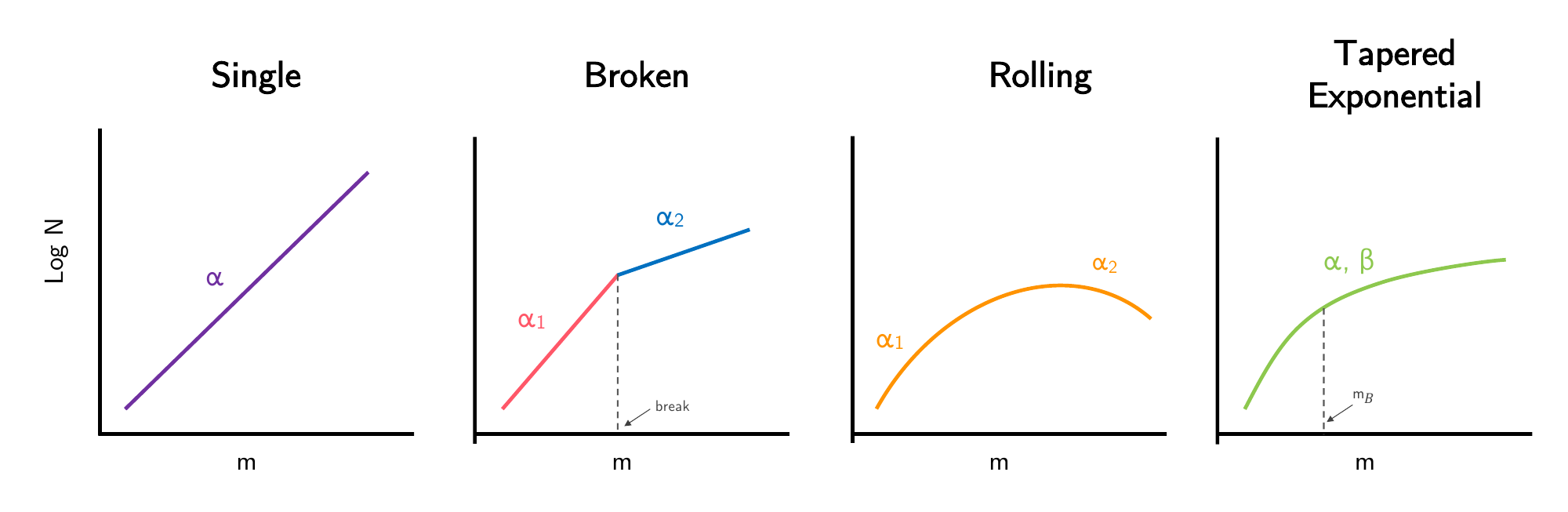}
    \caption{A schematic showing the four functional forms that describe the TNO size distribution. Each form describes the distribution of objects across different  magnitudes $m$ and are parameterized by specific slopes and transition points that reflect changes in object number density. Note that while our detections lie beyond the magnitude break, we include the broken power law form in this schematic simply for visual comparison.}
\label{fig:sd_shapes}
\end{figure}

We fit several functional forms forms of varying complexity to model our datasets (see Figure \ref{fig:sd_shapes}), and use Markov Chain Monte Carlo (MCMC) sampling to estimate parameter uncertainties and identify the best-fit solutions. For analysis, we define three dynamical groups in our survey:

\begin{enumerate}
    \item The \textit{full} sample includes all objects detected at barycentric distances $d > 30$ au. This cut excludes one centaur from our total detections, as its dynamical evolution likely differs from that of the current TNO population \citep{brasser2012oort, volk2013centaurs}.

    \item  The \textit{cold} population sample is a subset of the full object sample, defined by barycentric distances $38 < d < 55$ au and ecliptic inclinations $i \leq 5\degree$. 

    \item The \textit{hot} population is a complement of the cold sample in the full sample. These are objects with $d > 55$ au or within any distance range beyond Neptune but with inclinations $i > 5\degree$.  
    
\end{enumerate}

Our separation of the hot and cold classical components at $i=5\degree$ is not definitive.  It is the unforced component of the ecliptic inclination (\ie, free inclination) that shows a clear delineation between the dynamically cold, largely undisturbed population and the dynamically excited population of the Kuiper Belt \citep{van_laerhoven_ossos_2019, huang_free_2022}. While we do not have free inclinations for our detections, the hot and cold components share comparable apparent magnitude distributions, so a few misclassifications would not significantly change our results.

For the purposes of the $\Sigma(m)$ analysis, we transform our objects' $F150W2$ filter magnitudes to the conventional $r$-band magnitude. To estimate the color difference between the $F150W2$ filter (effective wavelength $\sim1.7\micro\text{m}$) and the $r$ filter (effective wavelength $\sim0.6\micro\text{m}$), we use the reflectance spectrum of the cold classical TNO Arrokoth \citep{grundy2020color, cruikshank_organic_2020}. We also calculate the $r-F150W2$ color of the Sun from values given in \cite{willmer_absolute_2018}. Taking all of these together, we adopt $r-F150W2 \approx 1.0 $. 

We point out that there remain possible systematic issues with inter-instrument calibrations between LEISA and MVIC that was the utilized for the Arrokoth spectrum in Figure \ref{fig:AlbedoSpec} \citep{howett_inflight_2017, protopapa_disk-resolved_2020, fayolle_testing_2021}. Such differences would directly manifest in the $r-F150W2$ estimate we present, and some appropriate caution is warranted.

\subsection{Single Power Law Fits}
The luminosity function of the Kuiper Belt as a whole, as well as its dynamically distinct cold and hot classical components, has been extensively studied over the past two decades \citep{gladman2001structure, bernstein2004size, fraser_kuiper_2008, fraser_derivation_2008, fuentes2009subaru, napier2024decam}. We build on the methodologies established in these previous works by applying functional forms of varying complexity to model our dataset and assess how well each captures the observed distribution. We begin by fitting a single power law model, described by

\begin{equation}
    \Sigma_{\text{single}}(m) = 10^{\alpha(m - m_0)}
    \label{eq:single_power}
\end{equation}
where $m_0$ is the magnitude at which the surface density equals one object per square degree, and $\alpha$ is the power law slope. 

Fitting this model to samples A and A+B yields best-fit slope of $\alpha = 0.27_{-0.07}^{+0.06}$ and $\alpha = 0.29_{-0.07}^{+0.06}$, respectively. For Sample A, we find $\alpha = 0.27_{-0.09}^{+0.08}$ for the cold population and $\alpha = 0.24_{-0.10}^{+0.09}$ for the hot population, with similarly consistent values obtained for Sample A+B. The agreement within uncertainties across all subsample indicates that the faint-end slope of the luminosity function does not differ significantly between dynamical groups (see Table~\ref{tab:bestfit_params}).

It is worth noting that earlier studies have already established that the luminosity function breaks or transition from a steep slope, for bright and large objects, to a shallower distribution for fainter and smaller objects. The transition or break magnitude occurs at $m_r \sim 25$ \citep{bernstein2004size,fuentes2009subaru, napier2024decam}. Our data consists exclusively of objects fainter than this break and therefore probes only the fainter end of the luminosity function. As a result, the observations do not provide meaningful constraints on the bright end behavior or the location of the break, and we therefore do not attempt to fit a broken power law model. Instead, a single power law is sufficient to characterize the faint-end slope sampled by the magnitude range of our observations.
 
\subsection{Rolling Power Law Fits}
As another level of complication, we fit a rolling power law model by Eq.~\ref{eq:likelihood}. Unlike the broken power law, which introduces a sharp transition between the bright-end and faint-end slopes ($\alpha_1$ and $\alpha_2$), the rolling power law allows for a gradual change in slope across magnitudes. This smoother transition is achieved by including a quadratic term in the exponent, allowing the logarithmic slope to change gradually with magnitude instead of shifting suddenly, and is given by

\begin{equation}
    \Sigma_{\text{rolling}}(m) = \Sigma_{23} 10^{\alpha_1(m - 23) + \alpha_2(m - 23)^2}
\end{equation}
where $\Sigma_{23}$ is the number of objects with $m_{r}=23$ per square degree, while $\alpha_1$ and $\alpha_2$ are the bright- and faint-end slopes, respectively, which control the shape of the function.

To account for the fact that our dataset provides no leverage on the luminosity function bright-ward of $m_r \sim 25$, we adopt the fitting results from \cite{napier2024decam} as informative priors for our MCMC sampling. We adopted Gaussian priors centered on their reported best-fit values because the distribution in those fits were close to being Gaussian (see Table \ref{tab:fitting_priors}). 
This approach incorporates existing knowledge and effectively constrains the parameter space explored during the sampling process and is done to both the current and subsequent functional fits. 


\subsection{Exponentially Tapered Power Law Fits}
\begin{figure}[p]
    \centering
    \includegraphics[width=0.75\textwidth]{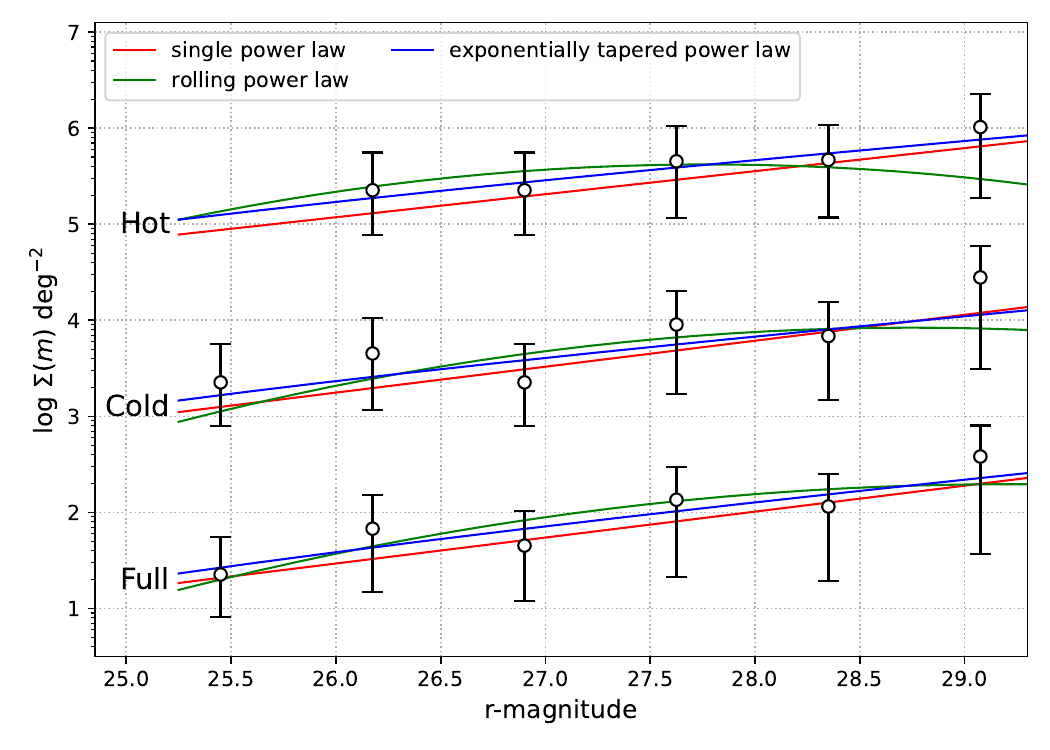}
    \caption{The best-fit differential distributions for the hot, cold, and full TNO of \textbf{Sample A}, with red, green and blue lines corresponding to the single, rolling and exponentially tapered power law models, respectively. The magnitude distributions are presented in $r$ band, assuming a assuming a nominal color $r - F150W2 = ~0.99$ mag. For visual clarity, the cold and hot samples have been vertically offset by 2 and 4 units, respectively. Error bars are the $2\sigma$ Poisson uncertainties on the number of objects in each magnitude bin. The best fitting parameters for each function are given in Table \ref{tab:bestfit_params}.}
\label{fig:sigma_all_curvesA}
\end{figure}

\begin{figure}[p]
    \centering
    \includegraphics[width=0.75\textwidth]{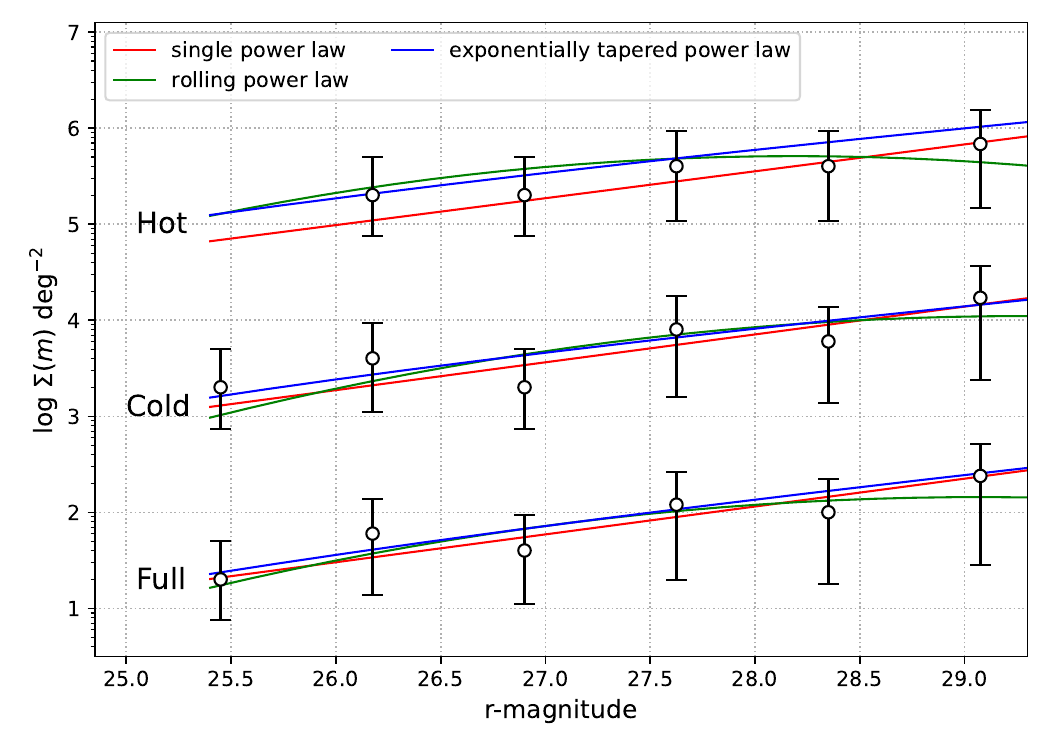}
    \caption{Same as the previous figure but using \textbf{Sample A+B}. The best fitting parameters for each function are given in Table \ref{tab:bestfit_params}.}
\label{fig:sigma_all_curvesAB}
\end{figure}

Finally, we fit an exponentially tapered power law to our differential distribution. This functional form is seen in numerical simulations of planet formation by SI+PCC \citep{schafer2017initial, li2019demographics} and is described by
\begin{equation}
    \Sigma_{
    \rm tapered} = \ln(10) \cdot \exp\left(-10^{\beta(m_B - m)}\right) \cdot 10^{(\alpha - \beta)m - \alpha m_0} \cdot \left( \alpha \cdot 10^{\beta m} + \beta \cdot 10^{\beta m_B} \right)
\end{equation}
Here,  $\alpha$ is the faint-end slope of the distribution, $\beta$ is the strength of the exponential bright-end behavior, $m_0$ is a normalization factor, and $m_B$ is the magnitude where the exponential term starts to overpower the power law term.
We again use the \cite{napier2024decam} results as Gaussian priors, but taking into consideration the phase curve effects at $m_0$ and  $m_B$ parameters as our JWST observation were done near quadrature\footnote{In this observing geometry, the larger phase angle reduces the apparent brightness of the object by $\sim0.35$ mag relative to typical near-opposition phase angles due to smaller illuminated area and enhanced surface scattering (see Figure 6 in \cite{verbiscer_phase_2019}).}, rather than near opposition where their fit was derived. We plot our fits results in Figures \ref{fig:sigma_all_curvesA} and \ref{fig:sigma_all_curvesAB}.

\begin{deluxetable}{c|cc|cccc|ccccc}[htbp]
\setlength{\tabcolsep}{2.5pt}
\tabletypesize{\small}
\tablecaption{Best-Fit Differential Surface Density Models}
\label{tab:bestfit_params}
\tablehead{
\multicolumn{1}{c|}{\textbf{Sample}} & 
\multicolumn{2}{c|}{\textbf{Single Power Law}} &
\multicolumn{4}{c|}{\textbf{Rolling Power Law}} &
\multicolumn{5}{c}{\textbf{Exponentially Tapered Power Law}} \\
\cline{1-12}
&
\colhead{$\alpha$} & \colhead{$m_0$} &
\colhead{$\alpha_1$} & \colhead{$\alpha_2$} & \colhead{$\Sigma_{23}$} & \colhead{$\Delta \mathrm{BIC}$} &
\colhead{$\alpha$} & \colhead{$\beta$} & \colhead{$m_0$} & \colhead{$m_B$} & \colhead{$\Delta \mathrm{BIC}$}
}
\startdata
Hot A   & $0.24^{+0.10}_{-0.09}$ & $21.53^{+2.10}_{-3.75}$
        & $0.86^{+0.09}_{-0.09}$ & $-0.09^{+0.02}_{-0.02}$ & $0.37^{+0.10}_{-0.10}$ & $0.9$
        & $0.18^{+0.06}_{-0.05}$ & $0.14^{+0.05}_{-0.05}$ & $16.29^{+3.85}_{-6.20}$ & $25.23^{+3.50}_{-6.25}$ & $3.8$ \\
Cold A  & $0.27^{+0.09}_{-0.08}$ & $21.38^{+1.79}_{-2.64}$
        & $0.92^{+0.14}_{-0.14}$ & $-0.08^{+0.02}_{-0.02}$ & $0.19^{+0.10}_{-0.09}$ & $1.9$
        & $0.20+^{+0.07}_{-0.06}$ & $0.18^{+0.10}_{-0.10}$ & $17.03^{+3.37}_{-5.58}$ & $25.10^{+3.14}_{-6.25}$ & $4.6$ \\
Full A  & $0.27^{+0.07}_{-0.06}$ & $20.56^{+1.53}_{-2.10}$
        & $0.87^{+0.08}_{-0.08}$ & $-0.07^{+0.01}_{-0.01}$ & $0.39^{+0.09}_{-0.09}$ & $0.2$
        & $0.21^{+0.07}_{-0.05}$ & $0.14^{+0.05}_{-0.05}$ & $16.07^{+3.61}_{-5.92}$ & $25.63^{+3.10}_{-6.19}$ & $3.9$ \\
Hot AB  & $0.28^{+0.11}_{-0.09}$ & $22.46^{+1.73}_{-2.58}$
        & $0.83^{+0.08}_{-0.08}$ & $-0.08^{+0.02}_{-0.02}$ & $0.35^{+0.11}_{-0.10}$ & 2.9
        & $0.19^{+0.07}_{-0.06}$ & $0.13^{+0.05}_{-0.05}$ & $16.17^{+3.61}_{-6.36}$ & $26.37^{+3.61}_{-6.65}$ & $3.8$ \\
Cold AB & $0.29^{+0.08}_{-0.07}$ & $21.61^{+1.51}_{-2.20}$
        & $0.88^{+0.13}_{-0.13}$ & $-0.07^{+0.02}_{-0.02}$ & $0.19^{+0.09}_{-0.08}$ & $2.8$
        & $0.21^{+0.06}_{-0.06}$ & $0.17^{+0.10}_{-0.10}$ & $17.11^{+3.02}_{-5.52}$ & $25.89^{+2.99}_{-6.25}$ & $6.1$ \\
Full AB & $0.29^{+0.07}_{-0.06}$ & $20.89^{+1.43}_{-1.74}$
        & $0.85^{+0.08}_{-0.07}$ & $-0.07^{+0.01}_{-0.01}$ & $0.38^{+0.09}_{-0.10}$ & $3.5$
        & $0.22^{+0.07}_{-0.06}$ & $0.14^{+0.05}_{-0.05}$ & $16.43^{+3.38}_{-6.06}$ & $26.34^{+2.93}_{-5.69}$ & $6.8$ \\
\enddata
\tablecomments{
$\Delta \mathrm{BIC} = \mathrm{BIC}_m - \mathrm{BIC}^*$, where $\mathrm{BIC}^*$ is the reference model (here, the single power law) and $BIC_m$ is any candidate model. The size of $\Delta \mathrm{BIC}$ shows how strongly the candidate model is disfavored compared to the single power law model.
}
\end{deluxetable}

\begin{deluxetable}{c|ccc|cccc}
\tablewidth{0pt}
\tabletypesize{\small}
\tablecaption{Informative Priors}
\label{tab:fitting_priors}
\tablehead{
\multicolumn{1}{c|}{\textbf{Sample}} & 
\multicolumn{3}{c|}{\textbf{Rolling Power Law}} &
\multicolumn{4}{c}{\textbf{Exponentially Tapered Power Law}} \\
\cline{1-8}
& \colhead{$\alpha_1$} & \colhead{$\alpha_2$} & \colhead{$\Sigma_{23}$} &
\colhead{$\alpha$} & \colhead{$\beta$} & \colhead{$m_0$} & \colhead{$m_B$}
}
\startdata
Hot  & $0.9\pm0.1$ & $-0.1\pm0.3$ & $0.4\pm0.1$ 
     & $0.2\pm1.0$ & $0.1\pm0.5$ & $14.0\pm11.4$ & $29.1\pm27.0$ \\
Cold & $1.0\pm0.2$ & $-0.1\pm0.4$ & $0.2\pm0.1$ 
     & $0.2\pm1.0$ & $0.2\pm1.0$ & $12.6\pm8.0$ & $28.4\pm25.0$ \\
Full & $0.9\pm0.1$ & $-0.1\pm0.3$ & $0.4\pm0.1$ 
     & $0.2\pm1.0$ & $0.1\pm0.5$ & $14.3\pm11.4$ & $29.1\pm27.0$ \\
\enddata

\caption{We adopt the best-fit values from \cite{napier2024decam} as informative priors for our rolling and exponentially tapered power law fits. Since the distributions in their analysis were approximately Gaussian, we employ Gaussian priors centered on these reported values. For parameters governing the bright-end of the luminosity function and normalization values ($\alpha_1, \Sigma_{23}, m_0$), we use the reported uncertainties. Alternatively, for parameters influencing the faint-end ($\alpha_2, \alpha, \beta, m_B$) of the function, where most of our data lies, we broaden the Gaussian widths by $10\sigma$ to allow for greater exploration of the parameter space.}
\end{deluxetable}

\subsection{Comparison of Fits}
To determine which $\Sigma$ function is a better representation of our data, we make use of the Bayesian Information Criterion (BIC; \cite{kass1995bayes}),

\begin{equation}
    BIC = k \: ln(n) - 2 \: ln \: (\mathcal{L})
\end{equation}

\noindent Here, $k$ denotes the number of parameters in each fit ($k=2$ for a power law, $k = 3$ for rolling power law, and $k=4$ for tapered power law), $n$ refers to the sample size, and $ln \: (\mathcal{L})$ is the natural logarithm of the likelihood function defined in Equation \ref{eq:likelihood}, evaluated at the best-fit (maximum likelihood) parameter values. $ln \: (\mathcal{L})$ reflects how well the model fits the data under those parameters, and can thus be interpreted as a goodness-of-fit measure. 

The BIC imposes penalty on models with more parameters to discourage overfitting and thereby balancing the goodness of fit with the complexity of the model. In general, the model with the lowest BIC value is preferred. To facilitate model comparisons, we consider the difference $\Delta BIC = BIC_m - BIC^*$ which directly measures the relative support for one model over another. Our reference model (i.e., single power law) is denoted by $BIC^*$, while $BIC_m$ is the candidate model. The value of $\Delta BIC$ indicates the strength of evidence against a candidate model being optimal. As a rule of thumb, if $\Delta BIC$ is less than 2, the difference is negligible. Values between 2 and 6 indicate positive evidence against the candidate model, while values between 6 and 10 suggest strong evidence \citep{fabozzi_basics_2014}. When $\Delta BIC$ exceeds 10, the evidence against the candidate model is considered very strong. We demonstrate our results in Figures \ref{fig:sigma_all_curvesA} and \ref{fig:sigma_all_curvesAB}, and summarize them in Table \ref{tab:bestfit_params}.

The single, rolling, and exponentially tapered power laws all fit our data reasonably well. However, based on the $\Delta BIC$ values, the single power law is preferred over the more complex rolling and exponentially tapered forms; within our data range, a single power law is sufficient. Adding extra parameters in the more complex models does not yield a significant improvement in fit quality when incorporating the prior from \citet{napier2024decam}.

\subsection{Color Estimates}

One so far unutilized measurement afforded by the NIRCam observations is the long- to short-wavelength flux ratio ($f_{322}/f_{150}$) listed in  Table \ref{tab:objects}. These fluxes can be estimated for objects observed by JWST-NIRSpec \citep{pinilla-alonso_jwstdisco-tnos_2025}, allowing inference of surface type for each of our detections. We adopt the broad 3-class scheme: Organics-, $\text{CO}_2$-, or $\text{H}_{2}\text{O}$-types \citep{holler_descriptive_2025}.


For each class, we adopt the corresponding average NIRSpec reflectance spectrum and derive its behavior in the $r$ band by fitting a linear continuum across $0.71$–$0.90,\mu$m. This continuum is then extended to shorter wavelengths to approximate the unobserved portion of the spectrum, from which an average $(r - F150W2)$ color is obtained for each compositional category (see Table \ref{tab:object_types}).

Since most orbital solutions derived from our two-epoch detections (Sample B) remain insufficiently constrained, we limit our compositional estimates only to the three-epoch detections (Sample A). Each sample A object is assigned an $(r - F150W2)$ color based on which spectral type closest matches the object's observed $f_{322}/f_{150}$ flux ratio. This provided $r$-band brightness estimates for all objects. We find that the cold classicals are consistent with being drawn from the organics-types and the excited objects are drawn from a mix of classes, consistent with findings from \citet{pinilla-alonso_jwstdisco-tnos_2025}. A companion paper, \cite{ana}, discusses the implications of these color measurements. The $r$-band luminosity function was fit using those values. Best fit values for each functional form are shown in Table \ref{tab:bestfit_params_Ana}, while the assigned classification for each object is listed in Appendix \ref{sec:object_class}. We find that the best fit slopes remain consistent regardless of the adopted colors for each detected object.

\begin{deluxetable}{lcc}[h]
\tablecaption{Compositional Types with $r$-F150W2 Values and Uncertainties \label{tab:object_types}}
\tablehead{
\colhead{Object Type} & \colhead{$r-F150W2$} & \colhead{$\sigma$}
}
\startdata
Cliff Organics-Type & 0.61 & 0.05 \\
DoubleDip CO2-Type  & 0.34 & 0.05 \\
Bowl H2O-Type       & 0.30 & 0.05 \\
\enddata
\tablecomments{
The color uncertainty $\sigma$ was derived from the range of slopes that changed the F150W2-magnitude by no more than the largest photometric uncertainty in our sample ($\pm0.07$ mag).}
\end{deluxetable}

\begin{deluxetable}{c|cc|cccc|ccccc}[h!]
\setlength{\tabcolsep}{2.5pt}
\tabletypesize{\small}
\tablecaption{Best-Fit Differential Surface Density Models}
\label{tab:bestfit_params_Ana}
\tablehead{
\multicolumn{1}{c|}{\textbf{Sample}} & 
\multicolumn{2}{c|}{\textbf{Single Power Law}} &
\multicolumn{4}{c|}{\textbf{Rolling Power Law}} &
\multicolumn{5}{c}{\textbf{Exponentially Tapered Power Law}} \\
\cline{1-12}
& \colhead{$\alpha$} & \colhead{$m_0$} &
\colhead{$\alpha_1$} & \colhead{$\alpha_2$} & \colhead{$\Sigma_{23}$} & \colhead{$\Delta \mathrm{BIC}$} &
\colhead{$\alpha$} & \colhead{$\beta$} & \colhead{$m_0$} & \colhead{$m_B$} & \colhead{$\Delta \mathrm{BIC}$}
}
\startdata
Hot A   & $0.24^{+0.11}_{-0.09}$ & $20.89^{+2.16}_{-3.64}$
        & $0.90^{+0.09}_{-0.09}$ & $-0.10^{+0.02}_{-0.02}$ & $0.40^{+0.10}_{-0.10}$ & 0.2
        & $0.18^{+0.07}_{-0.06}$ & $0.20^{+0.23}_{-0.16}$ & $16.39^{+3.67}_{-6.60}$ & $24.89^{+3.35}_{-6.18}$ & 3.1 \\
Cold A  & $0.27^{+0.09}_{-0.08}$ & $20.96^{+1.76}_{-2.78}$
        & $0.97^{+0.09}_{-0.08}$ & $-0.09^{+0.02}_{-0.02}$ & $0.20^{+0.09}_{-0.08}$ & 2.1
        & $0.21^{+0.08}_{-0.06}$ & $0.21^{+0.26}_{17}$ & $17.03^{+3.47}_{-5.47}$ & $24.44^{+4.23}_{-6.02}$ & 4.9 \\
Full A  & $0.28^{+0.07}_{-0.06}$ & $20.33^{+1.47}_{-1.95}$
        & $0.92^{+0.08}_{-0.08}$ & $-0.08^{+0.02}_{-0.02}$ & $0.41^{+0.09}_{-0.09}$ & 0.6
        & $0.22^{+0.07}_{-0.06}$ & $0.23^{+0.21}_{-0.17}$ & $16.72^{+3.12}_{-5.33}$ & $25.08^{+2.51}_{-5.37}$ & 4.3 \\
\enddata
\tablecomments{
$\Delta \mathrm{BIC}$ is similar to that described in Table~\ref{tab:bestfit_params}.
}
\end{deluxetable}

\section{The Absolute Magnitude Distribution of the Cold Classicals} \label{sec:H_dist}

 A streaming instability followed by pebble cloud collapse (SI + PCC) is the currently favored pathway to planet formation because it facilitates the rapid formation of larger planetesimals by overcoming growth barriers (\ie, collisional bouncing, fragmentation, inward drift) \citep{johansen2011planetesimal,robinson2020investigating,nesvorny_binary_2021}. Multiple independent studies have investigated the initial mass function of planetesimals formed through SI+PCC, but the lower-mass end of this distribution remains poorly constrained due to the limited mass resolution in simulations \citep{schafer2017initial, abod2019mass, li_demographics_2019}. Despite this, these studies consistently show similar mass distributions, which can be approximated by a steep exponential power law at the high-mass end and a gradual taper at small-mass range.

\begin{figure}
    \centering
    \includegraphics[width=\textwidth]{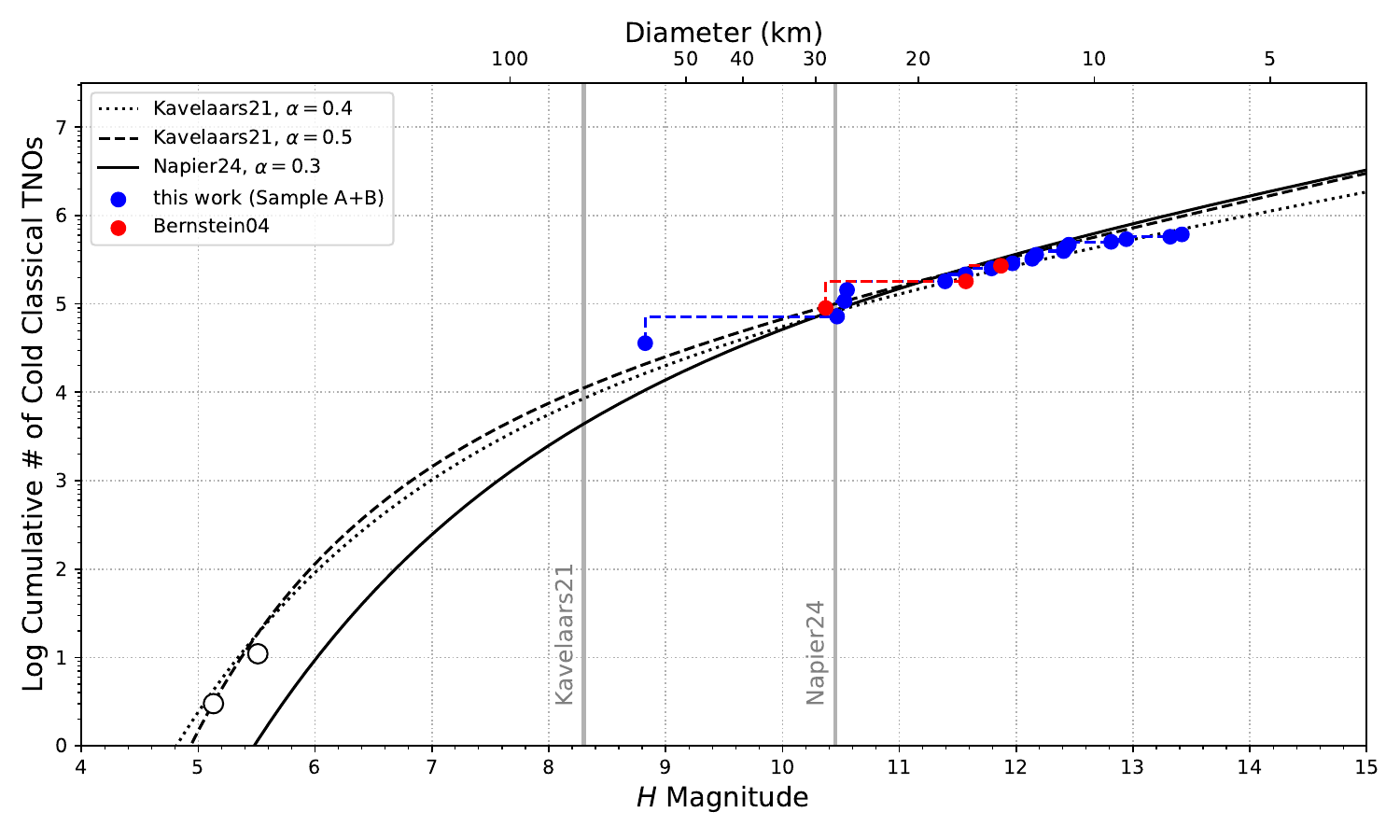}
    \caption{The $H_r$ cumulative distribution of the cold main Kuiper Belt. Blue circles show the scaled cumulative distribution of the cold classicals from Sample A+B of our survey while red circles represent the cumulative distribution of cold classicals detected by \cite{bernstein2004size}. The dotted and dashed lines indicate the best-fit power-law models from \cite{kavelaars2021ossos}, corresponding to slopes of $\alpha=0.4$ and $\alpha=0.5$, respectively. The solid black line shows the best-fit exponentially tapered function from \cite{napier2024decam}, scaled to match the full ecliptic population. White open circles mark the cumulative counts of 3 ($Hr \sim 5.13$) and 11 ($Hr \sim 5.51$) main-belt cold objects, as reported in the Minor Planet Center database. The vertical lines indicate the observing limit from each studies.}
\label{fig:Hmag}
\end{figure}

This section presents the cumulative distribution of the absolute magnitude\footnote{$H = m - 5\log_{10}(r(r-1))+2.5\log_{10}\phi(\alpha)$, where $r$ is the heliocentric distance in au and $\phi(\alpha)$ is the phase function. At quadrature, the geocentric and heliocentric distances are similar so the term $r(r-1)$ becomes $r^2$ and $\alpha=0\degree$, so $\phi=1$ \citep{bowell1989application, hughes2003absolute}.} $H$ for our sample of cold classical objects. Recent works demonstrate \citep{kavelaars2021ossos, napier2024decam} that the differential size distribution of CCs is consistent with exponentially tapered power-law form -- a feature of numerical simulations of planetesimal formation via SI + PCC. The depth of these surveys, however, were limited to $H_r \sim 8.3$ for OSSOS and $H_r \sim 10.5$ for DEEP. \cite{bernstein2004size} provided an anchor on the faint-end slope. This JWST survey is the first to add objects with $H_r > \sim 10.5$, and pushes the faint limit to nearly $H_r = 14$. While the OSSOS and DEEP results agree well at brighter magnitudes ($H \lesssim 11$), increasing uncertainties arise at fainter magnitudes due to limited small-object statistics. Our survey extends beyond these previous limits, and in Figure \ref{fig:Hmag}, we superpose our cumulative $H$ distribution on the extrapolated best-fit exponentially tapered power laws from \cite{kavelaars2021ossos} and \cite{napier2024decam}. We find that the cumulative distribution of our detected CCs is shallower than those reported in the two studies, suggesting that the previously proposed exponentially tapered power law model may not fully represent the size distribution of the smallest objects in this population.

\section{Mass of the Cold Kuiper Belt} \label{sec:MassOfKB}

Given the surface density $\Sigma(m)$, we are able to estimate the total mass of the cold classical belt as,

\begin{equation}
   M_{\text{tot}} = M_{23}  \frac{\Omega}{f} \int_{m=14}^{m=35} \Sigma(m) 10^{-0.6(m-23)} dr \,  \times \left\langle  \left( \frac{p}{p_{nom}} \right)^{-3/2} \left( \frac{d}{d_{nom}} \right)^{6} \left( \frac{\rho}{\rho_{nom}} \right) \right\rangle
    \label{eq:Mtot}
\end{equation}

\noindent where $\Omega$ is the surveyed solid angle, and $f$ is the fraction of the TNO population located within this surveyed area. Using the OSSOS++ model, which is the leading dynamical TNO model accounting for both the orbital structure and size distribution \citep{bannister_ossos_2018, petit2011canada, petit_canadafrance_2017, alexandersen_carefully_2016}, we estimate the fraction of cold classicals per square degree at the epoch and location of our observation to be $f \approx 3.8 \times10^{-4}$.  The normalization constant $M_{23} = 7.8 \times 10^{18} \, \text{kg}$ corresponds to the mass of a TNO with apparent magnitude $m_r = 23$, assuming canonical values for albedo, density, and distance.

The angle brackets denote the mean value of all TNOs at a given magnitude. For the cold classical population, the average heliocentric distance is  $d \sim 43.8$ au \citep{napier2024decam, bannister_ossos_2018}, while the average albedo is also found to be $p \sim 0.15$ \citep{fraser2014absolute, lacerda2014albedo}. The material density of an object is a function of its size, with smaller TNOs often having densities as low as $500 \ \text{kg m}^{-3}$ and larger objects being as dense as $3000 \ \text{kg m}^{-3}$ \citep{grundy_mutual_2015}. For simplicity, we assume a density corresponding to a sphere of pure water ice, $\rho \sim 900 \ \text{kg m}^{-3}$.

Using the best-fit parameters for the Cold AB sample and assuming a rolling power law size distribution, we find that the total mass of the cold classical belt is $0.01  \ M_{\oplus} \times \big( \frac{p}{0.15} \big)^{-3/2} \times \big( \frac{d}{43.8} \big)^{6} \times \big( \frac{\rho}{900 \ kg \  m^{-3}} \big)$. Note that we only calculate for the total mass of the CC population as the total mass of the HTNOs is more uncertain due to poor constraints on their vertical and radial distribution, although \cite{bernstein2004size} suggested that the mass of objects within $50\ \text{au}$ is comparable to that of the CKB. We also choose to report the mass of the rolling power law to provide direct comparison with other recent works \citep{bernstein2004size, napier2024decam}. With our assumptions, the integrated mass faintward of our detection limit is $\sim 7 \times10^{-5} \ M_{\oplus}$, much smaller than the mass contained in larger objects. This implies that the cold classical belt mass uncertainty is no longer driven by uncertainty in the luminosity function, but in the albedo and density assumptions \citep{fraser2014absolute}.

\section{Discussion and Conclusions} \label{sec:DiscussionsAndConclusions}

In this paper we have presented detections from the first TNO dedicated survey using JWST. By using the shift-and-stack technique we are able to achieve a limiting magnitude (40\% efficiency) of $m_{F150W2} = 28.6$ for Sample A and  $m_{F150W2} = 28.8$ for Sample A+B (assuming $F150W2-r \approx 1.0$, this translates to $m_{r} \approx 29.6$ mag and $m_{r} \approx 29.8$ mag, respectively) over a sky area of $0.05 \deg ^2$. This is \textit{by far} the deepest TNO survey to date, with roughly a visible magnitude deeper than the landmark HST survey by \cite{bernstein2004size}, which had a limiting magnitude of $m_r \sim 28.9$ and covered $2.5\times$ more survey area.  Our data yield 23 three-epoch detections and 7 two-epoch detections, with magnitudes $24.1 < m_{F150W2} < 29.3$ and heliocentric distances $28.3 \,\text{au} < r \lesssim 133.6 \,\text{au}$ --- all new discoveries except 2015 GK56, which was purposely targeted. In addition, we identify 13 single-epoch detections that aren't static transients. None of these new discoveries exhibit lightcurves indicative of binary TNO, in agreement with expectations scaled from the SSOLS dataset. Using these discoveries we determine the faint-end magnitude distributions for both the cold and hot TNO populations, as well as for the Kuiper Belt as a whole. 

The cold population of our nominal sample (Sample A+B) exhibits a shallow slope of $\alpha = 0.29_{-0.08}^{+ 0.07}$, comparable to that of the hot component, $\alpha = 0.28_{-0.11}^{+ 0.09}$. The observed luminosity functions suggest that both populations have similar size distributions. The slope derived for the full TNO sample is also consistent with those of the individual components. These consistently shallow slopes across all populations indicate that the majority of the Kuiper Belt’s mass is concentrated in objects near the break diameter, $\sim200$ km.

Multiple lines of evidence suggest that the cold and hot TNO populations are genetically distinct, having likely formed in different regions of the planet-forming disk and under different physical conditions. For instance, the CCs are predominantly very red in color, while HTNOs display a wide range of colors---a contrast that is generally interpreted as indicative of primordial compositional differences between the distinct regions in which they formed \citep{brown_hypothesis_2011, fraser_hubble_2012, schwamb_col-ossos_2019, buchanan_col-ossos_2022, nesvorny_ossos_2020, fraser_col-ossos_2023, bernardinelli_photometry_2025, henault_irradiation_2025, de_pra_widespread_2025, pinilla-alonso_jwstdisco-tnos_2025}. CCs are thought to have formed near their current locations, while HTNOs likely originated closer to the Sun before being scattered outward by the migration of the giant planets \citep{morbidelli2005chaotic, tsiganis2005origin, levison2008contamination, nesvorny2012migration, nesvorny2015jumping}. The region where CCs reside is also thought to have been relatively less dense, causing Neptune to halt its outward migration at 30 au \citep{gomes_planetary_2004, nesvorny_dynamical_2018}, and to have undergone relatively modest collisional evolution, thereby preserving surface volatiles responsible for their red coloration \citep{abedin_ossos_2022}. The prevalence of binaries in the cold classical belt further supports the idea that CCs are formed \textit{in situ} and are therefore primordial, as substantial dynamical excitation during the giant planet migration would likely have disrupted such systems \citep{parker_destruction_2010, noll_chapter_2020}. In contrast, the higher density of the inner disk from which HTNOs originated would have exposed them to more intense collisional grinding, further altering their surface properties \citep{nesvorny_dynamical_2018}. Despite these strikingly different formation environments and evolution histories, our results indicate that CCs and HTNOs nevertheless share a remarkably similar size distribution.

The similarity of slopes across the two populations we observe in this work ($H_r \simeq 8.3-13.4$ mag) is consistent with \cite{petit_hot_2023}, who reported comparable SD shapes for both populations in the absolute magnitude range $H_r \simeq 5.5-8.3$ mag. The very similar size distributions from $H_r\sim5$ down to the observable limit suggests that both populations likely underwent analogous formation mechanisms, resulting in nearly identical size distributions despite those processes occurring in very different conditions. That is to say, the underlying process responsible for planetesimal formation (potentially SI+PCC) appears to be largely insensitive to local disk conditions (e.g. temperature and density gradients), resulting in a consistent size distribution regardless of environments. Consequently, the indistinguishable shallow slopes imply that the hot population, much like the cold one, seems to have retained its primordial size distribution down to diameters of $\sim10$ km.

The shallow slopes observed in both populations are consistent with the slope ($\alpha \sim 0.4 \pm0.04$) of impactors above the  turnover in the size distribution inferred from the Pluto and Charon cratering record \citep{singer_impact_2019}. This behavior hints at the idea that the SI+PCC mechanism requires a minimum mass or size threshold for gravitational collapse to occur efficiently. That turnover may mark the point at which local pebble concentrations become dense enough to overcome turbulent diffusion, enabling the formation of bound planetesimals. Below this critical size, the collapse mechanism could become much less effective, leading to a relative deficit of smaller bodies. 

Moreover, the fact that we find the HTNOs slope to be shallow, despite originating from a denser and collision-heavy environment, places further constraint on the nature and timescale of its evolutionary process. In particular, this suggests that dynamical excitation and subsequent collisional evolution were not prolonged or intense enough to significantly alter the original distribution of bodies larger than $10$ km. 

In the absence of direct size measurements from stellar occultations, the geometric albedos ($\rho$) of smaller TNOs are inferred from the median albedos of their dynamical classes, which have been derived from observations of larger objects \citep{ stansberry_physical_2007, lacerda2014albedo, vilenius_tnos_2014}. This then introduces a bias in size estimates, particularly for the CC population, making them appear bigger\footnote{the absolute magnitude ($H$) scales as $H \propto -2.5 \log(\rho D^2)$, where $D$ is the object's diameter.} than they may actually be. The result that fainter CCs and HTNOs exhibit similar slopes may indicate that their albedos converge at smaller sizes, implying more comparable reflectivities in that regime. 

If the hot and cold populations exhibit different albedo distributions across the observable size range, then the similarity of their magnitude distributions implies a similar shape of the underlying size distributions, but that are offset by an amount comparable to the magnitude of reflectance difference driven by their albedos. It has been suggested that on average, smaller cold classical TNOs are redder than larger bodies \citep{benecchi_optical_2011}. If this suggestion is true, it is conceivable then that albedos may also have a size dependency such that the difference in average aledbos of the hot and cold populations vanishes at smaller sizes.  It is likely that evidence for such a trend will arise with further compositional and thermal observations with focus on smaller objects.

Although the hot and cold TNO populations currently appear to share the same size distribution across their entire observable range, this inference is drawn from a sample of only about 1,000 objects. Characterized TNO discoveries expected from the upcoming Legacy Survey of Space and Time (LSST; \cite{ivezic_lsst_2019}) will increase this sample by more than an order of magnitude (e.g., \cite{kurlander_predictions_2025}), enabling a detailed search for any differences between the SDs of the hot and cold populations. In particular, it will be possible to search for any small variations in the SD due to the enhanced collisional grinding expected of the hot population since its formation. However, despite its unprecedented sample size, LSST's limiting magnitude is $H_r \sim 8$ mag and will not constrain the faint end of the TNO size distribution as in this work.


Finally, we note the on-going efforts to assess the population of objects beyond the main Kuiper Belt. Recent work has suggested that there may be an increase in density of objects beyond $\sim80$ au as compared to objects at closer distances \citep{fraser2024candidate}. Scaling those results to our survey, and assuming a luminosity function similar to that observed here, we would expect roughly 1 detection at $\sim80$ AU; and we have detected one object at these distances. We point out however, that our current sample of distant such bodies remains much too small to provide meaningful constraints.

\section{Acknowledgments} \label{sec:Acknowledgments}

This research is based on observations obtained with the James Webb Space Telescope (JWST), a collaboration between the National Aeronautics and Space Administration (NASA), the European Space Agency (ESA), and the Canadian Space Agency (CSA). The observations correspond to Cycle 1 GO program \#1568, and the data were retrieved from the Mikulski Archive for Space Telescopes (MAST) at the Space Telescope Science Institute (STScI), which is operated by the Association of Universities for Research in Astronomy (AURA). Support for program \#1568 was provided by NASA through a grant from STScI under contract NAS 5-03127.

W.C.F. and M.R.E. acknowledge funding from the CSA via grants 222JWGO1-09 and 22EXPCO11. This work also used the facilities of the Canadian Astronomy Data Centre (CADC), operated by the National Research Council of Canada (NRC) with support from the CSA.

G.M.B. acknowledges support from the National Science Foundation (NSF) under grants AST-2205808 and AST-2407527, as well as from STScI through grants JWST-GO-01568.002-A and HST-GO-16720.002-A. 

M.J.H. and K.J.N. acknowledge funding from NSF grant AST-2206194 and NASA Yearly Opportunities for Research in Planetary Defense (YORPD) Program grant 80NSSC22K02.

Lastly, we thank Joshua Emery and Lucas McClure for their assistance with the DiSCo spectral data used in this work as well as Ryder Strauss for identifying 2015~GK$_{56}$ for inclusion in our observations and detection algorithm validation. 

\newpage
\bibliographystyle{aasjournal}
\bibliography{references}

\appendix
\section{ResNet configurations and Acceptance Threshold Selection}
\label{sec:ML_criteria}

\begin{figure}[h]
\centering
    \includegraphics[scale=0.43]{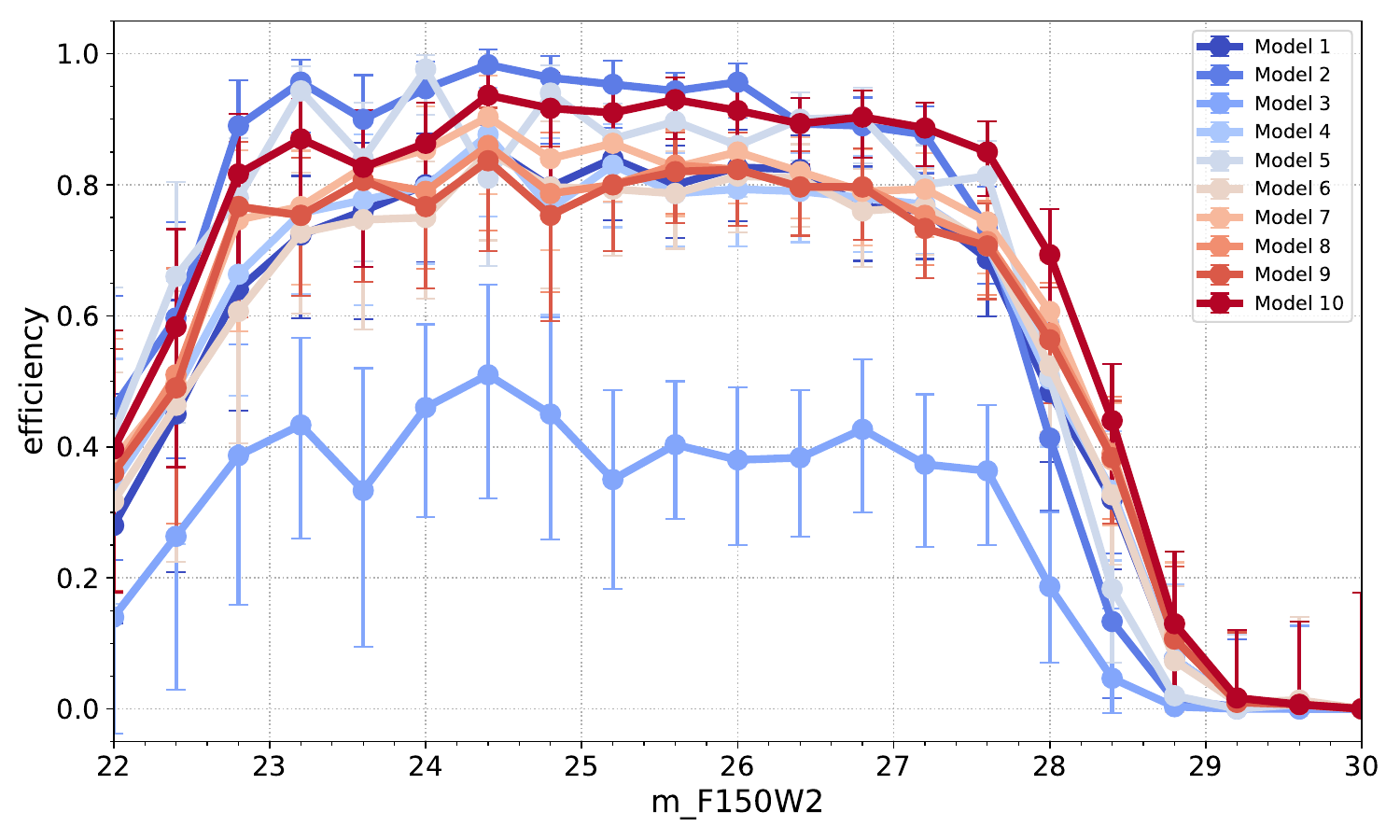} 
    \includegraphics[scale=0.43]{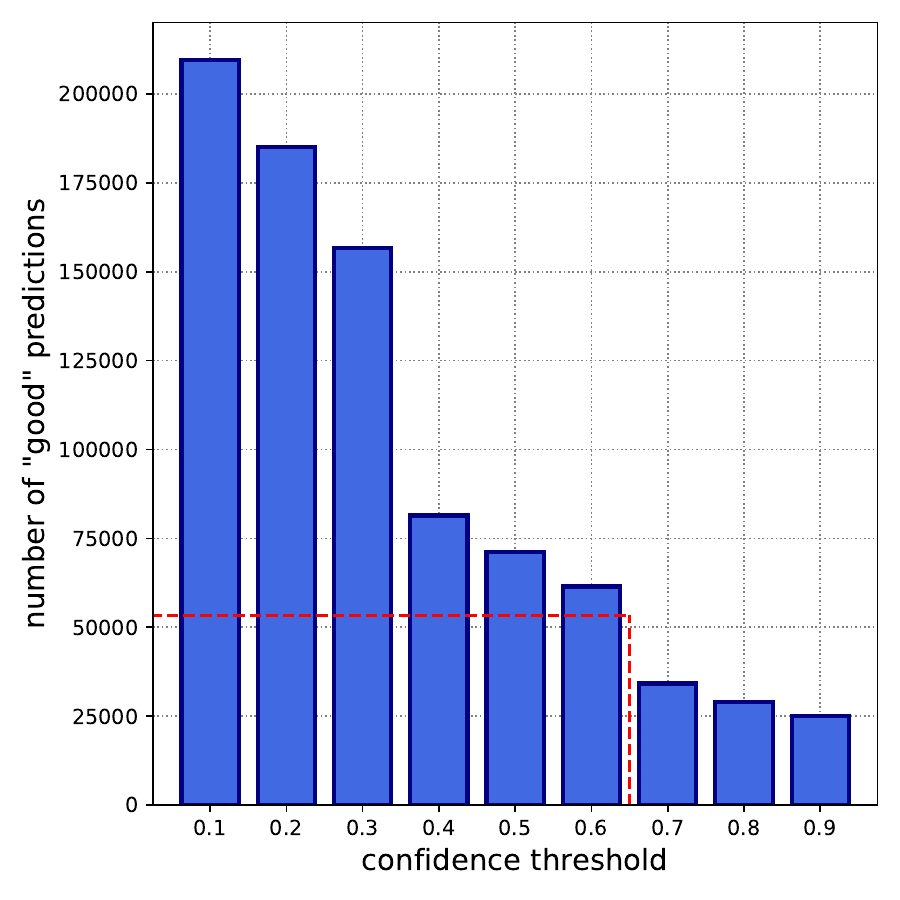} 
    \caption{\textbf{Left:} Detection efficiency of 10 ResNet models. We find Model 10 to be the most ideal. \textbf{Right:}  Number of objects vetted as "good" source by Model 10 as a function of acceptance threshold. We eventually settled with 0.65.}
\label{fig:model_comparison}
\end{figure}

We explored 10 different ResNet model configurations and ultimately settled with the model that achieved the faintest detection depth and highest peak detection efficiency. Comparison of the performance of these 10 models are shown left of Figure \ref{fig:model_comparison}. The model with the optimum performance is Model 10. We then investigate the best acceptance threshold for Model 10 by evaluating how the number of vetted predictions varies with the threshold (see right of Figure \ref{fig:model_comparison}), identifying the point where a lower threshold introduces excessive false positives while a higher threshold becomes overly stringent.



\section{Information on individual TNOs}
\label{sec:oddballs}
We describe here the steps that were used to discover and determine orbital elements for each of the 2- and 3-epoch TNOs listed in Table~\ref{tab:objects}.  Those that are not listed here were discovered through ``normal procedures,'' meaning that three epoch's tracks were found, linked, and fit to an orbit using the procedures described in Sections~\ref{sec:tnosearch} and \ref{sec:3epoch}.

\vskip5pt\noindent\textbf{JPB01:} The $S/N=51$ apparition in epoch 2 was mis-classified as spurious by the ResNet, probably because of overtraining on implants with a different PSF.  It is considered part of sample A.

\vskip5pt\noindent\textbf{JPB02:} The $S/N=7.8$ apparition in epoch 1 was classified as spurious by ResNet, and linked in the search that bypassed ResNet.  Since the implants used to determine detection efficiency and insure against false positives were filtered by ResNet, we cannot insure the reality of this source nor use it as part of sample A for population statistics.  Since the epoch 3 detection has $S/N<8,$ it also is excluded from sample B.  We retain it in the auxiliary files as an illustration of an uncertain detection.

\vskip5pt\noindent\textbf{JPB03:} The $S/N=5.9$ apparition in epoch 3 was classified as spurious by ResNet, and linked in the search that bypassed ResNet.  This is even fainter than JPB02 and is excluded from both samples A and B for the same reasons.

\vskip5pt\noindent\textbf{JPB04:} The duplicate detection at $S/N=178$ in epoch 1 was misclassified by ResNet.  The object remains in sample A since it was discovered in 3 epochs via normal procedures.

\vskip5pt\noindent\textbf{JPB05:} The $S/N=40$ apparition in epoch 3 was mis-classified as spurious by the ResNet.  Like JPB01, this is probably a ResNet blunder and the object is considered part of sample A.

\vskip5pt\noindent\textbf{JPB23:} This is \gk, a previously known TNO that was placed in the field of view intentionally.  It is therefore not a random sampling of the sky population and is omitted from the population-statistics samples.

\vskip5pt\noindent\textbf{JPB24:} The epoch 2 apparition at $S/N=7.4$ was not found by KBMOD because of residuals from the subtraction of the bright galaxy in its background.  This source is \emph{not} included in sample A, because the calibrating implanted sources were subject to the same sort of failure.  It does, however, enter sample B, since its other two epochs were found at $S/N>8.$  The distance is well determined from the recovery of epoch 2 using SMP.

\vskip5pt\noindent\textbf{JPB25:} This object has pulled a ``vanishing act.''  It was found once in epoch 2 and twice in epoch 3 by normal procedures at $S/N=8.6,12,$ and $8.7.$  Linking these two epochs leaves a very wide range $55<d_{\rm bary}<280$~\au\ as possible (meaning that this is not a cold source).  At any such distance, the object was imaged by an unobscured region of the the NIRCAM detector B2 in epoch 1, but all attempts to recover the source failed, requiring this apparition to have $S/N<4.5.$  We conclude that either this is a TNO that has flux variations of a factor $\gtrsim 2,$ or the linkage is spurious.  We choose to omit this TNO from sample A (since it was not found in 3 epochs), but include it in sample B (since the other 2 epochs have $S/N>8$), despite some doubt that it is real.  The distance is, in any case, too large to be a cold classical.

\vskip5pt\noindent\textbf{JPB26:} This source was found twice in epoch 1 and once in epoch 2 at $11.3<S/N<15.2,$ but leaves no trace in epoch 3.  It could avoid detection by having $40.3<d_{\rm bary}<48.6$~\au, in which case it would have been in a gap between detectors.  Otherwise it would have appeared in unobscured regions of a NIRCAM SW detector, where it would have needed to be $\gtrsim3\times$ fainter than in the other epochs to avoid detection, which we consider unlikely.  
Under the former interpretation, this object belongs to sample B.
This distance range and its inclination $i<2\degree$ place it nominally among the cold sources, albeit with $e_{\rm min}>0.25.$

\vskip5pt\noindent\textbf{JPB27:} The object is missing from epoch 2 despite $S/N\approx90$ detections in epochs 1 and 3.  The only viable explanation is that the object has $42.4<d_{\rm bary}<42.6$~\au, which causes it to fall into a gap between two SW detectors.  Indeed the object is detectable in HST WFC3 exposures \citep{ana}, which allow a precision determination of $d_{\rm bary}=42.49,$ placing this source into sample B and also among the dynamically cold bodies.

\vskip5pt\noindent\textbf{JPB28:} Detections at $S/N\approx15$ in epochs 1 and 2 indicate that the object would have been in a detector gap in epoch 3 if $43.2<d_{\rm bary}<48.0$~\au.  We take the absence of detection to exclude other distances.  This places the source in sample B, with cold dynamics.

\vskip5pt\noindent\textbf{JPB29:}  Detections at $S/N\approx19$ in epochs 1 and 3 indicate that this TNO 
was outside the NIRCAM fields of view in epoch 2
for any viable distance.  Observed motion rates of motion imply $41.2<d_{\rm bary}<46.9$~\au, placing this source in sample B, with cold dynamics.

\vskip5pt\noindent\textbf{JPB30:}  Detections at $S/N\approx14$ in epochs 2 and 3 indicate that this TNO 
was outside the NIRCAM fields of view in epoch 1 
for any viable distance.  Observed rates of motion constrain the distance poorly: $42.8<d_{\rm bary}<60.2$~\au, although the inclination is always too high to be a cold classical.  The object is in sample B.

\section{Available data}
\label{sec:products}
The NIRCAM images collected by the program are available from the STScI MAST archive under program GO-1568 which can be acced via \href{https://doi.org/10.17909/ekq3-rz42}{10.17909/ekq3-rz42}. We also make available at \href{https://www.canfar.net/citation/landing?doi=26.0008}{10.11570/26.0008} additional files that facilitate further analyses of the moving/transient sources that we have discovered in these images.  A full inventory of the auxiliary files and the meaning of all table columns is in the \texttt{readme.md} file include in the repository.  Briefly, the auxiliary files include:
\begin{itemize}
    \item Tables including cutouts from the individual integrations' images around all of the detected objects in Table~\ref{tab:objects}, plus the positions, rates, and fluxes, and pixelized models that best fit these images in each epoch.  Both SW and LW data are included.
    \item A library of figures giving the key statistics and showing the cutouts and stacked cutouts for each track.
    \item A table giving the results of orbit-level SMP fitting (Section~\ref{sec:OrbitSelection}) to each TNO appearing in 3 epochs.  This includes the individual integrations' cutouts with the models jointly fit to all epochs; the best-fit orbital parameters and fluxes, plus their uncertainties; the epoch-averaged and fully averaged fluxes/magnitudes and LW/SW flux ratios; and their classifications into our samples A and B, and dynamically cold samples.
    \item A library of figures for each 3-epoch TNO detection in the format of Figure~\ref{fig:jpb06}.
    \item A table giving the results of the limiting orbit fitting to the 2-epoch TNOs, plus summarization of their SW and LW fluxes in their observed epochs.
    \item A table summarizing the fluxes and rates of motion of the single-epoch detections listed in Table~\ref{tab:singlets}.
\end{itemize}

\section{Derivation of Likelihood Function}
\label{sec:derivation_likelihood}

For an assumed sky surface density of TNOs vs mean magnitude
\begin{equation}
  \Sigma(m,\theta) = \frac{dN}{dm\,d\Omega}
  \end{equation}
with free parameters $\theta,$ we wish to express the Bayesian posterior
probability $p(\theta | \{\hat m_i\}$ of the parameters conditioned on the survey's
detection of TNOs with measured magnitudes $m_i, i=1,\ldots,M.$ Expressions for
this have been derived---with different results---by \citet{bernstein2004size} and
\citet{loredo_accounting_2004}, among others.  Here we derive what we hope are clarifying
results.

Under Bayes' theorem we will have
\begin{equation}
  p(\theta | \{\hat m_i\}) \propto \likeli(\{\hat m_i\} | \theta) p(\theta)
\end{equation}
with the last term being a prior on the parameters of $\Sigma.$  We concentrate
on deriving the likelihood term $\likeli.$

In our survey and many others, there is a detection/selection step followed by a
measurement step.  For any TNO within the survey area with true mean magnitude
$m$, we can denote successful detection/selection by the condition $s.$  One
outcome, $\tilde s,$  is that the object is not found at all, or is found and
fails the selection criteria.  We denote at $\eta(m) = p(s|m)$ the probability
of successful selection.  For every \emph{selected} source, we also measure a
magnitude $\hat m$ for the source, so there is a joint probability
\begin{equation}
  p(\hat m, s | m) = p(\hat m | s, m) \eta(m).
\end{equation}
\citet{loredo_accounting_2004} assumes that selection is made on the basis of the $\hat m$ value
alone, so that for any value $\hat m$ in some region $S$, we can write $p(s |
\hat m)=1,$  and we can create an unambiguous function $p(\hat m | m)$ for any
$\hat m \in S.$   But in many surveys, there are stochastic characteristics of
astronomical observations that influence selection without changing $\hat m,$ such as whether the orbit brings the source into functioning areas of the detectors, or atop bright sources, and how image noise fluctuations orthogonal to $\hat m$ affect our ResNet outputs.  In general, there is no universal expression as $p(\hat m | m)$ since at least \emph{some} TNOs at $m$ will never get a measured magnitude.

Assuming each TNO's selection and measurement to be independent of other TNO's
existence or outcome, and a surveyed solid angle $\Omega$ with uniform expected
density, we obtain the standard Poisson formulation of the
likelihood:
\begin{align}
  \likeli( \{\hat m_i, s_i\}  | \theta) & = e^{-\bar N(\theta)} \prod_i \Omega\int
                                          dm_i\, p(\hat m_i, s_i | m_i) 
                                          \Sigma(m_i, \theta) 
  \label{eq:like1} \\
   & \propto  e^{-\bar N(\theta)} \prod_i \Omega \int
                                          dm_i\, p(\hat m_i | s_i,m_i) \eta(m_i) 
     \Sigma(m_i, \theta),   \label{eq:like2} \\
   \bar N(\theta) & \equiv \int d\hat m \int dm\, p(\hat m, s | m) \Omega
                    \Sigma(m,\theta) \\
    & = \Omega \int dm \,\Sigma(m,\theta) \int d\hat m \,p(\hat m, s | m) \\
   & = \Omega \int dm \,\Sigma(m,\theta) \eta(m).
\end{align}

We calculated the selection rate $\eta(m)$ in Section~\ref{sec:completeness} using implanted
sources, as is typical for most analyses, and enables calculation of $\bar
N(\theta)$ for any choice of $\theta$ using the last equation.

Knowledge of $\eta(m)$ is, however, formally insufficient to calculate
$\likeli.$  There are two possible approaches.  One is to use implanted sources
to characterize the joint distribution $p(\hat m, s | m),$  and proceed using
Equation~(\ref{eq:like1}). This however, requires a larger number of
simulated sources than estimating the univariate function $\eta(m),$ so we have
not done this.

A second approach is to know or approximate the conditional distribution $p(\hat m | s, m)$ and use Equation~(\ref{eq:like2}).  Injected sources could be used for
this, with the same caveat the joint distribution requires more injections.
We choose instead to make the approximation that $p(\hat m | s,m)$ is narrow,
\ie\ the measurement noise on magnitude is small compared to the scale of
variation of $\Sigma$ against which it is being integrated.  This yields our adopted form of the likelihood,

\begin{equation}
   \likeli( \{\hat m_i, s_i\}   | \theta) \propto  e^{-\bar N(\theta)} \prod_i
 \eta(m_i) \Sigma(m_i, \theta).
 \label{eq:like3}
\end{equation}

\subsection{Comparison with Loredo (2004)}

We note that neither of our exact forms in (\ref{eq:like1}),
(\ref{eq:like2}), nor our approximate form in (\ref{eq:like3}), match the
results of \citet{loredo_accounting_2004}.  He derives his Equation~(8) 
\begin{equation}
  \likeli(\{\hat m_i\} | \theta) = e^{-\bar N(\theta)} \prod_i dm\,p(\hat m_i |
  m) \Sigma(m, \theta).
\label{eq:loredo}
\end{equation}

This form lacks any specific conditioning on $s$ because of it is assumed that 
selection is a deterministic Heaviside function of $\hat m,$ and every TNO has
some measured value $\hat m$ with known $p(\hat m | m)$, such that:
\begin{equation}
  p(\hat m, s | m) = \left\{ \begin{array}{cl}
                               p(\hat m | m) & \hat m < m_{\rm lim} \\
                               0 & \hat m > m_{\rm lim}.
                             \end{array} \right. .
\end{equation}
With this substitution, our Equation~(\ref{eq:like1}) reduces to Loredo's form.

As noted earlier, there is the possibility that the selection is not binary over the space of $\hat m$; $p(\hat m | m)$ is not always well-defined without being joint with or conditioned on selection $s.$   But if it is well defined, it is not the appropriate quantity for the likelihood integration.  As an example, consider the absence of measurement errors so that $p(\hat m | m) = \delta(\hat m-m).$  In this case the distribution of observed magnitudes will be $p(\hat m) = \eta(m) \Sigma(m),$ and the likelihood clearly takes the Poisson form

\begin{equation}
  \likeli(\{\hat m_i\} | \theta) = e^{-\bar N(\theta)} \prod_i \Omega
  \eta(\hat m_i)\Sigma(\hat m_i,\theta),
\end{equation}
containing a factor of $\eta(m)$ in the integrand that is missing from Loredo's
form, but is present in Equation~(\ref{eq:like2}).  
The missing factor can be traced to a small error in \citet{loredo_accounting_2004} where when generalizing their Equation~(5), a factor of $\eta(m)$ is added for the mean $\bar N$ but is not also added to under the product part of the Poisson likelihood in his Equation~(8).

To summarize, the correct form of the observational likelihood (and hence the Bayesian posterior derived from it) is given by either Equation~(\ref{eq:like1}) using the joint $p(\hat m, s | m)$ or Equation~(\ref{eq:like2}) using the equivalent quantity $p(\hat m | s,m) \eta(m).$  If the joint or conditional distribution is not available, Loredo's form, Equation~(\ref{eq:loredo}), is equivalent in the limit where the selection probability is a binary function of the observed $\hat m,$ \ie\ the measurement noise on $m$ determines the selection.  A different limit, when the measurement noise is small compared to the range of $m$ over which selection $\eta(m)$ and $\Sigma(m)$ vary, admits the approximation yielding Equation~(\ref{eq:like3}), used by \citet{bernstein2004size}.  We point out that in the current work, these subtleties in the treatment of the joint distribution make insignificant difference to the inference, since there are few counts in the magnitude regime where $\eta(m)$ transitions from 0 to its maximum value.

\section{Object Classification and Spectral Types}
\label{sec:object_class}

Table \ref{tab:object_class} lists the objects orbital and spectral classification. We limit our compositional estimates to objects with three epoch detections, since their orbital solutions are well constrained.

\begin{deluxetable}{lcc}
\tablecaption{Object Classification and Spectral Types \label{tab:object_class}}
\tablehead{
\colhead{Object} & \colhead{Cold Classical?} & \colhead{Spectral Type}
}
\startdata
JPB01 & Yes & Organics \\
JPB04 & Yes & Organics \\
JPB05 & Yes & Organics \\
JPB06 & Yes & Organics \\
JPB07 & Yes & Organics \\
JPB08 & Yes & Organics \\
JPB09 & No  & $\text{H}_{2}\text{O}$/$\text{CO}_2$ \\
JPB10 & No  & $\text{H}_{2}\text{O}$ \\
JPB11 & No  & $\text{H}_{2}\text{O}$/$\text{CO}_2$ \\
JPB12 & Yes & Organics \\
JPB13 & Yes & Organics \\
JPB14 & No  & $\text{CO}_2$  \\
JPB15 & Yes & Organics \\
JPB16 & Yes & Organics \\
JPB17 & No  & Organics \\
JPB18 & No  & $\text{H}_{2}\text{O}$ \\
JPB19 & No  & $\text{H}_{2}\text{O}$ \\
JPB20 & Yes & Organics \\
JPB21 & Yes & Organics \\
JPB22 & No  & $\text{H}_{2}\text{O}$ \\
JPB23 & Yes & Organics \\
JPB27\tablenotemark{a} & Yes & Organics \\
\enddata
\tablecomments{
Objects labeled as $\text{H}_{2}\text{O}$/$\text{CO}_2$ have uncertainties large enough that they could either be $\text{H}{2}\text{O}$ or $\text{CO}_2$-type. Therefore, we use the mean $r - F150W2$ value of the two types for those objects.
}

\tablenotetext{a}{This object was detected in two JWST epochs and independently in HST WFC3 images (see Section \ref{sec:elements}), allowing us to constrain its orbital solution; nevertheless, we exclude it from our luminosity function fits.
}

\end{deluxetable}



\end{document}